\pdfoutput=1
\documentclass[twocolumn]{aastex62}

\usepackage{natbib}
\usepackage{footnote}
\usepackage{tabularx}
\usepackage{longtable, tabu}
\usepackage{tablefootnote}
\usepackage{graphicx}
\usepackage{amsmath}
\usepackage{subfigure}
\usepackage{lipsum}
\usepackage{float}
\usepackage{color}
\usepackage{array,booktabs}
\usepackage[toc,page]{appendix}

\newcounter{ltfootnote}

\makeatother

\newcolumntype{n}{>{\centering\arraybackslash}X}

\def\K2{{\slshape K2}}
\def\spitzer{{\slshape Spitzer}}
\def\most{{\slshape MOST}}
\def\HST{{\slshape HST}}
\def\WFC3{{\slshape WFC3}}
\def\STIS{{\slshape STIS}}
\def\JWST{{\slshape JWST}}
\def\TESS{{\slshape TESS}}
\def\NIRSpec{{NIRSpec}}
\def\NIRISS{{\slshape NIRISS}}
\def\Gaia{{\slshape Gaia}}
\def\Chandra{{\slshape Chandra}}
\def\XMM{{\slshape XMM-Newton}}

\newcommand{\HL}[1]{{#1}} 

\begin{document}

\title{Updated Parameters and a New Transmission Spectrum of HD~97658b}

\correspondingauthor{Ian J.\ M.\ Crossfield}
\email{ianc@ku.edu} 

\author{Xueying Guo}
\affiliation{MIT Kavli Institute for Astrophysics \& Space Research, Cambridge, MA 02139, USA}

\author{Ian J.\ M.\ Crossfield}
\affiliation{MIT Kavli Institute for Astrophysics \& Space Research, Cambridge, MA 02139, USA}
\affiliation{The University of Kansas, Department of Physics and Astronomy, Malott Room 1082, 1251 Wescoe Hall Drive, Lawrence, KS, 66045, USA}

\author{Diana~Dragomir}
\affiliation{MIT Kavli Institute for Astrophysics \& Space Research, Cambridge, MA 02139, USA}
\affiliation{Department of Physics and Astronomy, University of New Mexico, Albuquerque, NM, USA}

\author{Molly~R.~Kosiarek}
\altaffiliation{NSF Graduate Research Fellow}
\affiliation{Department of Astronomy and Astrophysics, University of California, Santa Cruz, CA, USA}

\author{Joshua~Lothringer}
\affiliation{Department of Physics and Astronomy, Johns Hopkins University, 3400 North Charles Street, Baltimore, MD 21218, USA}

\author{Thomas~Mikal-Evans}
\affiliation{MIT Kavli Institute for Astrophysics \& Space Research, Cambridge, MA 02139, USA}

\author{Lee~Rosenthal}
\affiliation{California Institute of Technology, Department of Astronomy, 1200 E California Blvd., Pasadena, CA 91125, USA}

\author{Bjorn~Benneke}
\affiliation{Departement de Physique, and Institute for Research on Exoplanets, Universite de Montreal, Montreal, H3T J4, Canada}

\author{Heather~A.~Knutson}
\affiliation{Division of Geological and Planetary Sciences, California Institute of Technology, Pasadena, CA 91125, USA}

\author{Paul~A.~Dalba}
\altaffiliation{NSF Astronomy and Astrophysics Postdoctoral Fellow}
\affiliation{Department of Earth and Planetary Sciences, UC Riverside, 900 University Ave., Riverside, CA 92521, USA}

\author{Eliza~M.~R.~Kempton}
\affiliation{Department of Astronomy, University of Maryland, College Park, MD 20742, USA}
\affiliation{Department of Physics, Grinnell College, 1116 8th Avenue, Grinnell, IA 50112, USA}

\author{Gregory~W.~Henry}
\affiliation{Center of Excellence in Information Systems, Tennessee State University, Nashville, TN  37209, USA}

\author{P.~R.~McCullough}
\affiliation{Department of Physics and Astronomy, Johns Hopkins University, 3400 North Charles Street, Baltimore, MD 21218, USA}
\affiliation{Space Telescope Science Institute, 3700 San Martin Drive, Baltimore, MD 21218, USA}

\author{Travis~Barman}
\affiliation{Lunar \& Planetary Laboratory, University of Arizona, 1629 E. University Blvd, Tucson, AZ 85721, USA}

\author{Sarah~Blunt}
\altaffiliation{NSF Graduate Research Fellow}
\affiliation{California Institute of Technology, Department of Astronomy, 1200 E California Blvd., Pasadena, CA 91125, USA}

\author{Ashley~Chontos}
\altaffiliation{NSF Graduate Research Fellow}
\affiliation{Institute for Astronomy, University of Hawaii, 2680 Woodlawn Drive, Honolulu, HI, USA}

\author{Jonathan~Fortney}
\affiliation{Department of Astronomy and Astrophysics, University of California, Santa Cruz, CA, USA}

\author{Benjamin~J.~Fulton}
\affiliation{California Institute of Technology, Department of Astronomy, 1200 E California Blvd., Pasadena, CA 91125, USA}

\author{Lea~Hirsch}
\affiliation{Kavli Institute for Particle Astrophysics and Cosmology, Stanford University, Stanford, CA, USA}

\author{Andrew~W.~Howard}
\affiliation{California Institute of Technology, Department of Astronomy, 1200 E California Blvd., Pasadena, CA 91125, USA}

\author{Howard~Isaacson}
\affiliation{Astronomy Department, University of California, Berkeley, CA, USA}

\author{Jaymie~Matthews}
\affiliation{Department of Physics and Astronomy, University of British Columbia, Vancouver, Canada}

\author{Teo~Mocnik}
\affiliation{Department of Earth and Planetary Sciences, UC Riverside, 900 University Ave., Riverside, CA 92521, USA}

\author{Caroline~Morley}
\affiliation{Department of Astronomy, University of Texas, Austin, TX 78712, USA}

\author{Erik~A.~Petigura}
\affiliation{California Institute of Technology, Department of Astronomy, 1200 E California Blvd., Pasadena, CA 91125, USA}

\author{Lauren~M.~Weiss}
\altaffiliation{Beatrice Watson Parrent Fellow}
\affiliation{Institute for Astronomy, University of Hawai‘i, Honolulu, HI, USA}


\begin{abstract}

Recent years have seen increasing interest in the characterization of sub-Neptune sized planets because of their prevalence in the Galaxy, contrasted with their absence in our solar system. HD~97658 is one of the brightest stars hosting a planet of this kind, and we present the transmission spectrum of this planet by combining four \HST\ transits, twelve \spitzer\ IRAC transits, and eight \most\ transits of this system. 
Our transmission spectrum has higher signal to noise ratio than that from previous works, and the result suggests that the slight increase in transit depth from wavelength 1.1 to 1.7 microns reported in previous works on the transmission spectrum of this planet is likely systematic. Nonetheless, our atmospheric modeling results are not conclusive as no model provides an excellent match to our data. Nonetheless we find that atmospheres with high C/O ratios ($\rm C/O\gtrsim 0.8$) and metallicities of $\gtrsim 100\times$ solar metallicity 
are favored. We combine the mid-transit times from all the new \spitzer\ and \most\ observations and obtain an updated orbital period of $P=9.489295 \pm 0.000005$, with a best-fit transit time center at $T_0 = 2456361.80690 \pm 0.00038$~(BJD). 
No transit timing variations are found in this system. We also present new measurements of the stellar rotation period ($34\pm2$~d) and stellar activity cycle (9.6~yr) of the host star HD~97658. 
Finally, we calculate and rank the Transmission Spectroscopy Metric of all confirmed planets cooler than 1000~K and with sizes between 1~$\rm R_{\oplus}$ and 4~$\rm R_{\oplus}$. We find that at least a third of small planets cooler than 1000~K can be well characterized using \JWST, and of those, HD~97658b is ranked fifth, meaning it remains a high-priority target for atmospheric characterization.

\end{abstract}


\bigskip

\section{Introduction}

With abundant candidate planets and confirmed planets being identified through various exoplanet surveys, efforts have been made to measure their mass and density, and to detect and characterize their atmospheres. Among all confirmed planets, sub-Neptune sized planets ($2-4~R_{\oplus}$)  are of great interest because of their absence in the solar system yet abundance in the galaxy \citep{Fressin2013, Howard2012}, and their role of connecting the formation scenario of larger gaseous planets and smaller terrestrial-sized planets \citep{Crossfield2017}.

HD~97658b is a sub-Neptune of 2.4~$R_{\oplus}$ radius discovered  with Keck-HIRES in the NASA-UC Eta-Earth Survey \citep{Howard2011}, and later found to transit by \cite{Dragomir2013} using the \most\ telescope. It orbits a bright ($V=7.7$) K1 star with a 9.5~day period, and was ranked the 6th best confirmed planet for transmission spectroscopy with $R_{\rm p}<5~R_{\oplus}$ in \cite{Rodriguez2017}. HD~97658b was also monitored by the \spitzer\ Space Telescope, and \cite{VanGrootel2014} reported the photometric analysis result, as well as a global Bayesian analysis result combing the \spitzer, \most, and Keck-HIRES data. They found that HD~97658b has an intermediate density of $\rm 3.90^{+0.70}_{-0.61}~g/cm^3$, indicating a rocky composition of at least 60\% by mass, around 0\%--40\% of water and ice, and a H-He dominated envelope of at most 2\% by mass \citep{VanGrootel2014}.

Transmission spectroscopy is one of the most effective ways of constraining planet atmospheres, along with emission spectroscopy and phase curve analysis. This method has been widely applied to the atmospheric characterization of large gaseous planets, yet to this day, no more than half a dozen of the planets smaller than $4~R_{\oplus}$ have had published transmission spectra \citep{Kreidberg2014a, Dragomir2013, Southworth2017, Benneke2019}. \cite{fu2017} and \cite{Crossfield2017} proposed linear relationships between measured spectral amplitudes and planet equilibrium temperatures and H/He mass fractions, which could be better evaluated and constrained by decreasing the uncertainty amount of each transmission spectrum with more observations. 

Based on Wide Field Camera 3 (\WFC3\ hereafter) observations during two \HST\ visits in 2013 and 2014, \cite{Knutson2014} reported the first transmission spectrum of HD~97658b from 1.1~$\mu$m to 1.7~$\mu$m. By comparing a range of atmospheric models to the transmission spectral data, they ruled out clear atmospheres with 50$\times$solar or lower metallicity and pure water+hydrogen atmospheres with $\leq$10\% fraction of water.

We obtained two more \HST/\WFC3\ observations of HD~97658b and eleven additional \spitzer\ transits of the planet. In addition, ten transits were observed with the Direct Imaging mode of the \most\ telescope, and three transits were observed with the \it Space Telescope Imaging Spectrograph \rm (\STIS\ hereafter) on \HST\ using its G750L grism. In this work, we analyze all these datasets. 
With the extracted transit depths and the combined transmission spectrum from 1.1~$\mu$m to 1.7~$\mu$m, where molecular features including water, methane, carbon dioxide and carbon monoxide can be present, we test atmosphere retrievals as well as forward modeling methods to explore plausible atmospheric models and discuss their statistical significance. 

Data reduction and transit analysis of \HST/\WFC3, \spitzer, \most, and \STIS\ observations are described in section~\ref{sec:HST_analysis} and section~\ref{sec:other_bandpass}. In section~\ref{sec:RV_ephemeris}, we present TTV analysis results using multiple ephemerides and updated RV measurements, as well as a discussion of the system's newly-identified stellar activity cycle. Atmosphere property retrievals and forward modeling are discussed in section \ref{sec:atm_retrieve}, and we conclude our findings and discuss future prospects in section \ref{sec:discussion}.

\bigskip

\section{\HST/\WFC3\ Data Analysis}\label{sec:HST_analysis}

\subsection{Raw Data Reduction}

\begin{figure}
\begin{center}
    \subfigure
	{%
	\label{fig:first}
	\includegraphics[width=0.22\textwidth]{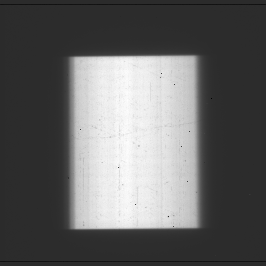}
	}
    \subfigure
	{%
	\label{fig:first}
	\includegraphics[width=0.22\textwidth]{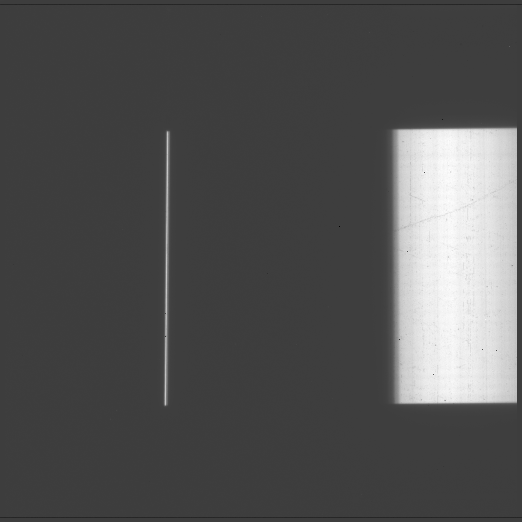}
	}
\end{center}
\caption{Left: a typical raw image from visit1 and visit2 (256$\times$256 pixel subarray). Right: a typical raw image from visit3 and visit4 (512$\times$512 pixel subarray). The zeroth order image is the bright thin line at the center, and the right edge of the dispersion image of the latter two visits goes out of the subarray. This is an unexpected observation error, which results in the loss of two channels in the transmission spectra extraction from visit3 and visit4.}
\label{fig:typical_ima}
\bigskip
\end{figure}

HD~97658b was observed on 12/19/2013 (visit1) and 01/07/2014 (visit2) under the \HST\ program 13501 (PI: Knutson), and then on 04/12/2016 (visit3) and 01/31/2017 (visit4) under the \HST\ program 13665 (PI: Benneke). A spatial scan mode is adopted for all four visits to accumulate abundant photons on the detector. We use the ``round trip'' scan method, which means the scan is carried out in two opposite directions alternatively (one direction per exposure) during the observations, and during each scan (one exposure), the image on the detector is read out several times. Each visit of program 13501 was observed with a 256$\times$256 pixel subarray, consisting of around 100 14-sec exposures in each scan direction and 3 read outs in each exposure, while each visit of program 13665 was observed with a 512$\times$512 pixels subarray, consisting of around 100 23-sec exposures in each scan direction and 2 read outs in each exposure. Two typical raw images from program 13501 and program 13665 are shown in Figure \ref{fig:typical_ima}, and raw data from all visits are processed with a standard pipeline described as follows. 

\begin{figure*}
\begin{center}
    \subfigure
	{%
	\label{fig:first}
	\includegraphics[width=0.48\textwidth]{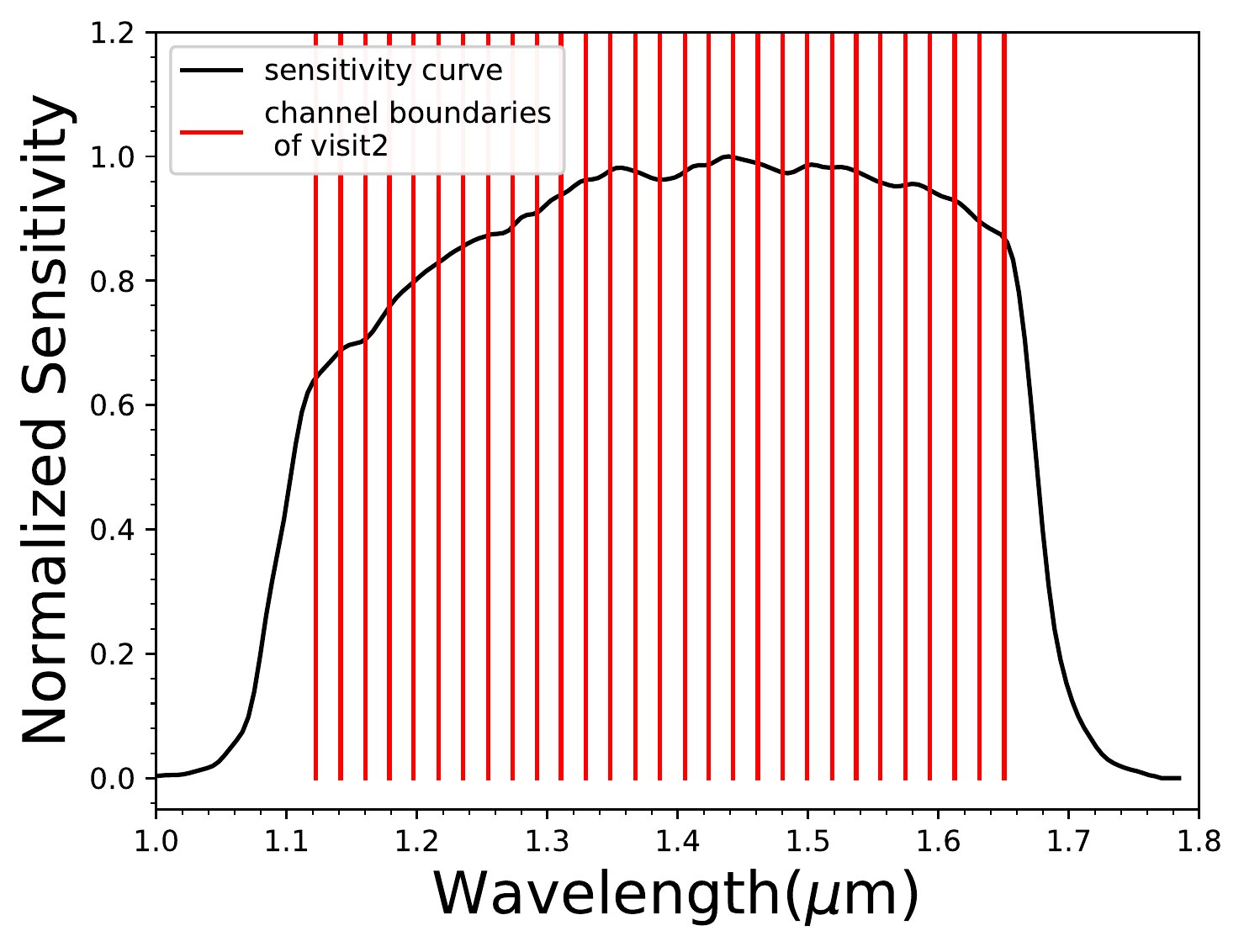}
	}
    \subfigure
	{%
	\label{fig:first}
	\includegraphics[width=0.48\textwidth]{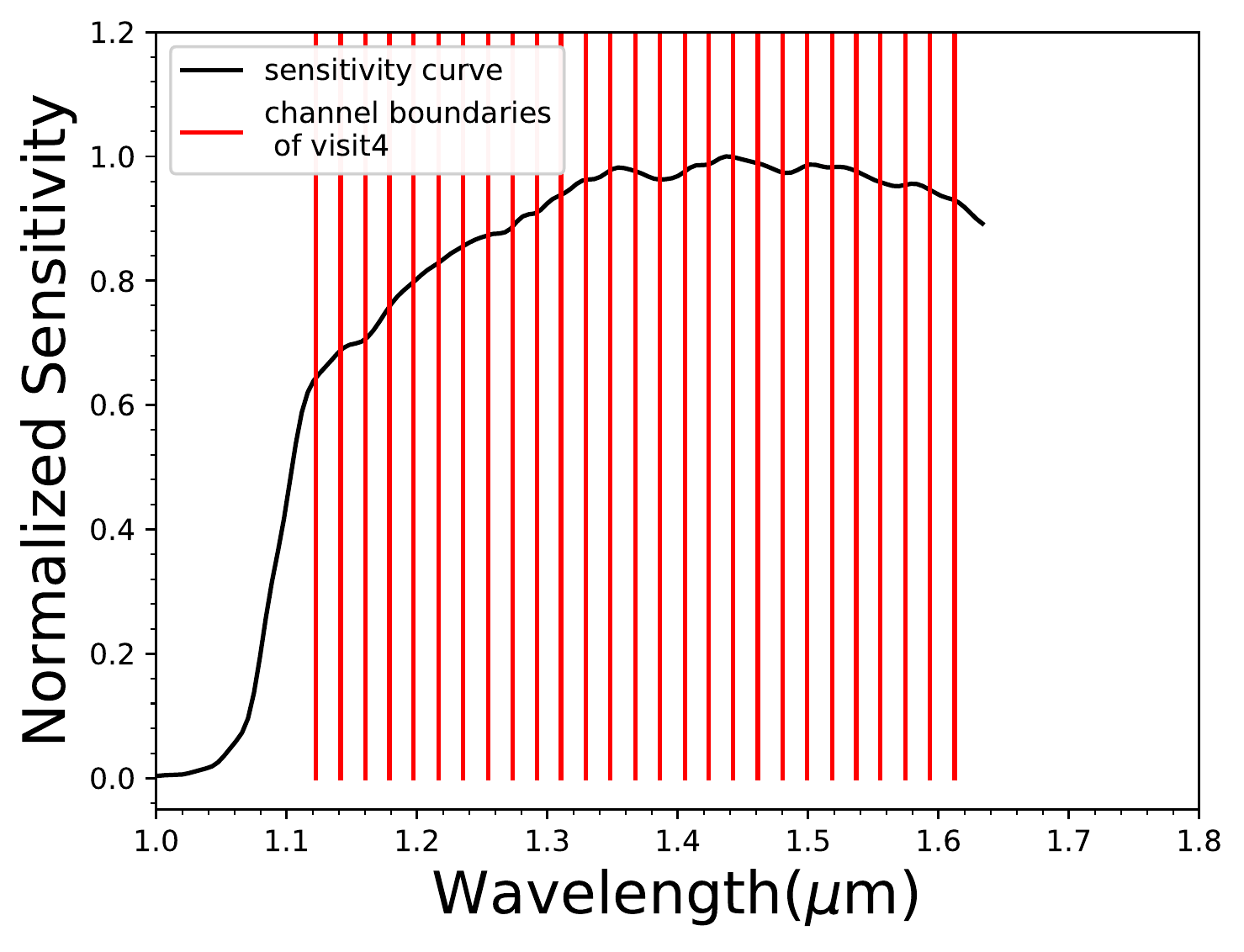}
	}
\end{center}
\caption{\HST/\WFC3\ sensitivity curves and our channel cuts for transmission spectra extraction. All channel boundaries are selected to be identical to those in \cite{Knutson2014}. The left panel shows the 28 channels applied on visit1 and visit2, and in the right panel, the two reddest channels are dropped because the observed flux goes out of the CCD border in visit3 and visit4.}
\label{fig:channels}
\bigskip
\end{figure*}

\HL{We start the data analysis from the \HST/\WFC3\ IMA files.} For each exposure, \HL{we mask out bad pixels that were identified with flags from the \WFC3/IR bad pixel table \citep{WFC3_Badpixel_table}}. Then we select a ``clean'' rectangle on the image -- where detected flux roughly flattens spatially -- as the background and take the median flux in that rectangle as our background flux level $f_{bkg}$. Next, we subtract $f_{bkg}$ from the whole image and correct the image with the data cube from STScI.

Next, mean flux along the scanning direction is calculated for each wavelength pixel in each exposure, thus a raw spectrum is produced for each exposure. And finally we correct for the wavelength shift on the detector over time by picking the spectrum from the first exposure as the template, and shifting all subsequent exposures along the dispersion direction to match the template.

The wavelength solutions, which translate pixel values on the detector to wavelengths, are calculated using the wavelength calibration coefficients from STScI, and we select the range of pixels in the dispersion direction such that our wavelength range coincides with that chosen by \citet{Knutson2014} for easy comparison. We then obtain the raw white light curve of each visit by summing up all flux in the dispersion range in each exposure. 

\begin{figure*}
\begin{center}
    \subfigure
	{%
	\label{fig:first}
	\includegraphics[width=0.48\textwidth]{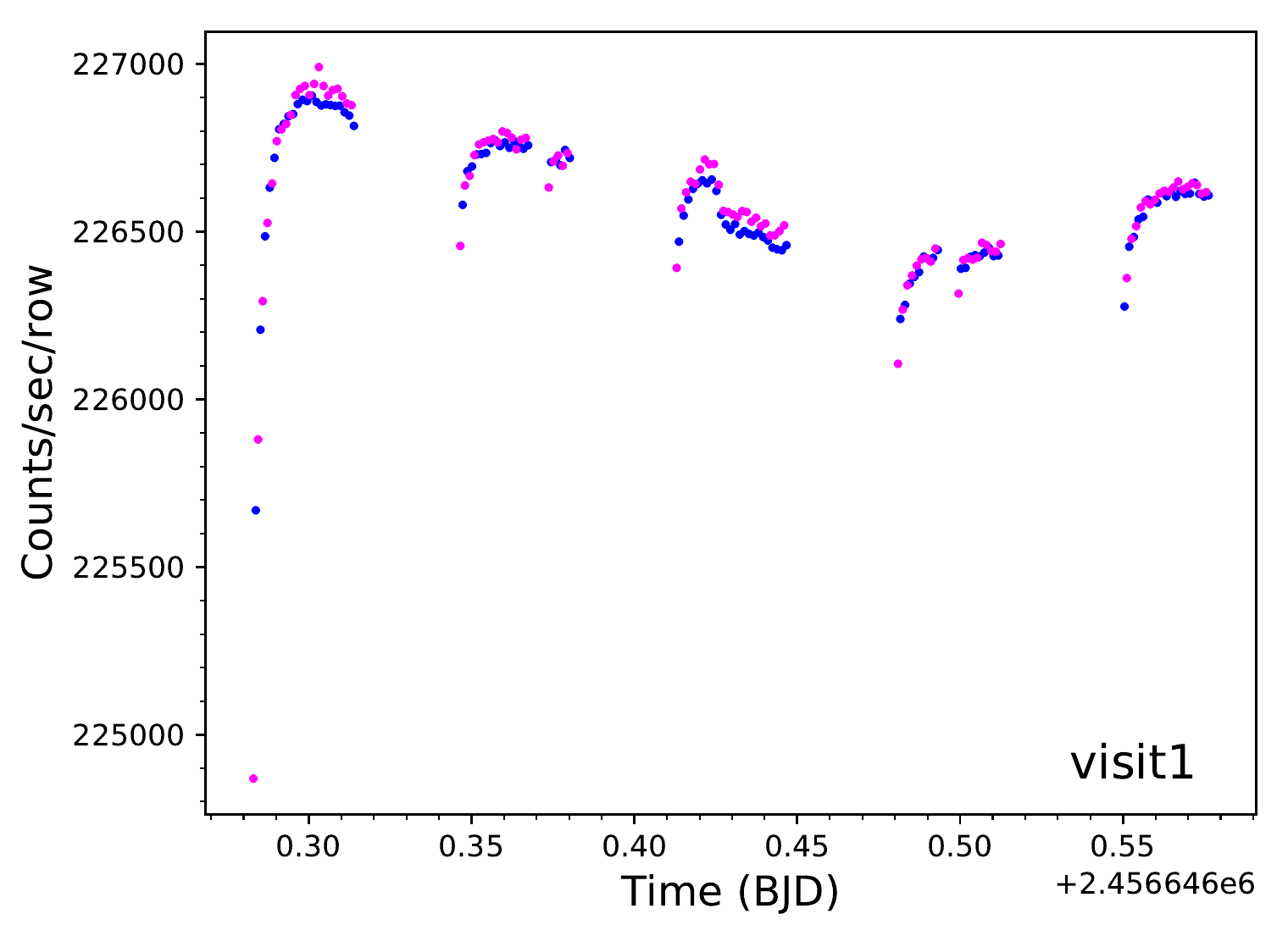}
	}%
    \subfigure
	{%
	\label{fig:first}
	\includegraphics[width=0.48\textwidth]{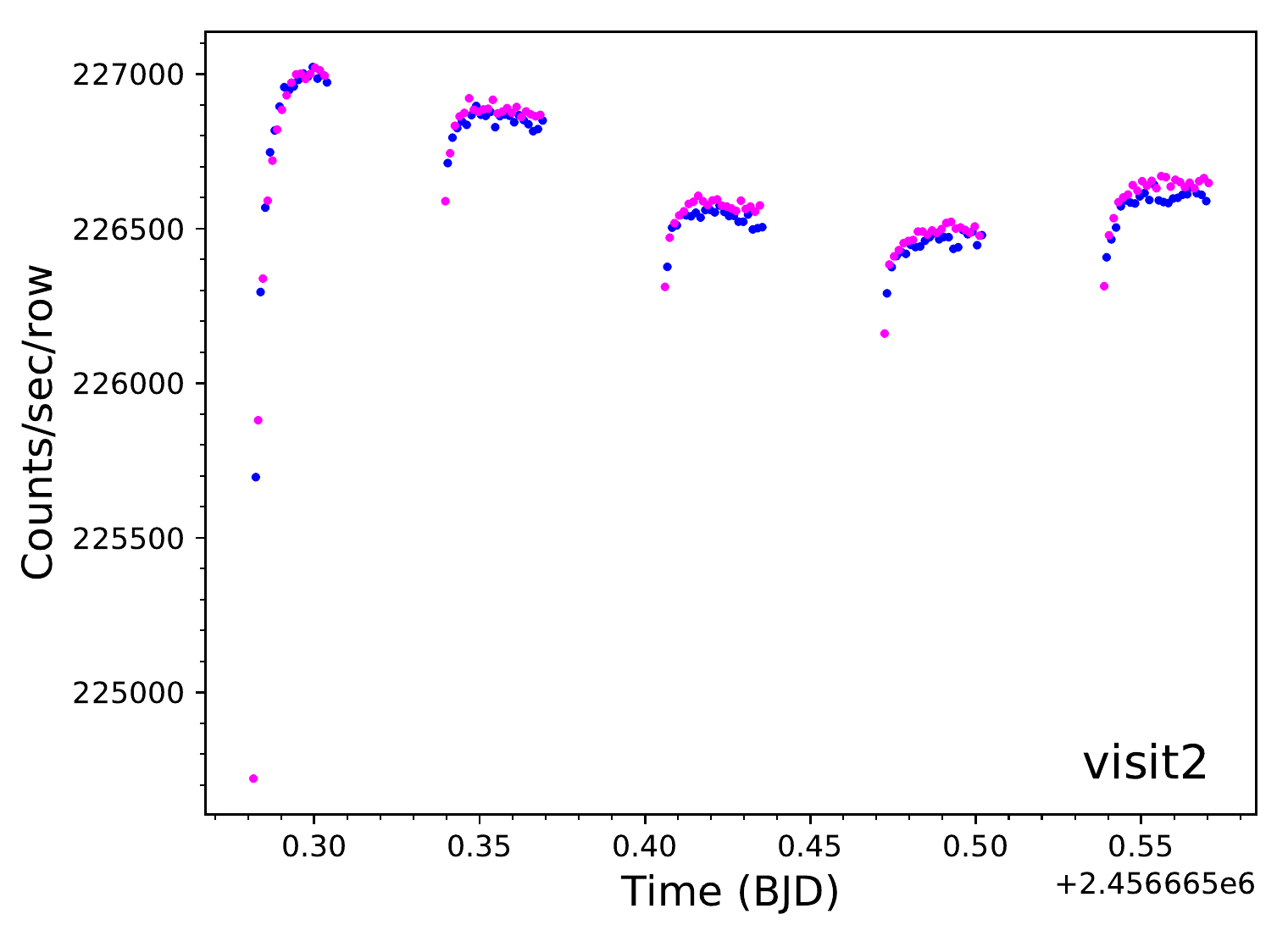}
	}\\
	\subfigure
	{%
	\label{fig:first}
	\includegraphics[width=0.48\textwidth]{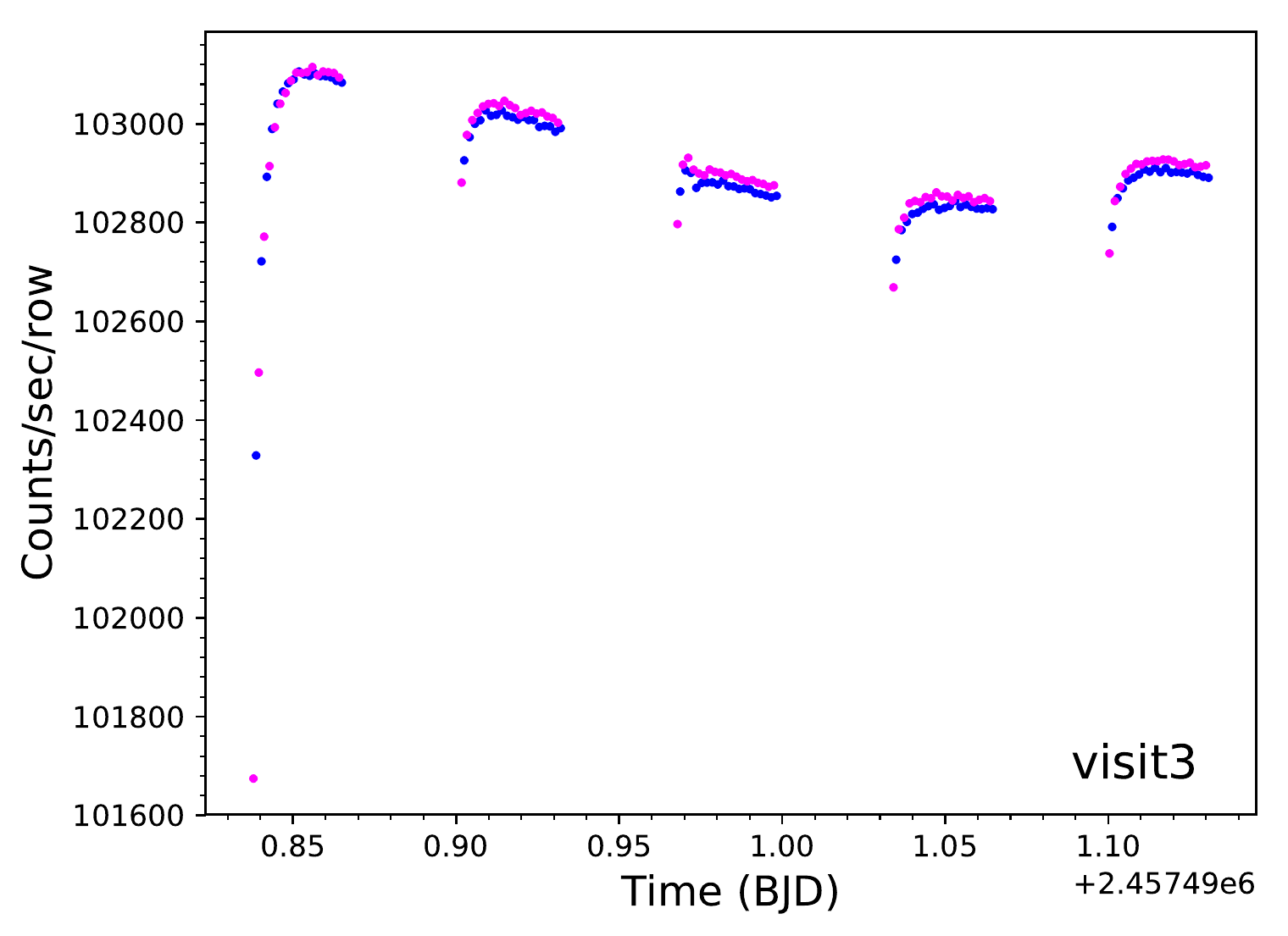}
	}%
    \subfigure
	{%
	\label{fig:first}
	\includegraphics[width=0.48\textwidth]{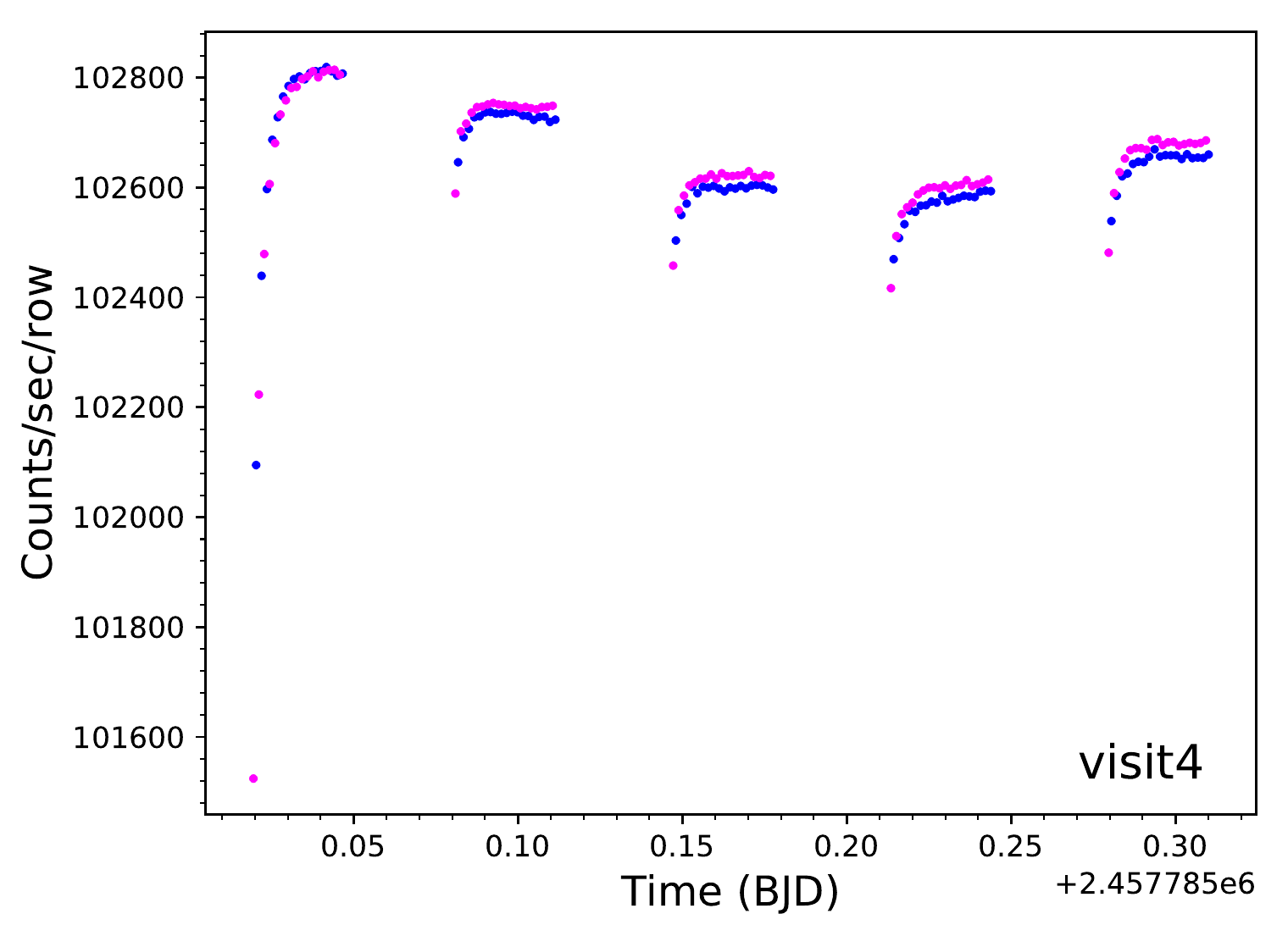}
	}
\end{center}
\caption{Raw white light curves observed during the four visits. Each visit is labeled at the lower right corner of its panel. The blue and magenta data points represent forward and reverse scan directions respectively.}
\label{fig:lcs}
\bigskip
\end{figure*}

Lastly, we correct for the difference in flux levels between the two scan directions of each visit, which results from the dependence of the total flux on the vertical position of the spectrum relative to the middle row of the detector (the ``up-stream/down-stream'' effect; \citet{McCullough2012}). Following a similar process to the one described in \citet{Tsiaras2016}, we fit the flux profile in the scan direction of each exposure with a box shape to extract its read-out length and mid-position, which should be related by two different linear relations for the two scan directions. Then we fit the linear relations and choose the first exposure as our reference point to scale the flux of all subsequent exposures to the level where they should be if they all had the same read-out mid-position as the first exposure. Our final white light curve of each visit is shown in Figure \ref{fig:lcs}.

\subsection{White Light Curve Analysis}

Major systematic effects in the \HST/\WFC3\ observation data include: (1) the ``ramp'' effect in each \HST\ orbit, which is thought to be caused by free charge carriers trapped in the depletion regions of the detector; (2) visit-long trends, which is a quasilinear trend across the entire observation period; and (3) ``\HST\ breathing effect'', caused by the spacecraft temperature variation during each orbital period of \HST\ \citep{Wakeford2016}.

\subsubsection{Orbital Ramp Effects}

The orbital ramp is one of the hardest \HST\ observational systematic effects to characterize. A traditional method to correct for this effect is to apply an empirical exponential model to the \HST/\WFC3\ data sets, proposed by \citet{Berta2012}, and this method has been used in a number of previous works \citep{Line2013, Knutson2014, Kreidberg2014a, Kreidberg2014b}. Subsequently, other empirical methods (such as a polynomial model correction) have also been proposed \citep{Wakeford2013}. 

Since empirical models are not based on a good understanding of the physical processes that are causing the systematics, they are hard to compare and evaluate. A marginalization method proposed by \citet{Gibson2014} was implemented by \citet{Wakeford2016} to remove \HST/\WFC3\ systematics. The method combines the best-fit results from 52 polynomial/exponential models by calculating and assigning a weight to each one. And from another perspective, \citet{Zhou2017} described a method named \it Ramp Effect Charge Trapping Eliminator \rm (RECTE hereafter), which models the intrinsic physical process of charge trapping on the \WFC3\ detector with a set of equations and parameters. The RECTE model has been successfully applied to the \HST/\WFC3\ observations of a range of exoplanets and brown dwarfs  \citep{Zhou2017}, and since RECTE is computationally efficient we  adopt this method to correct for the orbital ramp effects in our \HST/\WFC3\ data set.

An idealized charge carrier trapping process can be described by equations (1)-(4) in \citet{Zhou2017}, which contain 11 free parameters. Although the original paper commented that 6 of the 11 parameters can be fixed to their best-fit values for all \HST/\WFC3\ observations, our tests on a range of parameter settings show that 2 of the 6 parameters, $\eta_{\rm s}$ and $\eta_{\rm f}$, which describe the efficiencies with which charge carriers can fill the traps, converge to different values from those provided in \citet{Zhou2017}, and the best-fit values change for different transit visits. Table \ref{tab:eta_values} shows the comparison of best-fit $\eta_{\rm s}$ and $\eta_{\rm f}$ values of each \HST\ visit against the best-fit values presented in \citet{Zhou2017}. Therefore, we decide that $\eta_{\rm s}$ and $\eta_{\rm f}$ should be set as free parameters when analyzing this data set. 

Although \citet{Zhou2017} states that the RECTE method can model the ramp effect well in all orbits in a visit, including the first orbit, our tests show that for our HD~97658b observations, the ramp effect in the first orbits cannot be well modelled with RECTE. The fact that the brightness of HD~97658 is approaching the saturation limit of the \WFC3\ detector may be a reason. Therefore, we apply RECTE only to the rest of the orbits of each visit, and the first orbit of each visit is removed from the ramp-effect modeling. 

\begin{table*}
\renewcommand*{\arraystretch}{1.2}
    \caption{Best-fit $\eta_{\rm s}$ and $\eta_{\rm f}$ from the White Light Curve of Each \HST/\WFC3\ Visit}
    \centering
    \begin{tabularx}{\textwidth}{cnnnnn}
    \hline\hline
     & \citet{Zhou2017} & visit1 & visit2 & visit3 & visit4 \\
    \hline
    $\eta_{\rm s}$ & $0.01320\pm0.00003$ & $0.019\pm 0.001$ & $0.011\pm0.001$ & $0.0030\pm 0.004$ & $0.015\pm 0.002$\\
    $\eta_{\rm f}$ & $0.00686\pm0.00007$ & $0.0004\pm 0.0003$ & $0.0033\pm 0.0002$ & $0.0044\pm 0.0003$ & $0.0044\pm0.0002$\\
    \hline
    \end{tabularx}
    \label{tab:eta_values}
\bigskip
\end{table*}

\subsubsection{Residual Systematics and Noise Modeling}\label{sec:ResidualModel}

We present our treatment and discussions of visit-long trend corrections in the next section, along with the best-fit white light curve transit signals.  

In addition to orbital ramp effects removal and visit-long-trend corrections, we treat any residual systematics, including the ``\HST\ breathing effect'' and other red-noise sources that do not have a certain functional form, with a Gaussian process (GP hereafter). GP has been successfully applied to \WFC3\ data analysis in previous works \citep{Evans2016, Evans2017, Evans2018, Gibson2012HD189733b} by fitting posteriors with a multivariate Gaussian distribution. In this work, we model the noise with a Gaussian process assuming a Matern-3/2 kernel using the Celerite package \citep{celerite}. The kernel amplitude and correlation timescale are set as free parameters, complemented with another free parameter to describe \HL{the magnitude of the white noise component}. The noise model is fitted simultaneously with the mean photometric model to extract posteriors of all parameters. 

To ensure a realistic parameter uncertainty level, we scale the photometric uncertainty associated with each data point in each transit visit so that the final best-fit model gives a reduced chi-square of 1.

\subsubsection{Visit-long Trend}\label{sec:visit_long_trend}

Visit-long trend corrections can be combined into the base-level flux parameter $f$ in the RECTE model, which becomes $(f_0 + bt)$ in a linear-shape visit-long trend case, where $f_0$ and $b$ are free parameters, and $t$ represents the time stamp of each exposure. Although most previous works have used a simple linear trend \citep{Knutson2014, Evans2018}, in this data set of HD~97658b, we observe evidence of a long-term trend that deviates from a linear form from visual examination of the white light curves (Figure \ref{fig:lcs}), which may have been induced or magnified by the high brightness of the host star that is close to the \WFC3\ detector saturation limit. 

Therefore, we try four common functional forms to correct for the visit-long trend: \\
\noindent
(1) linear, with the form $(f_0 + bt)$;\\
(2) quadratic, with the form $(f_0 + bt + ct^2)$; \\
(3) exponential, with the form $(f_0 - b\exp(-t/c))$; \\
(4) logarithmic, with the form $(f_0 - b\log(t+c))$. \\

\noindent
When applying these functional forms, we make use of the first orbit to help anchor the  trend by modeling the second half of its data points, which are less affected by the ramp effect because all charge traps should have been filled. To sum up, the orbits after the first orbit are modelled with function $F_{1} = f_{ramp} \times f_{trend} \times f_{transit}$, where $f_{ramp}$, $f_{trend}$ and $f_{transit}$ represent the ramp effect model, the visit-long trend model, and the transit model respectively, while the second half of the first orbits are modelled with the function $F_{0} = f_{trend} \times f_{transit}$. The data from the first half of the first orbits are discarded. \\

Simultaneously with instrumental systematic models, we fit the transit signal with models generate using the BAsic Transit Model cAlculatioN (BATMAN) package provided by \cite{Kreidberg2015}. In our joint fit of the white light curves,  the transit depth, inclination and orbital semi-major axis are tied to be the same free parameters for all four visits, while the mid-transit time of each visit is a separate free-floating parameter. We adopt fixed stellar parameters reported in Gaia Data Release 2 \citep{GaiaDR2}, and the 2-parameter limb-darkening coefficients are extracted from \citet{Claret2011} assuming these stellar parameters are fixed in our models. 

\begin{figure}
\begin{center}
    \subfigure
	{%
	\label{fig:first}
	\includegraphics[width=0.49\textwidth]{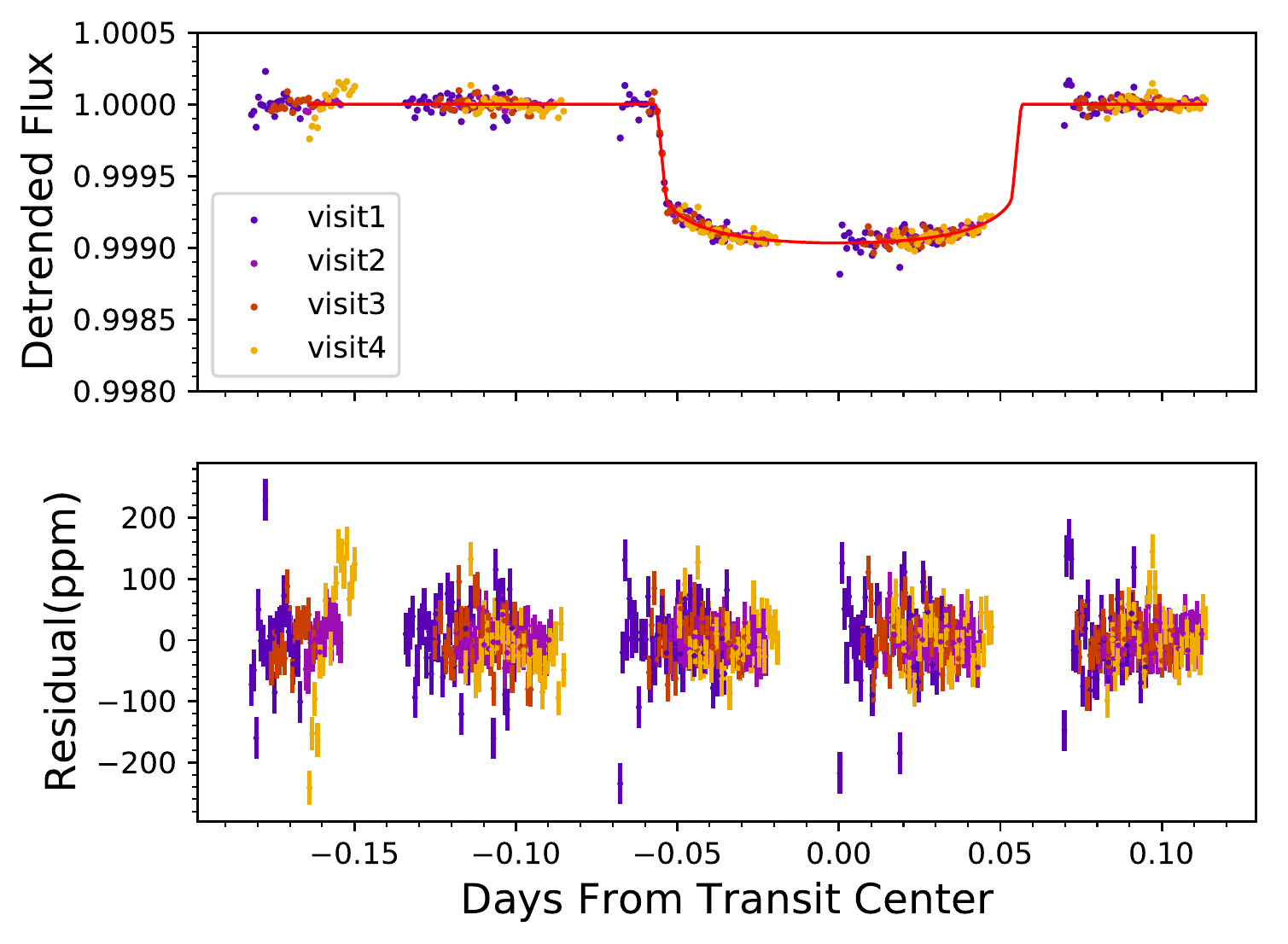}
	}
\end{center}
\caption{Best-fit transit light curves from all four visits stacked together and shifted so that the the mid-transit times are at 0. Here we only show the results when assuming a logarithmic visit-long trend.}
\label{fig:bestfit}
\bigskip
\end{figure}

The best-fit transit model and detrended white light curves of all four visits assuming a logarithmic shape visit-long trend are shown in Figure \ref{fig:bestfit}, with input parameters and best-fit output parameters shown in Table \ref{tab:whitelc_params}. \HL{We can see the model does not fit the second half of the first orbit very well, which is expected because the ramp effect component was excluded from the modeling of this section of the data, and this section of the data was only in the analysis to provide an anchor for modeling the visit-long trends. Apart from this, there are a few outliers in the detrended visit1 light curve, but they 
do not bias our overall science results of this planet.} The best-fit white light curves assuming other forms of visit-long trends are of similar quality, and we discuss those results as follows.

\begin{deluxetable*}{cccc}
\renewcommand*{\arraystretch}{1.2}
\tablecaption{Input and best-fit parameters from \HST/\WFC3\ white light curves fitting\label{tab:whitelc_params}}
\tablehead{
  \colhead{Parameter} & 
  \colhead{Symbol} & 
  \colhead{Value} & 
  \colhead{Unit}
}
\startdata
\sidehead{\bf{Input fixed Parameters}}
  Orbital period & $P$ & 9.4903 & days \\
  Eccentricity & $e$ & 0.078 &  \\
  Argument of periapsis & $\omega$ & 90.0 & degree \\
  Quadratic limb-darkening coefficients & $u$ & [0.246, 0.203] &  \\
\hline
\sidehead{\bf{Best-fit of output parameters}}
  Radii ratio & $R_{\rm p}/R_{\rm s}$ & $0.0293\pm 0.0001$ &  \\
  Mid-transit time visit1 & $T_{\rm 0, visit1}$ & $2456646.4829\pm 0.0011$ & BJD \\
  Mid-transit time visit2 & $T_{\rm 0, visit2}$ & $2456665.4621\pm 0.0012$ & BJD \\
  Mid-transit time visit3 & $T_{\rm 0, visit3}$ & $2457491.0312\pm 0.0011$ & BJD \\
  Mid-transit time visit4 & $T_{\rm 0, visit4}$ & $2457785.2021\pm 0.0011$ & BJD \\
  Semi-major axis ratio & $a/R_{\rm s}$ & $26.7\pm 0.4$ &  \\
  Inclination & $i$ & $89.6\pm 0.1$ & 
\enddata
\end{deluxetable*}

\begin{figure} 
\begin{center}
\includegraphics[width=0.45\textwidth]{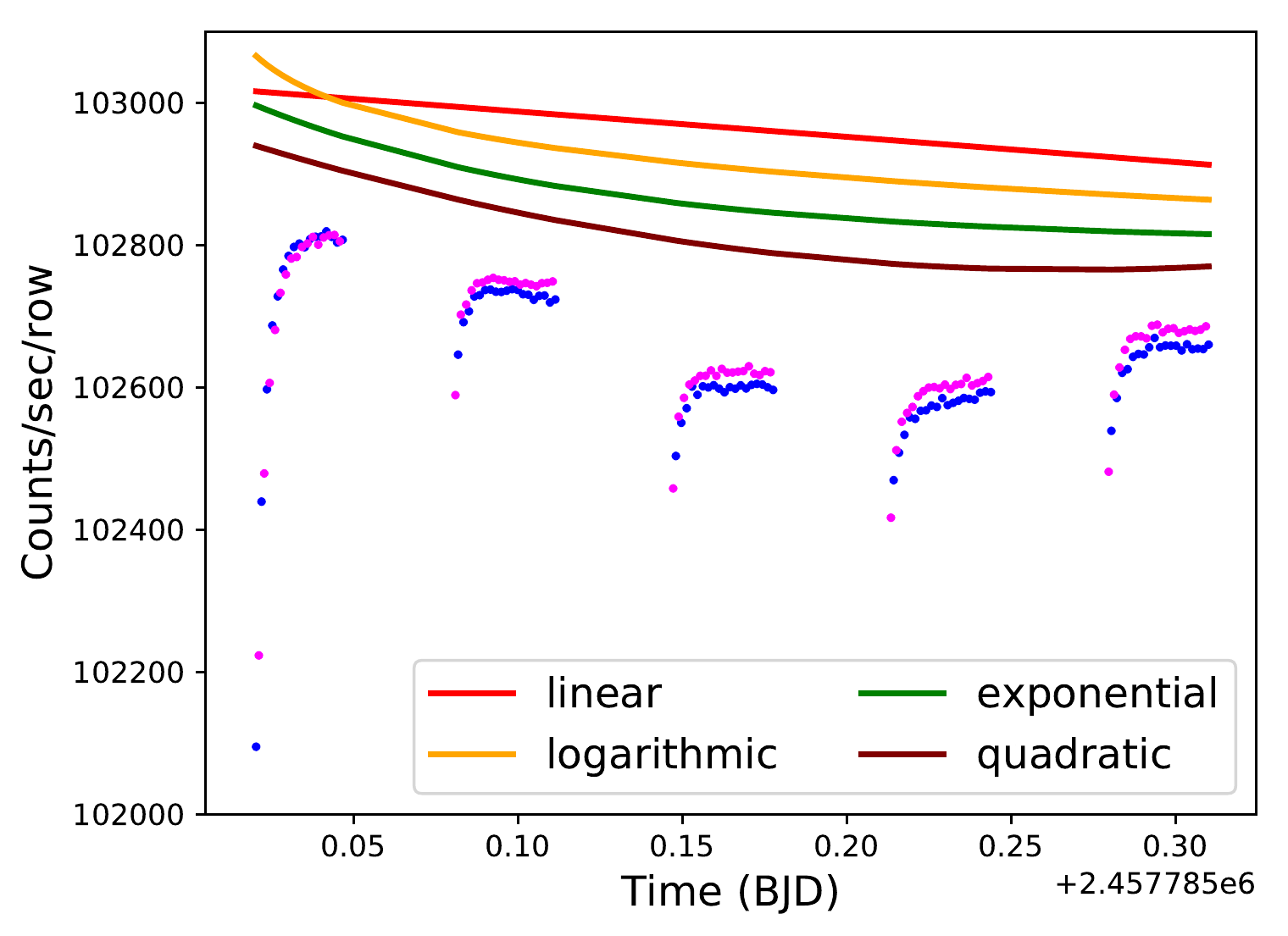}
\end{center}
\caption{As an example, this plot shows the four best-fit visit-long trend models of the light curve from visit4, and they are shifted vertically for clarity.}
\label{fig:compareTrends}
\end{figure}

\begin{figure}
\begin{center}
\includegraphics[width=0.45\textwidth]{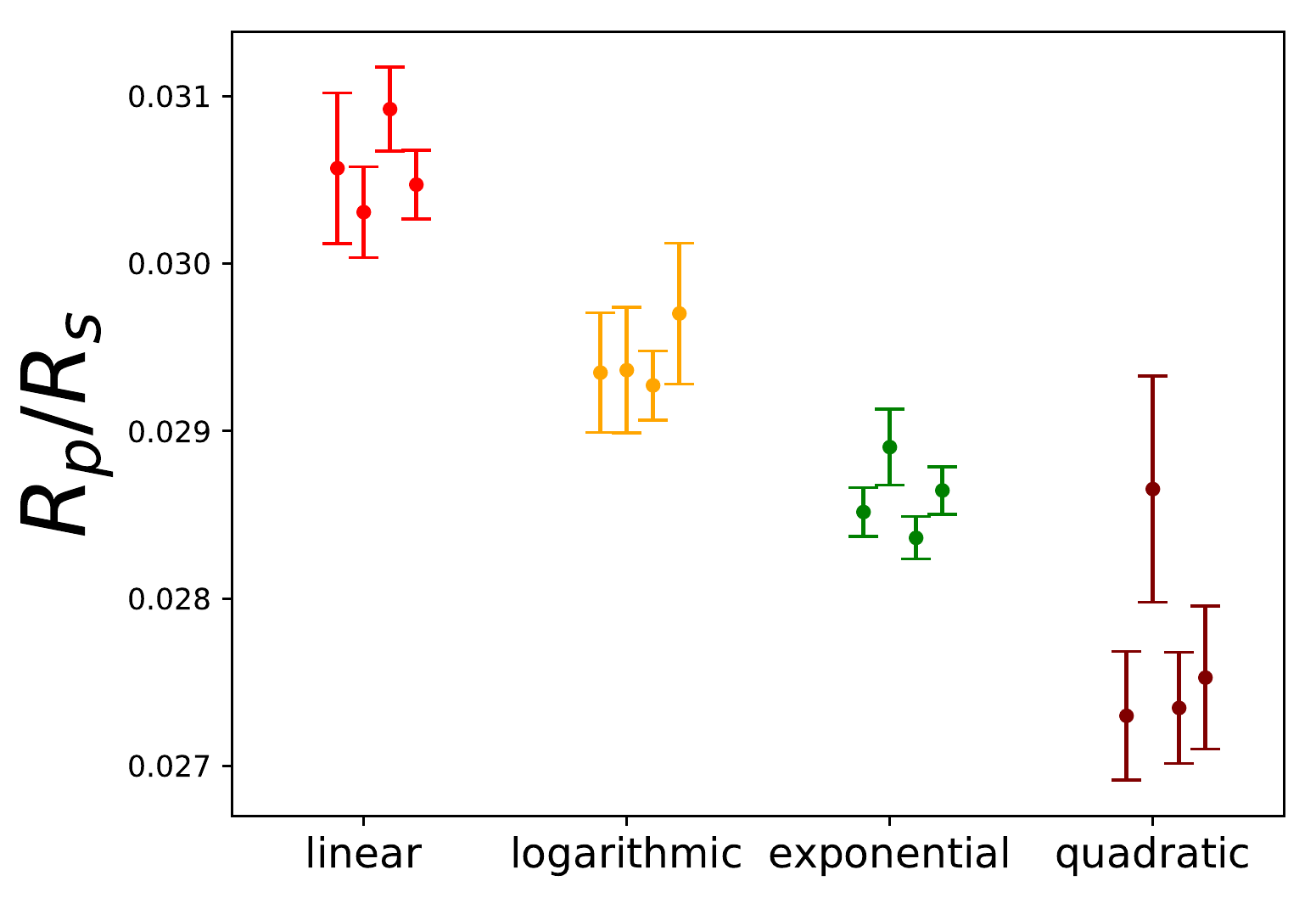}
\end{center}
\caption{Comparison of white light curve best-fit $R_{\rm p}/R_{\rm s}$ of each visit, assuming four different visit-long trends. The four visits are listed from left to right in chronological order for each model. The quadratic trend model produces the largest transit depth uncertainties and the most discrepant results among visits. }
\label{fig:compareTrends_whiteDepths}
\bigskip
\end{figure}

Using the raw white light curve from visit4 as an example, we show the four best-fit visit-long trend models in Figure \ref{fig:compareTrends}. The first thing we examine is whether those models produce consistent transit depths for different visits. Figure \ref{fig:compareTrends_whiteDepths} shows the best-fit $R_{\rm p}/R_{\rm s}$ of each \HST\ visit assuming four different visit-long systematic trends. The logarithmic trend produces the most internally consistent white light transit depths -- all well within 1$\sigma$ uncertainty with respect to each other. And while the other three cases all have maximum transit depth differences around 1$\sigma$ uncertainty, the quadratic trend model produces the largest transit depth uncertainties and the most discrepant results among visits. This result is expected since the curvature of a quadratic model is much more sensitive to its parameters than that of the three other models, and a transit signal itself could approximately be fitted with a quadratic shape if it is blended with a high systematic level. Similar to this work, \citet{Agol2010} compared long-term systematic trend functions to be used to model \spitzer\ light curves, and found that a quadratic function could bias transit depth measurements. 

The white light curve fitting result comparison does not strictly rule out any of the four visit-long trend models, even though the quadratic model is disfavored. Hence we proceed to calculate the transmission spectra assuming these four different models and make further comparison of their corresponding spectra in the following sections. 

\subsection{Spectral Light Curve Analysis}\label{sec:spectral_lc}

We group pixels along the dispersion direction into 28 channels with the same wavelength boundaries as those in \citet{Knutson2014} for easy comparison, and extract raw light curves in each channel. One thing to notice is that the long wavelength end of dispersed flux ran over the edge of the detector in visit3 and visit4 because of an observaion error, as is shown in Figure \ref{fig:typical_ima}. Therefore, the last two channels of these two visits were not observed. The channel boundaries compared against the sensitivity curve are shown in Figure \ref{fig:channels}.

We use the divide-by-white technique to analyze the spectral light curves. This technique was introduced by \citet{Kreidberg2014a} and has been widely used in \HST/\WFC3\ transmission spectrum measurements \citep{Knutson2014, Tsiaras2018, Damiano2017}. First, a common-mode signal is generated by dividing the white light curve by the best-fit transit signal, and then this common-mode is injected as a systematic component of spectral light curves in each channel. This systematic component is multiplied by a new transit model with the mid-transit time fixed to the best-fit transit time of the white light curve. And since we notice that the light curves in different channels have different visit-long trends, we also multiply the model of spectral light curves by a parameterized visit-long trend with the same form (linear, exponential, logarithmic, or quadratic) as that of the white light curve. 

After extracting the best-fit $R_{\rm p}/R_{\rm s}$ from each channel in each visit, we calculate the $R_{\rm p}/R_{\rm s}$ spectra over four visits using the weighted mean on each spectral channel. 
The uncertainty of $R_{\rm p}/R_{\rm s}$ in each channel is calculated by taking the larger value \HL{of either $\sigma_1(k)$ or $\sigma_2(k)$, where $k$ represents the channel number, $\sigma_1(k)$ represents the standard deviation of transit depth values from four visits weighted by their uncertainties, and $\sigma_2(k)$ represents the standard deviation of the mean transit depth of four visits. Formulae to calculate $\sigma_1(k)$ and $\sigma_2(k)$  are shown as follows:}
\begin{equation}
\begin{aligned}
\HL{\sigma_1(k) = \sqrt{\frac{\sum^N_{i=1}w_i(k)(D_i(k) - \bar{D}(k))^2}{\frac{(N-1)\sum^N_{i=1}w_i(k)}{N}}},}
\end{aligned}
\end{equation}\label{eq:err1}

\noindent
\HL{and}
\begin{equation}
\begin{aligned}
\HL{\sigma_2(k) = \sqrt{\frac{\sum^N_{i=1}{\sigma_i(k)}^2}{N(N-1)}},}
\end{aligned}
\end{equation}\label{eq:err2}

\noindent
\HL{where $N$ is the total number of visits, $D_i(k)$ is the transit depth in channel $k$ from visit $i$, $\sigma_i(k)$ is the uncertainty in $D_i(k)$, $\bar{D}(k)$ is the weighted average of $D_i(k)$ defined as $\bar{D}(k) = \frac{\sum^N_{i=1}w_i D_i(k)}{\sum^N_{i=1}w_i}$, and $w_i(k)$ is the weight on $D_i(k)$ defined as $w_i(k) = \frac{1}{{\sigma_i(k)}^2}$.} 

The final averaged $R_{\rm p}/R_{\rm s}$ result in each channel, assuming linear, quadratic, logarithmic and exponential visit-long trend models, are presented in Table \ref{tab:rp_rs}.

\begin{table*}
\renewcommand*{\arraystretch}{1.2}
    \caption{$R_{\rm p}/R_{\rm s}$ Averaged over Four Visits in Each \WFC3\ Bandpass assuming Four Different Visit-long Trends}
    \centering
    \begin{tabularx}{\textwidth}{cnnnnnnnn}
    \hline\hline
    Bandpass Center  & Linear &  & Logarithmic &  & Exponential &  & Quadratic &  \\
    ($\mu$m) & $R_{\rm p}/R_{\rm s}$ & $\sigma_{R_{\rm p}/R_{\rm s}}$ & $R_{\rm p}/R_{\rm s}$ & $\sigma_{R_{\rm p}/R_{\rm s}}$ & $R_{\rm p}/R_{\rm s}$ & $\sigma_{R_{\rm p}/R_{\rm s}}$ & $R_{\rm p}/R_{\rm s}$ & $\sigma_{R_{\rm p}/R_{\rm s}}$ \\
    \hline
     1.132 & 0.0308 & 0.0003 & 0.0292 & 0.0003 & 0.0285 & 0.0003 & 0.0286 & 0.0004 \\
     1.151 & 0.0306 & 0.0004 & 0.0290 & 0.0004 & 0.0280 & 0.0009 & 0.0283 & 0.0006 \\
     1.170 & 0.0305 & 0.0005 & 0.0293 & 0.0004 & 0.0287 & 0.0003 & 0.0282 & 0.0005 \\
     1.188 & 0.0308 & 0.0005 & 0.0294 & 0.0004 & 0.0290 & 0.0003 & 0.0274 & 0.0003 \\
     1.207 & 0.0300 & 0.0004 & 0.0286 & 0.0002 & 0.0278 & 0.0003 & 0.0273 & 0.0003 \\
     1.226 & 0.0303 & 0.0004 & 0.0287 & 0.0003 & 0.0283 & 0.0003 & 0.0277 & 0.0004 \\
     1.245 & 0.0302 & 0.0003 & 0.0288 & 0.0002 & 0.0283 & 0.0003 & 0.0276 & 0.0003 \\
     1.264 & 0.0299 & 0.0004 & 0.0286 & 0.0003 & 0.0282 & 0.0005 & 0.0268 & 0.0006 \\
     1.283 & 0.0301 & 0.0005 & 0.0289 & 0.0003 & 0.0281 & 0.0003 & 0.0278 & 0.0007 \\
     1.301 & 0.0307 & 0.0002 & 0.0293 & 0.0002 & 0.0286 & 0.0002 & 0.0276 & 0.0004 \\
     1.320 & 0.0303 & 0.0002 & 0.0290 & 0.0002 & 0.0282 & 0.0002 & 0.0272 & 0.0005 \\
     1.339 & 0.0300 & 0.0003 & 0.0287 & 0.0002 & 0.0280 & 0.0003 & 0.0268 & 0.0004 \\
     1.358 & 0.0302 & 0.0002 & 0.0291 & 0.0002 & 0.0284 & 0.0003 & 0.0268 & 0.0007 \\
     1.377 & 0.0302 & 0.0003 & 0.0288 & 0.0002 & 0.0280 & 0.0003 & 0.0273 & 0.0005 \\
     1.396 & 0.0308 & 0.0004 & 0.0294 & 0.0005 & 0.0289 & 0.0005 & 0.0275 & 0.0006 \\
     1.415 & 0.0314 & 0.0004 & 0.0302 & 0.0002 & 0.0298 & 0.0004 & 0.0285 & 0.0006 \\
     1.433 & 0.0308 & 0.0002 & 0.0294 & 0.0002 & 0.0288 & 0.0002 & 0.0276 & 0.0003 \\
     1.452 & 0.0306 & 0.0003 & 0.0292 & 0.0002 & 0.0283 & 0.0010 & 0.0281 & 0.0003 \\
     1.471 & 0.0303 & 0.0004 & 0.0291 & 0.0004 & 0.0285 & 0.0002 & 0.0273 & 0.0003 \\
     1.490 & 0.0302 & 0.0002 & 0.0290 & 0.0003 & 0.0280 & 0.0003 & 0.0273 & 0.0007 \\
     1.509 & 0.0305 & 0.0003 & 0.0291 & 0.0003 & 0.0282 & 0.0006 & 0.0267 & 0.0006 \\
     1.528 & 0.0302 & 0.0002 & 0.0288 & 0.0004 & 0.0284 & 0.0004 & 0.0270 & 0.0008 \\
     1.546 & 0.0302 & 0.0002 & 0.0288 & 0.0002 & 0.0283 & 0.0004 & 0.0262 & 0.0003 \\
     1.565 & 0.0304 & 0.0002 & 0.0290 & 0.0003 & 0.0280 & 0.0003 & 0.0258 & 0.0006 \\
     1.584 & 0.0302 & 0.0005 & 0.0295 & 0.0002 & 0.0290 & 0.0003 & 0.0279 & 0.0006 \\
     1.603 & 0.0314 & 0.0002 & 0.0298 & 0.0002 & 0.0297 & 0.0010 & 0.0288 & 0.0006 \\
     1.622 & 0.0311 & 0.0009 & 0.0299 & 0.0007 & 0.0294 & 0.0007 & 0.0277 & 0.0005 \\
     1.641 & 0.0312 & 0.0008 & 0.0299 & 0.0005 & 0.0291 & 0.0005 & 0.0294 & 0.0005 \\
    \hline
    \end{tabularx}
    \tablecomments{\HL{We list the results using all four different visit-long trend formulae, but only the transit depth results using the logarithmic visit-long trend are adopted in our atmospheric retrieval and modeling analysis. The reason for this choice and comparisons between results using different visit-long trends are presented in Section \ref{sec:visit_long_trend}.}}
    \label{tab:rp_rs}
\bigskip
\end{table*}

\begin{figure*}
\begin{center}
\hspace*{-0.5cm}
\includegraphics[width=\textwidth]{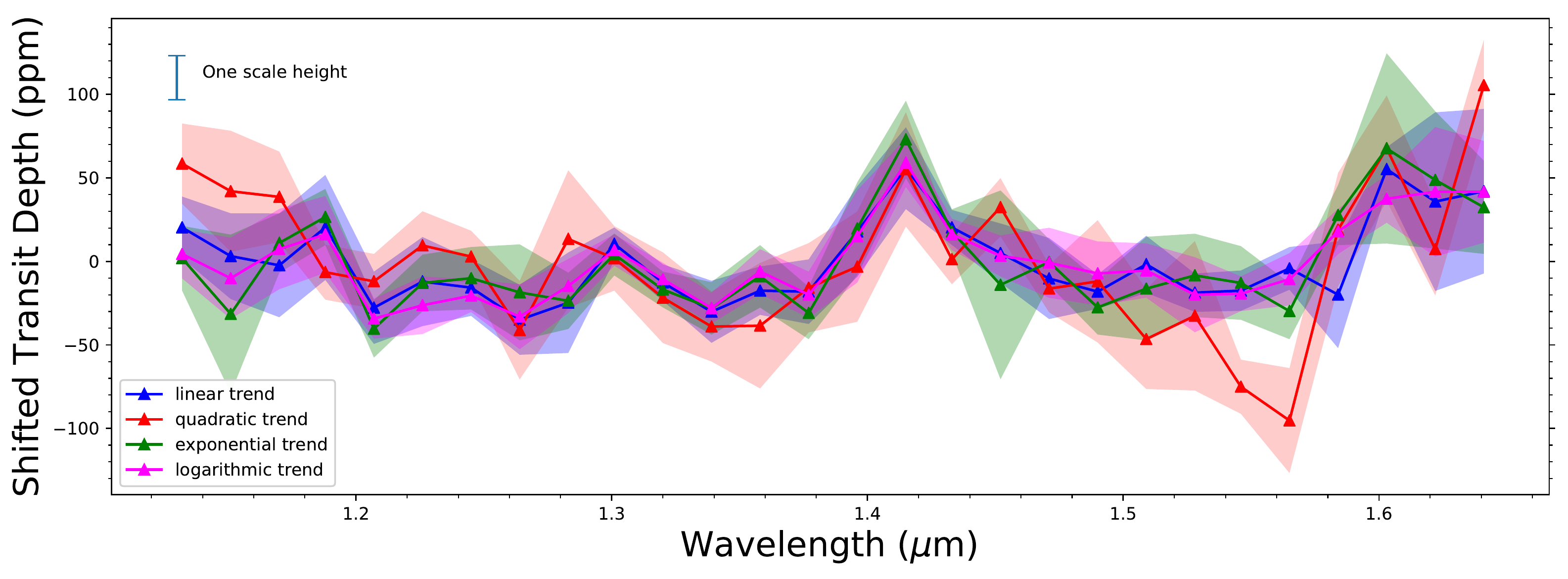}
\end{center}
\caption{Comparison of the shapes of 4-visits combined transmission spectra assuming four different visit-long trend functions. Spectra are shifted to have zero mean depth for easy comparison. It is shown here that the spectral shape produced with a linear, exponential and logarithmic visit-long functions are highly consistent with each other, while the spectral shape resulting from a quadratic visit-long trend is distinct from the other three. The size of one atmosphere scale height is shown for comparison, which assumes an equilibrium temperature of 900~K and mean molecular weight of 2.3.}
\label{fig:compareTrends_spectraShape}
\bigskip
\end{figure*}

We plot the transmission spectra for all four visit-long trends in Figure \ref{fig:compareTrends_spectraShape}, where all spectra are shifted to have zero mean for easy comparison. It is apparent that the spectra assuming three different visit-long trends have highly consistent shapes, with the only difference between them being their mean (white light) transit depth, whereas the spectral shape resulting from a quadratic visit-long trend is distinctive from the other three. Taking into consideration the white light curve transit depth comparisons shown in Figure \ref{fig:compareTrends_whiteDepths}, we decide that using a quadratic shape to model the \HST/\WFC3\ visit-long trend systematics of this dataset could have deformed the resulting transmission spectrum shape, so we discard this model in the rest of this work.
On the other hand, transmission spectral features on the $1.1\mu$m--$1.7\mu$m wavelength range extracted by assuming the other three visit-long trend models are consistent with each other. Therefore we adopt the spectrum assuming a logarithmic visit-long trend, which shows the most consistent white light curve transit depths among different visits (Figure \ref{fig:compareTrends_whiteDepths}), but we set the mean depth of the transmission spectrum on this wavelength range as a free parameter when performing atmosphere retrieval, so that uncertainties from the visit-long trend model selection are included in the final error budget. 

\subsection{Comparison with Previous Works}

\begin{figure}
\begin{center}
\includegraphics[width=0.48\textwidth]{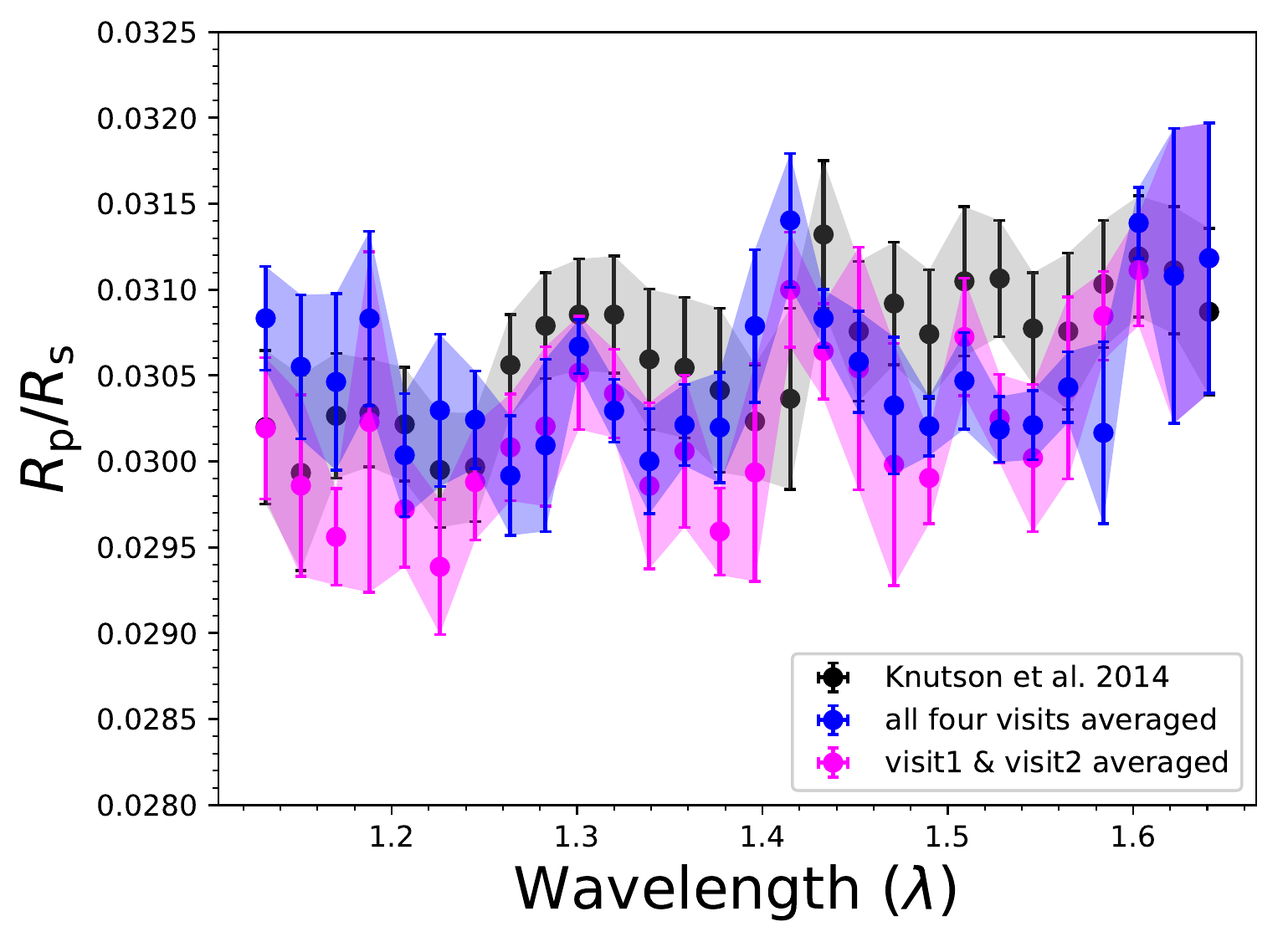}
\end{center}
\caption{Compare the transmission spectrum reported in \cite{Knutson2014} (grey), which uses a linear visit-long trend model, with our spectrum assuming a linear visit-long trend. The pink spectrum is the result using only visit1 and visit2, the same as in \cite{Knutson2014}. The two spectra are consistent with around 1$\sigma$ uncertainty, which is expected given that we use different methods to process the \HST/\WFC3\ data. The blue spectrum is the combined result using all four visits, shown here for comparison.}
\label{fig:comparewithPrevious}
\bigskip
\end{figure}

\citet{Knutson2014} fitted the light curves from visit1 and visit2 with an exponential orbital ramp model and a linear visit-long model. In Figure \ref{fig:comparewithPrevious}, we compare our transmission spectra (assuming a linear trend) averaged over visit1 and visit2 (pink shaded region) and averaged over all four visits (blue shaded region) with the final spectrum from \citet{Knutson2014} (grey shaded region). We can see that our two-visits-averaged spectrum shows a similar rising trend to the spectrum from \citet{Knutson2014}. Although there are some shifts in transit depths between our spectra and the previous one, their differences are mostly within $1\sigma$, and the reduced $\chi^2$ between our two-visits-averaged spectrum and the previous work is 0.98, showing the high consistency between the two results. However for the spectrum averaged over all four visits, the rising trend is mitigated, except for the possible feature redward of 1.6$\mu$m. Therefore it is reasonable to speculate that the rising trend of the HD~97658b spectrum presented in \citet{Knutson2014} results from an unidentified systematic effect. Nonetheless there could be astrophysical features remaining in the spectrum, and we discuss  atmospheric retrieval results in Section \ref{sec:atm_retrieve}.

\bigskip

\section{Observation and Analysis in Other Bandpasses}\label{sec:other_bandpass}

Aside from \HST/\WFC3 spectroscopy, the transit of HD~97658b was also observed with \STIS\ on \HST, with the \spitzer\ \it Space Telescope\ \rm in the 3.6~$\mu$m channel and 4.5~$\mu$m channel, and with the \most\ \it Space Telescope\ \rm in its 0.5~$\mu$m bandpass. We describe data reduction processes and transit depth results from these datasets in this section. 

\subsection{\STIS\ Observations and Data Reduction}\label{sec:stis_analysis}

We observed three transits of HD~97658b with \HST/\STIS\ using the G750L grism (0.524--1.027~$\mu$m) as part of \HST\ program 13665. Like \HST/\WFC3, an initial orbit is required to settle the telescope. The detector was purposefully saturated by about a factor of 3 in the brightest part of the spectrum to increase SNR \citep{Gilliland1999}. Data from each of the three visits was reduced using the same steps as \citet{Lothringer2018}. Counts were added along columns to reconstruct the observed flux at each wavelength. The observations were then split into ten bins of approximately 0.05~$\mu$m covering the G750L bandpass, and a transit and noise model were fit to the data assuming orbital parameters from \citet{Knutson2014}. Unlike the systematics marginalization procedure that was used in \citet{Lothringer2018}, we instead fit transit models to the data using Gaussian processes in a similar fashion to \citet{Bell2017} and \citet{Evans2013, Evans2018}, using a squared-exponential kernel. Using GPs instead of a parametric approach allows us to include time as a non-parametric covariate, as it was found that assuming a linear slope in time (as was done in \citet{Lothringer2018}) produced worse fits to the data and unrealistically small error bars. By including time as a covariate in the GP, we can better reflect our uncertainty in the baseline flux, which becomes more apparent over 4 science orbits rather than the 3 in most other \STIS\ datasets (also see \citet{Demory2015}). \HST's orbital phase was also used as a covariate to account for orbit-to-orbit systematics. 

\begin{table*}
\renewcommand*{\arraystretch}{1.2}
    \caption{Best-fit Spectroscopic Transit Depths for Each Transit Observed with \STIS\ }
    \centering
    \begin{tabularx}{\textwidth}{cnnnnnn}
    \hline\hline
    Bandpass~Center & Observation~1 &  & Observation~2 &  & Observation~3 & \\
    ($\mu$m) & $R_{\rm p}/R_{\rm s}$ & $\sigma_{R_{\rm p}/R_{\rm s}}$ & $R_{\rm p}/R_{\rm s}$ & $\sigma_{R_{\rm p}/R_{\rm s}}$ & $R_{\rm p}/R_{\rm s}$ & $\sigma_{R_{\rm p}/R_{\rm s}}$ \\
    \hline
0.553 & 0.0273 & 0.0015 & 0.0255 & 0.0011 & 0.0252 & 0.0028 \\
0.601 & 0.0282 & 0.0006 & 0.0272 & 0.0007 & 0.0277 & 0.0009 \\
0.650 & 0.0281 & 0.0004 & 0.0282 & 0.0004 & 0.0290 & 0.0005 \\
0.699 & 0.0278 & 0.0009 & 0.0281 & 0.0004 & 0.0290 & 0.0020 \\
0.748 & 0.0282 & 0.0010 & 0.0302 & 0.0011 & 0.0308 & 0.0019 \\
0.797 & 0.0268 & 0.0015 & 0.0286 & 0.0032 & 0.0315 & 0.0011 \\
0.845 & 0.0265 & 0.0014 & 0.0294 & 0.0009 & 0.0312 & 0.0008 \\
0.894 & 0.0246 & 0.0016 & 0.0275 & 0.0014 & 0.0315 & 0.0014 \\
0.943 & 0.0260 & 0.0026 & 0.0294 & 0.0023 & 0.0326 & 0.0017 \\
0.992 & 0.0213 & 0.0032 & 0.0346 & 0.0034 & 0.0378 & 0.0049 \\
    \hline
    Mid-transit Times (BJD) & $2457149.4191\pm 0.0005$ &  & $2457196.8642\pm 0.0003$ &  & $2457206.3537\pm 0.0008$ &  \\
    \hline\hline
    \end{tabularx}
    \label{tab:STIS_depths}
\bigskip
\end{table*}

In Table \ref{tab:STIS_depths}, we report the spectroscopic transit depths and the mid-transit time from each of the three transits observed with \STIS, along with their uncertainties for future reference. The \STIS\ data quality limits our ability to achieve consistent transit depths among different observations. As Table \ref{tab:STIS_depths} shows, the transit depths measured with data from three \STIS\ observations are highly discrepant in the long wavelength part ($\lambda > 0.79~\mu$m) of the \STIS\ bandpass. And with the same uncertainty calculation procedure as described in section \ref{sec:spectral_lc}, we find that the transit depth uncertainties can be as large as 100--200~ppm in red end channels of the \STIS\ bandpass. This uncertainty level is far from adequate for providing any meaningful constraint to atmospheric properties. Therefore we do not include the \STIS\ dataset in our atmospheric analysis.

\subsection{\spitzer\ Observations and Data Reduction}\label{sec:spitzer_analysis}

Six transits were observed with the IRAC 3.6~$\mu$m channel and five transits were observed with the 4.5~$\mu$m channel from July 2014 to April 2016 under \spitzer\ program 11131 \citep{Dragomir2013SpitzerProp}. Each transit was observed with  0.08 seconds exposure per frame and approximately 0.13 seconds per frame cadence. 

We analyze the raw light curves using the ``Pixel-Level Decorrelation'' (PLD hereafter) technique \citep{Deming2015}, following the same procedure as described in \citet{Guo2018}. Before fitting the light curves, we manually remove the first 50 to 80 minutes of each observation so that the drastic systematic ramp at the beginning of each \spitzer\ observation does not bias our fitting result. Using the PLD technique, fractional contributions to the total flux at the observational time points from each selected pixel are treated as an eigenvector, and we set the weight of each eigenvector as a free parameter and combine them together to model the total flux variations. Since the PLD technique assumes that the incoming stellar flux is falling on the same set of pixels throughout the entire time series, we include all pixels around the star that contribute more than 1\% of the total flux to ensure that the PLD technique is valid to apply. The same pixel selection procedure is also successfully used in \citet{Guo2018}. Simultaneously, we fit a transit model generated using BATMAN and defined with a free-floating transit depth and mid-transit time. The rest of the transit parameters are fixed in the same way as described in Section \ref{sec:ResidualModel}. 

\begin{figure*}
\begin{center}
    \subfigure
	{%
	\includegraphics[width=0.49\textwidth]{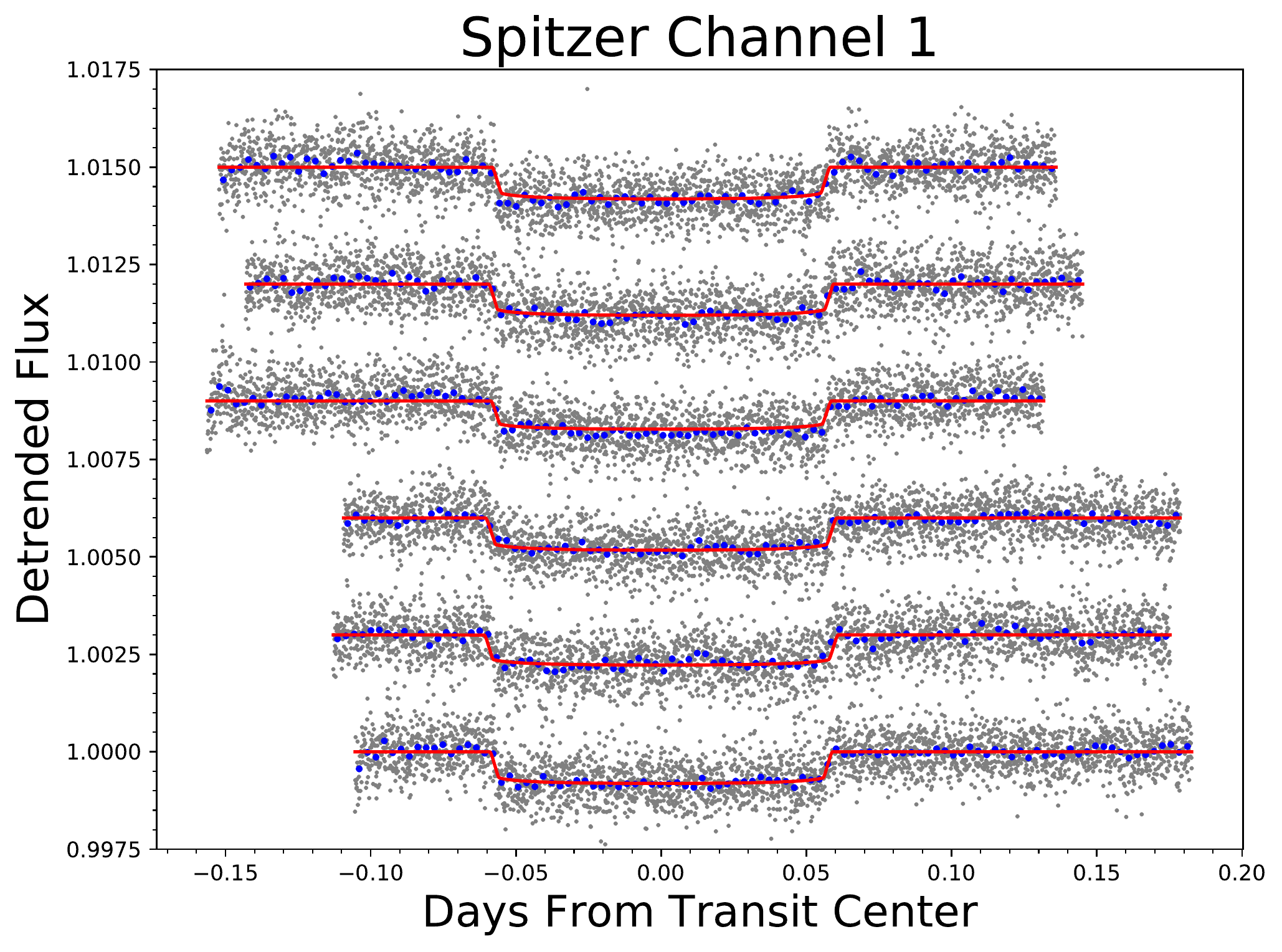}
	}%
	\subfigure
	{%
	\includegraphics[width=0.49\textwidth]{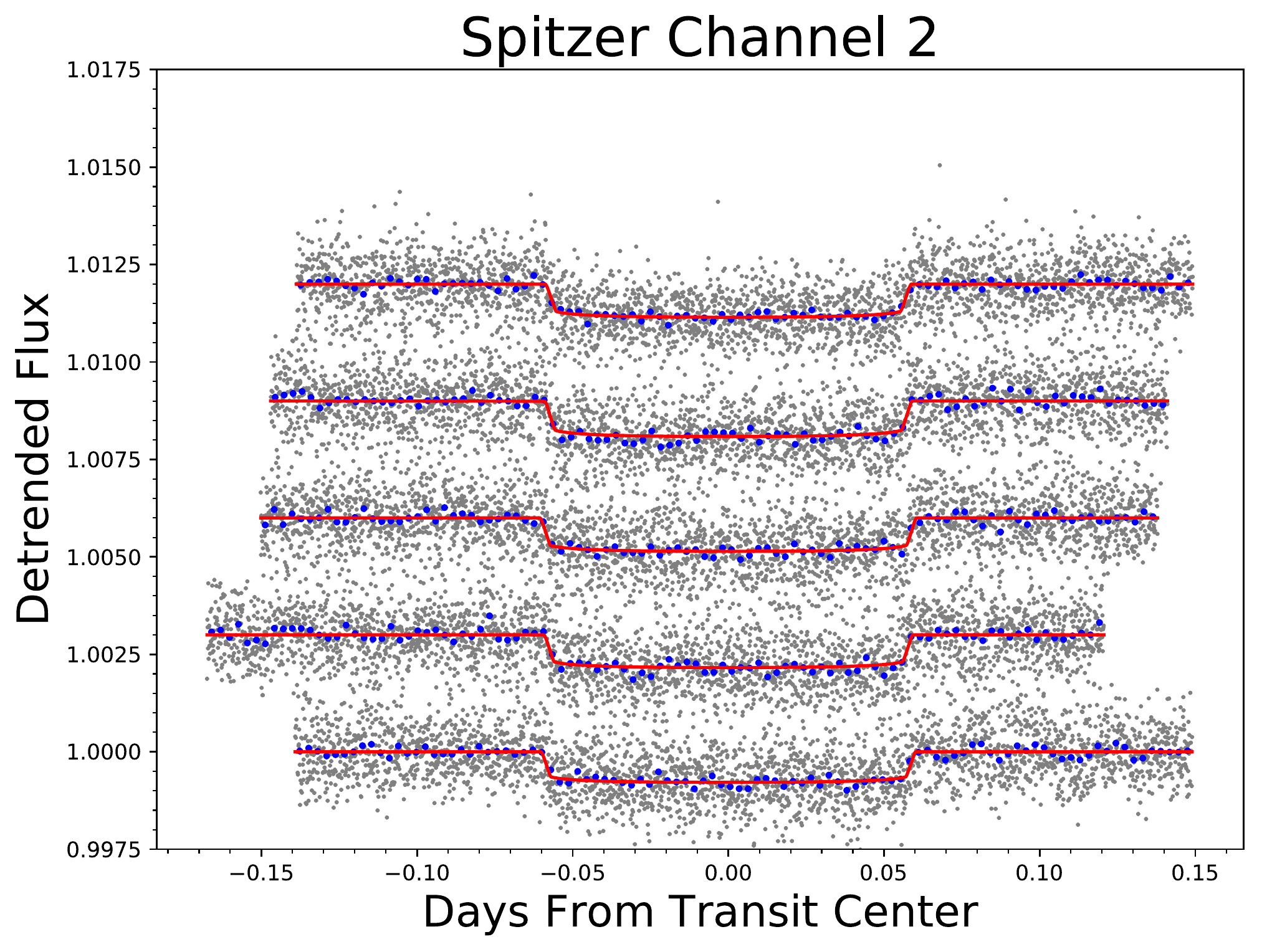}
	}%
\end{center}
\caption{Detrended transit light curves with their best-fit transit models of each \spitzer\ observation. Left panel: channel 1 ($3.6\mu$m); right panel: channel 2 ($4.5\mu$m). Light curves are shifted vertically for the purpose of display.}
\label{fig:spitzer_transitLC}
\bigskip
\end{figure*}

The best-fit transit models and detrended light curves of all \spitzer\ transits are shown in Figure \ref{fig:spitzer_transitLC}. The best-fit parameters of all transits are presented in Table \ref{tab:spitzer}.

\begin{table*}
\renewcommand*{\arraystretch}{1.2}
    \caption{Best-fit $R_{\rm p}/R_{\rm s}$ and mid-Transit Times of Each Transit Observed By \spitzer}
    \centering
    \begin{tabularx}{\textwidth}{nnnnnn}
    \hline\hline
    AOR ID & channel & $T_{0}$ (BJD-2400000.5) & $\sigma_{T_0}$ (days) & $R_{\rm p}/R_{\rm s}$ & $\sigma_{R_{\rm p}/R_{\rm s}}$ \\
    \hline
49696512 & 3.6$\mu$m & 56864.23938 & 0.00027 & 0.0276 & 0.0005 \\ 
49697536 & 3.6$\mu$m & 56883.21892 & 0.00027 & 0.0270 & 0.0004 \\ 
49698048 & 3.6$\mu$m & 56892.70742 & 0.00026 & 0.0280 & 0.0004 \\ 
52197888 & 3.6$\mu$m & 57091.98243 & 0.00034 & 0.0261 & 0.0004 \\ 
52198144 & 3.6$\mu$m & 57082.49324 & 0.00032 & 0.0275 & 0.0006 \\ 
52198656 & 3.6$\mu$m & 57101.47164 & 0.00034 & 0.0277 & 0.0005 \\ 
53908736 & 4.5$\mu$m & 57481.04364 & 0.00031 & 0.0272 & 0.0005 \\ 
53909248 & 4.5$\mu$m & 57471.55445 & 0.00025 & 0.0281 & 0.0004 \\ 
53909504 & 4.5$\mu$m & 57253.30001 & 0.00045 & 0.0284 & 0.0005 \\ 
53909760 & 4.5$\mu$m & 57243.81157 & 0.00022 & 0.0293 & 0.0004 \\ 
53910016 & 4.5$\mu$m & 57234.32179 & 0.00026 & 0.0284 & 0.0004 \\ 
    \hline
    \end{tabularx}
    \label{tab:spitzer}
\bigskip
\end{table*}

\begin{figure}
\begin{center}
    \subfigure
	{%
	\includegraphics[width=0.48\textwidth]{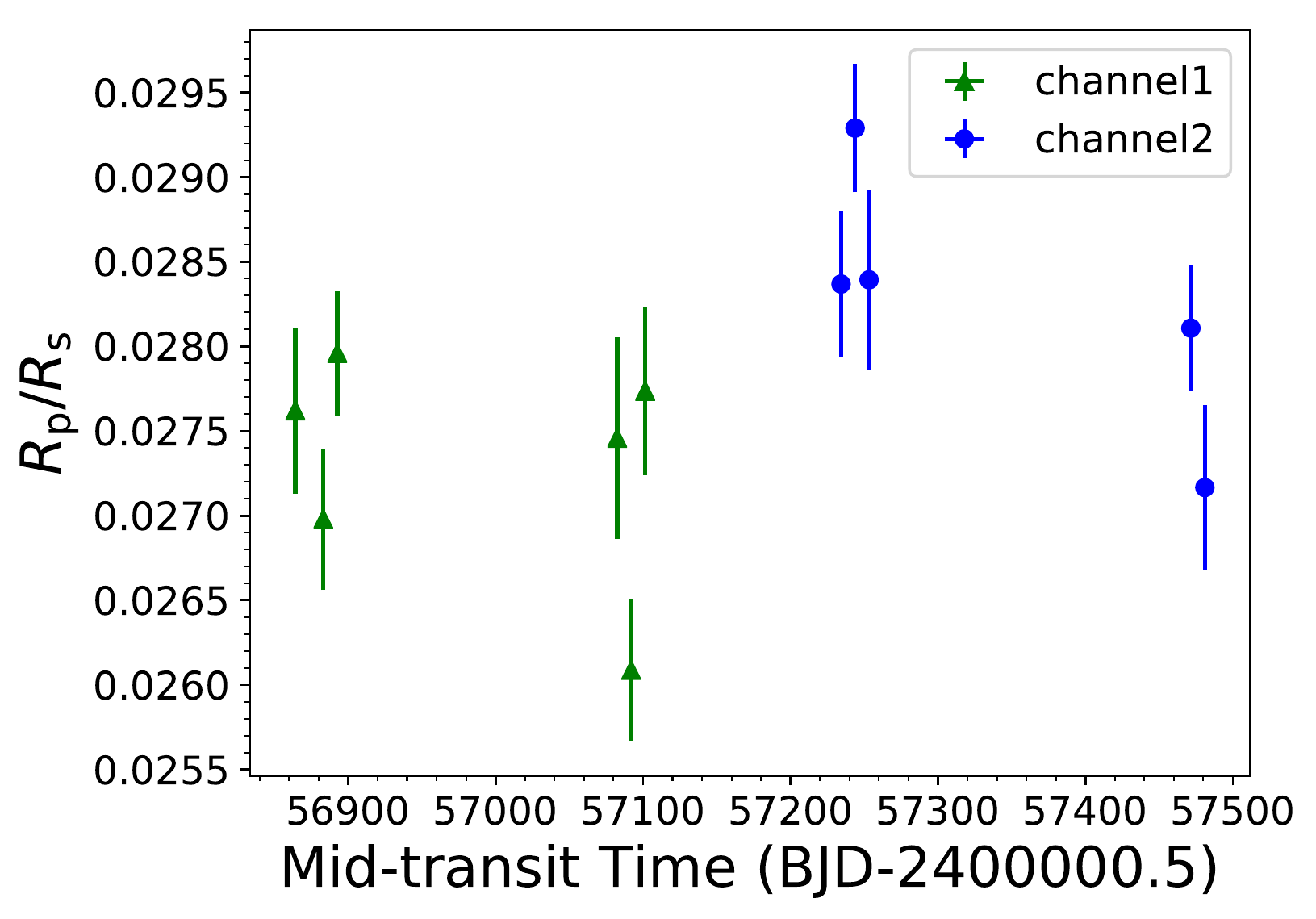}
	}%
\end{center}
\caption{$R_{\rm p}/R_{\rm s}$ from each \spitzer\ transit observation. The green data points represent channel~1 (3.6~$\mu$m) transits, and the blue data represent channel~2 (4.5~$\mu$m) transits.}
\label{fig:spitzer_Rpcompare}
\bigskip
\end{figure}

Figure \ref{fig:spitzer_Rpcompare} shows the $R_{\rm p}/R_{\rm s}$ of each transit arranged according to their mid-transit time. Discrepancies among transits of a same channel is around $1\sigma$. 
To ensure that the scatter in transit depths and the associated error bars  of each transit are properly included in our error budget, we again take the larger one between the standard deviation of the weighted average of center values and the weighted reduced error bars as our uncertainties in $R_{\rm p}/R_{\rm s}$. We find that the best-fit $R_{\rm p}/R_{\rm s}$ in the $3.6~\mu$m  and  $4.5~\mu$m channels are 0.0273$\pm$0.0003 and 0.0284$\pm$0.0003, respectively.

\subsection{\most\ Observation and Data Reduction}\label{sec:most_analysis}

MOST (Microvariability and Oscillations in STars; \citealt{Wal03, Mat04}) is a micro-satellite carrying a $15$ cm optical telescope that acquires light through a broadband filter spanning the visible wavelengths from $350$ to $700$ nm. It is in a Sun-synchronous polar orbit with a period of $101.4$ minutes, which allows it to monitor stars in a Continuous Viewing Zone (CVZ) without interruption for up to $8$ weeks. The CVZ covers a declination range of $+36^\circ>\delta>-18^\circ$. 

HD 97658 was observed by \most\ in Direct Imaging mode, in which defocused images of the stars were projected directly onto the CCD \citep{Row06}. One, four and five transits were observed in 2012, 2013 and 2014, respectively. The 2012 and 2013 transits have been previously published in \citet{Dragomir2013}, while the 2014 observations are unpublished. For the analysis performed for this paper, we used the three transits observed in 2013 that do not show interruptions (March 10, 19 and 29; see \citealt{Dragomir2013}), and all five transits that we observed in 2014 (all of which are also continuous). 

The exposure times were 1.5 s, and the observations were stacked on board the satellite in groups of 21 for a total integration time of 31.5 s per data point. Raw light curves were extracted from the images using aperture photometry \citep{Row08}. Outlier clipping and de-trending from the sky background and position on the CCD were performed as described in \citet{Dragomir2013}. After these steps, a straylight variation at the orbital period of the satellite remained. This variation was filtered by folding each light curve on this 101.4-minute period, computing a running average from this phased photometry, and removing the resulting waveform from the corresponding light curve.

\begin{figure}
\begin{center}
    \subfigure
	{%
	\includegraphics[width=0.48\textwidth]{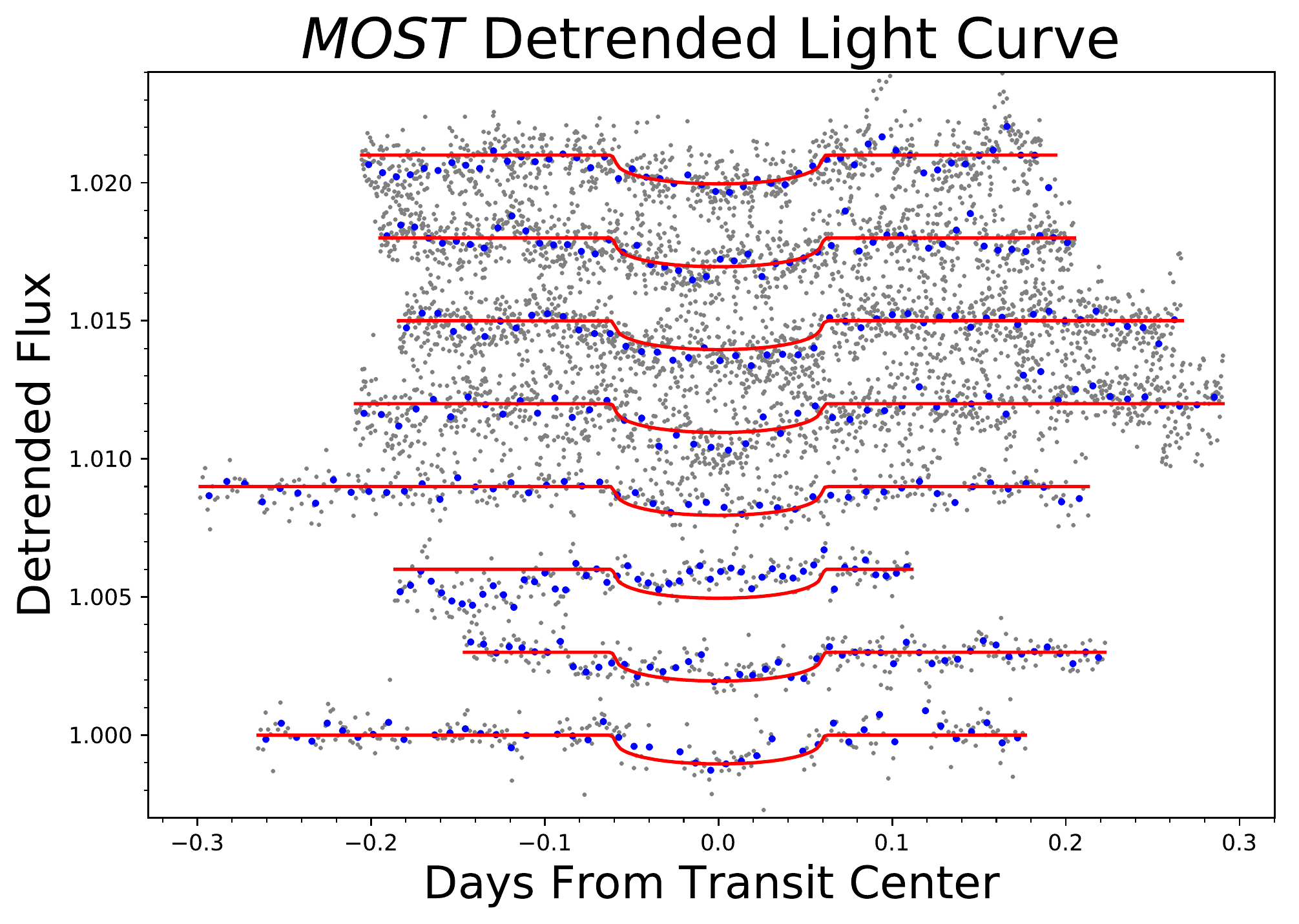}
	}%
\end{center}
\caption{The detrended transit light curve of each \most\ observation, along with the best-fit transit model from EXOFASTv2 joint-fit. The observations are listed in chronological order, with the bottom one being the latest observation. Light curves are shifted for the purpose of display.}
\label{fig:MOST}
\bigskip
\end{figure}

We fit the eight transits simultaneously using EXOFASTv2 \citep{Eas17}, a differential evolution Markov Chain Monte Carlo algorithm that uses error scaling, and obtained a best-fit $R_p/R_s$ value of $0.02866^ {+0.00054}_{-0.00056}$. We summarize the ephemeris of these eight transits into two best-fit mid-transit times: $2456361.8050 \pm 0.0033$~(BJD) in year 2013 and $2456712.9096 \pm 0.0024$~(BJD) in year 2014. The detrended light curves and the best-fit transit models are plotted in Figure \ref{fig:MOST}.

\bigskip

\section{Ephemeris and Radial Velocity Analysis}\label{sec:RV_ephemeris}

\begin{figure}
\begin{center}
    \subfigure
	{%
	\includegraphics[width=0.48\textwidth]{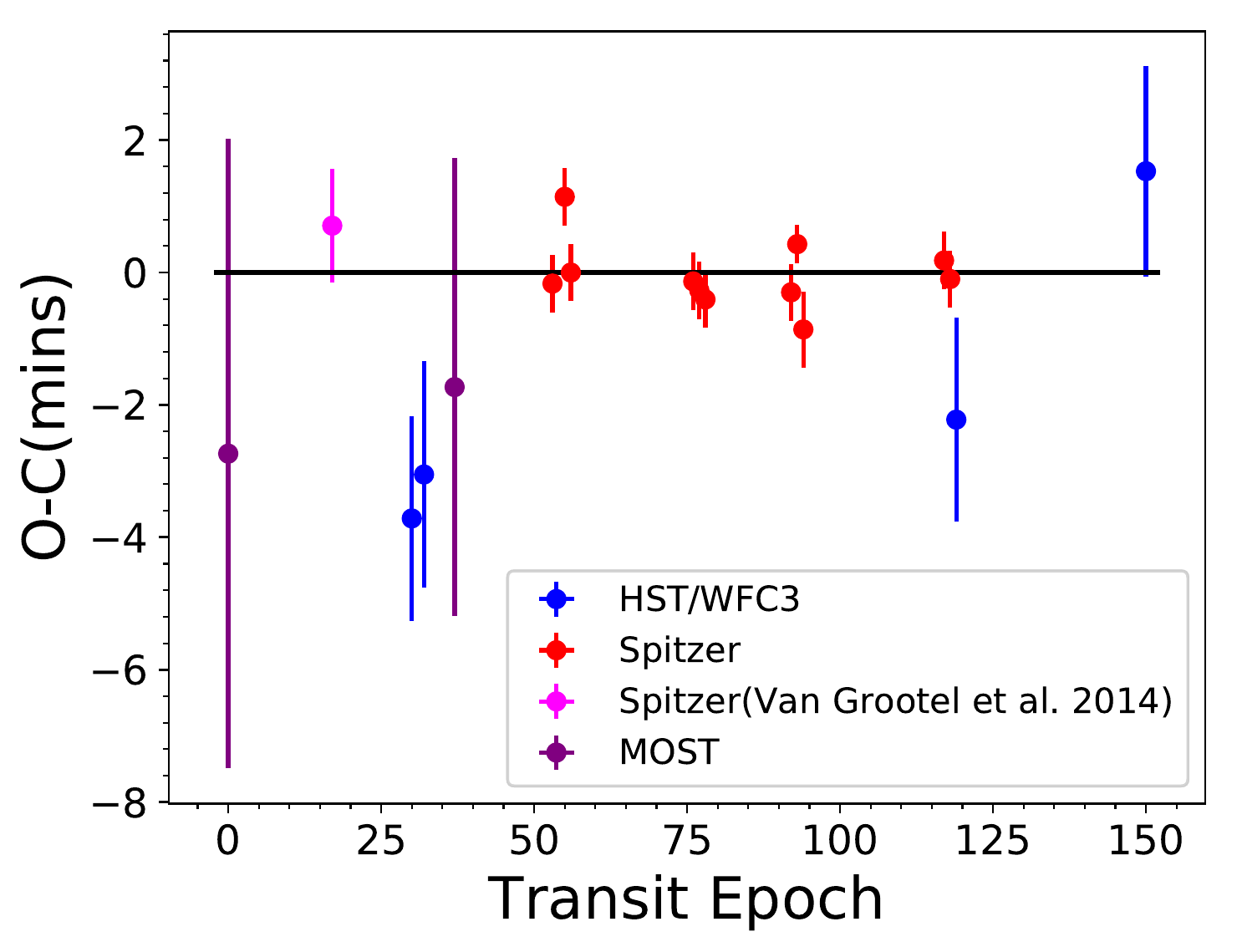}
	}%
\end{center}
\caption{The ephemeris of HD~97658b. A linear fitting shows the period to be $P=9.489295 \pm 0.000005$, with a best-fit transit time center at $T_0 = 2456361.80690 \pm 0.00038$~(BJD). The best-fit reduced $\chi^2$ is 1.7, which shows that no TTV is detected.}
\label{fig:ephemeris}
\bigskip
\end{figure}

We collect all HD~97658b mid-transit times from previous works, in combination with transit times measured in this work, and analyze the overall ephemeris variation of this planet. With a least squares fit, we find the best-fit period of HD~97658b to be $P=9.489295 \pm 0.000005$ days. The deviations from a linear ephemeris are shown in Figure \ref{fig:ephemeris}. Most observed transit times are consistent with a periodic orbit with $1-2\sigma$ confidence, and we obtain a best-fit reduced $\chi^2$ of 1.7, corresponding to only a $2\sigma$ difference between model and observation. This means that no transit time variation (TTV) is detected, which is consistent with our non-detection of additional planets in the RV data, as discussed below. 

\subsection{Keplerian RV Analysis}

Since January 1997 we have collected 553 radial velocity measurements with the High Resolution Echelle Spectrometer (HIRES, \citealp{Vogt1994}) on the Keck I Telescope on Maunakea and 215 measurements with the Levy spectrograph on the Automated Planet Finder at Lick Observatory (APF, \citealp{Radovan2014}, \citealp{Vogt2014}). These data were all collected through an iodine cell for wavelength calibration and point spread function reference \citep{Butler1996}. One set of iodine-free spectra were collected with each instrument to use as a model of the intrinsic stellar spectrum. The HIRES data were often taken in sets of three due to the short $\sim$2~minute exposures  to mitigate the effects of stellar oscillations, this was not necessary for the APF due to the smaller aperture and longer exposure times ($\sim$10-20 minute exposures). The HIRES data from January 2005 to August 2010 were previously analyzed in the discovery paper of HD 97658 b \citep{Howard2011}. The data reduction and analysis followed the California Planet Search method described in \citet{Howard2010}. The resultant radial velocities are presented in Table~\ref{tab:rvs} and in Figure~\ref{fig:rvfit}. 

\begin{figure}
\begin{center}
\hspace{-0.5cm}
\includegraphics[width=0.5\textwidth]{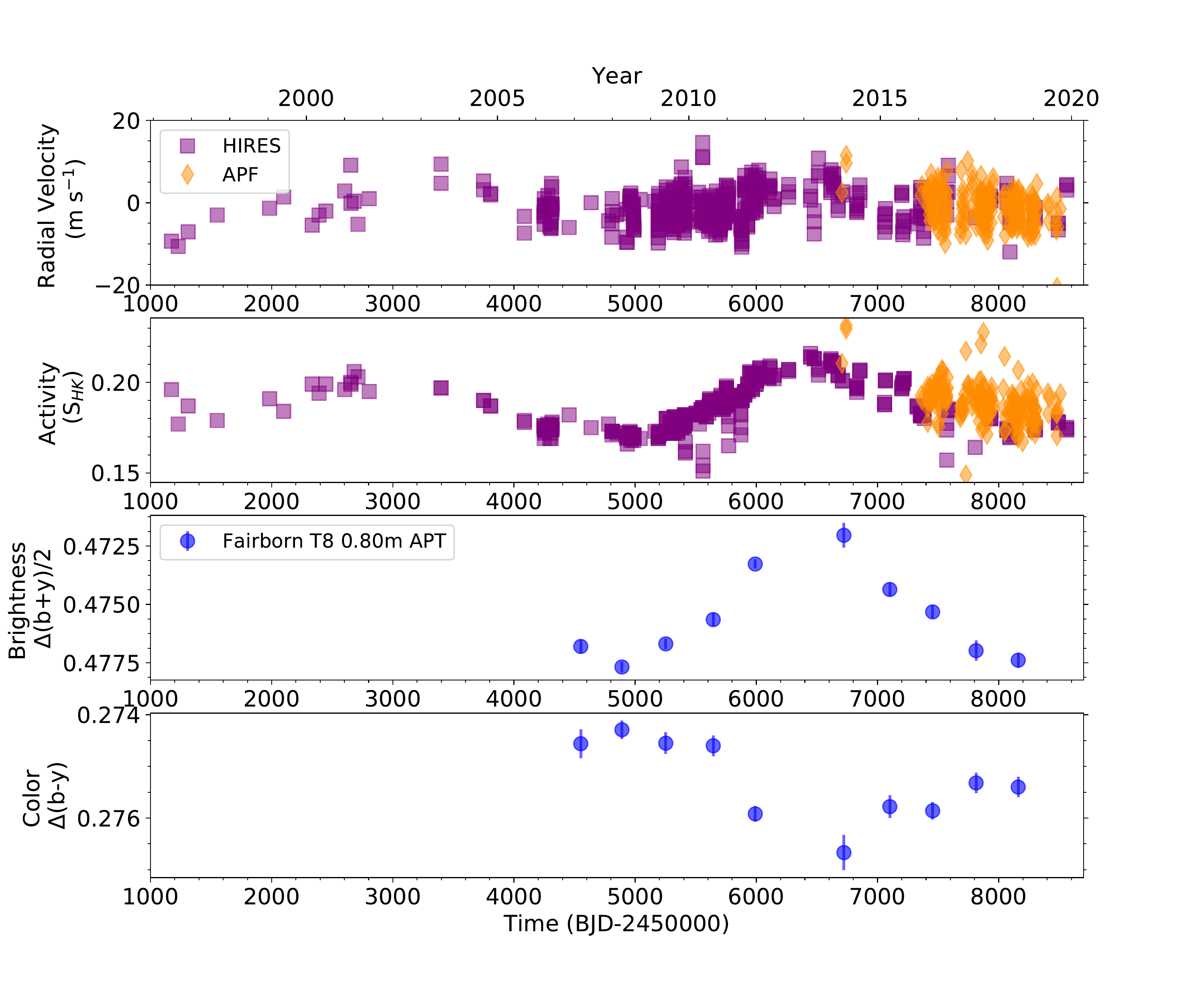}
\caption{\label{fig:cahk} Time series of our radial velocity and Calcium H and K activity (S$_{\rm HK}$) data from HIRES and APF, and photometry from the Fairborn T8 0.80m APT including both brightness and color information. There is a clear variation in the radial velocity data matched by the activity data, brightness, and color all without a phase offset. This relation implies that the long-term radial velocity variation is stellar activity.} 
\end{center}
\end{figure}

\begin{figure}
\begin{center}
\hspace{-0.5cm}
\includegraphics[width=0.5\textwidth]{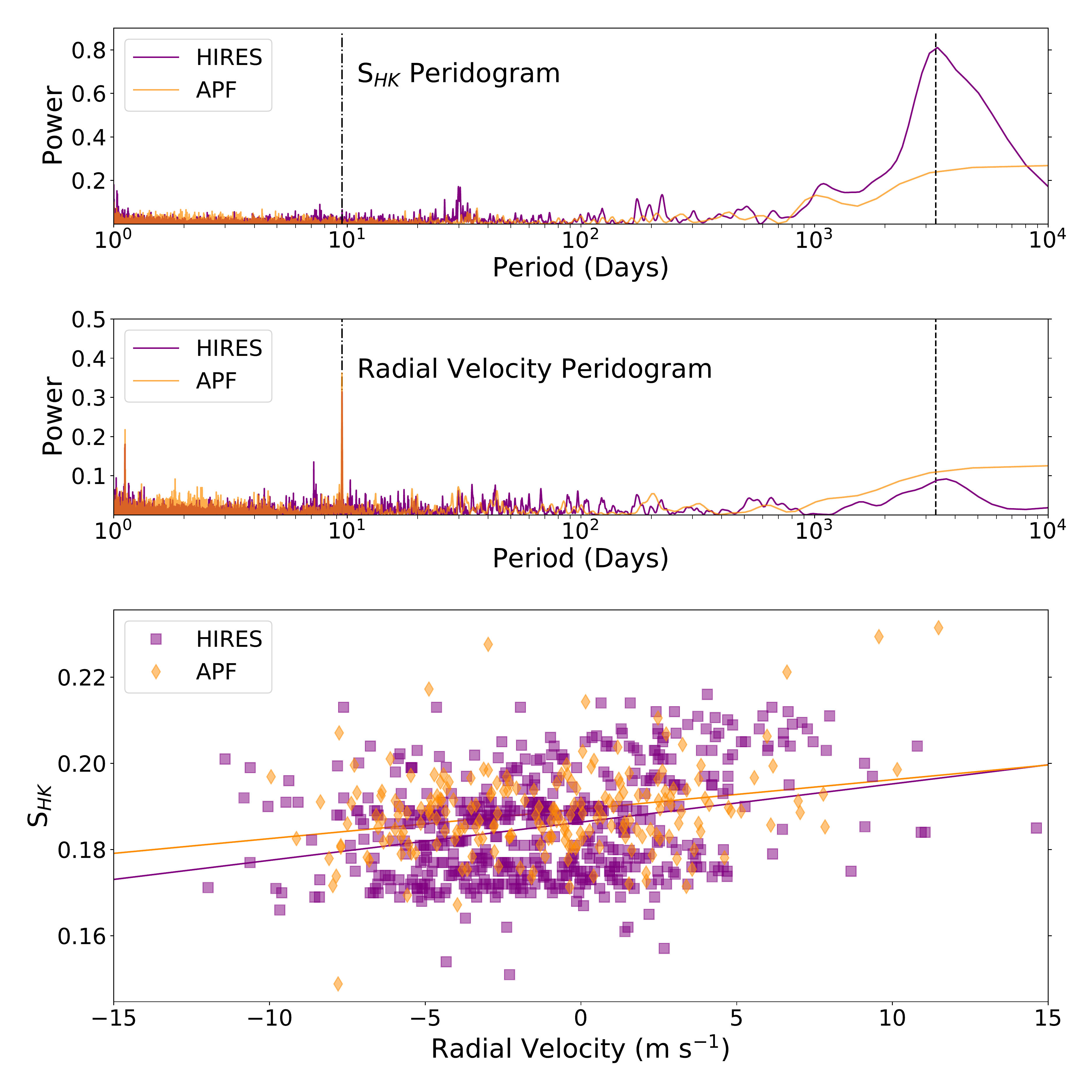}
\caption{\label{fig:strrotperiod} Periodograms of S$_{\rm HK}$ (top) and radial velocity (middle), and S$_{\rm HK}$ vs. radial velocity (bottom). In both periodograms, the stellar activity cycle (Figure~\ref{fig:cahk}, $\sim$ 3500 days) is represented by a dashed line and the planet's orbital period (9.49 days) is represented by a dash-dot line. There is a strong radial velocity signal and S$_{\rm HK}$ signal at the stellar activity cycle timescale in the HIRES data. There is a correlation between the S$_{\rm HK}$ and radial velocity data in both datasets, shown as the solid lines in the bottom panel.}  
\end{center}
\end{figure}

\begin{figure}
\begin{center}
\hspace{-0.5cm}
\includegraphics[width=0.50\textwidth]{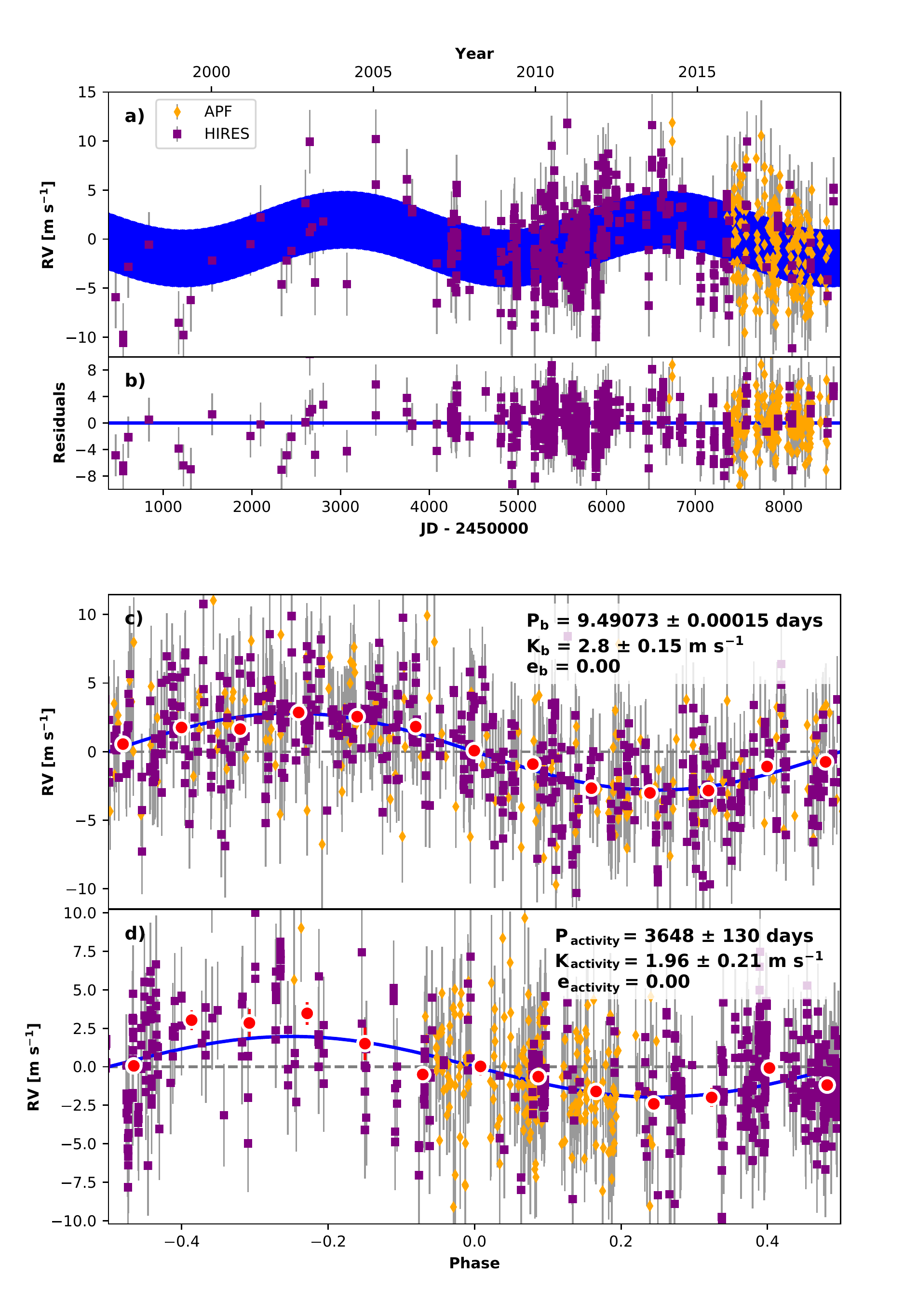}
\caption{\label{fig:rvfit} Best-fit 1-planet Keplerian orbital model
  for HD 97658 including stellar activity. The maximum likelihood model is plotted while the orbital parameters listed in Table \ref{tab:rvfit} are the
  median values of the posterior distributions.  The thin blue line is
  the best fit 1-planet model. We add in quadrature
  the RV jitter term(s) listed in Table \ref{tab:rvfit} with the
  measurement uncertainties for all RVs.  {\bf b)} Residuals to the
  best fit 1-planet model. {\bf c)} RVs phase-folded
  to the ephemeris of planet b. The Keplerian orbital model for the stellar activity has been subtracted.  The small point colors
  and symbols are the same as in panel a.  Red circles are the same velocities binned in 0.08 units of orbital
  phase.  The phase-folded model for planet b is shown as the blue
  line. {\bf d)} RVs phase-folded to the ephemeris of the stellar activity. All details are the same as panel c.} 
\end{center}
\end{figure}

We first investigate the star for signs of stellar activity by examining the Calcium H and K lines (S$_{\rm HK}$, \citet{Isaacson2010}, Figure~\ref{fig:cahk}) in the HIRES and APF data. There is a clear periodicity in both the S$_{\rm HK}$ and the radial velocity data around 3500 days in the HIRES dataset (Figure~\ref{fig:strrotperiod}). The APF data does not have a long enough baseline to detect such a long signal. In addition, we compare this long term variation in S$_{\rm HK}$ and radial velocity signals with the brightness and color variation of HD~97658 which was measured with the Fairborn T8 0.80m automatic photoelectric telescope \citep{Henry1999}. As is shown in Figure~\ref{fig:cahk}, there is a clear correlation in the variations seen in radial velocities, stellar activity data, stellar brightness, and stellar color. This relation implies that the long-term radial velocity variation is actually caused by stellar activity. 
The length of the signal indicates that it is likely the star's 9.6~yr magnetic activity cycle (slightly shorter than our Sun's eleven-year cycle); we discuss the stellar activity in more detail in Section \ref{sec:activity}.

We analyze the radial velocity data using RadVel\footnote{Available at \url{https://github.com/California-Planet-Search/radvel}} \citep{Fulton2017}, which  models Keplerian orbits to fit radial velocity data by performing a maximum-likelihood fit to the data and then subsequently determining the uncertainties through a Markov-Chain Monte Carlo (MCMC) analysis. We use the default MCMC parameters for RadVel of: 50 walkers, 10000 steps, and 1.01 as the Gelman-Rubin criteria for convergence, as described in \citet{Fulton2017}. 

We model this system in RadVel as a two-Keplerian system for planet b and the stellar activity. We include priors on the transit parameters of planet b from \citet{VanGrootel2014}. 
We incorporate this stellar activity signal at around 3500 days into our radial velocity fit as an additional Keplerian signal because it has a sinusoidal shape and only two cycles of this signal are captured by the data. We use a Gaussian prior on the period (3424$\pm$41 days) and reference phase of this signal (2457372$\pm$21 BJD) from a RadVel 1-Keplerian fit of the HIRES S$_{\rm HK}$ data. 
Our radial velocity fit is shown in Figure~\ref{fig:rvfit}, and the output parameters are listed in Table~\ref{tab:rvfit}. Note the planet mass is calculated assuming our best-fit inclination ($i = 89.6$) with the \HST/\WFC3\ dataset, which shows $\sin(i)\approx 1$. 

\HL{We also test a non-zero planet eccentricity for completeness;
the resulting eccentricity is small, consistent with zero to two sigma (e$_{b}$=0.030$^{+0.034}_{-0.021}$), and results in consistent planet parameters to the circular case. Therefore we adopt the circular fit results.}
We also test including a Gaussian process to model the stellar activity signal with the hyperparameters constrained from a fit of the HIRES S$_{\rm HK}$ data. The results are consistent with the Keplerian fit; the baseline covers only two cycles of the activity therefore the deviation from a simple sinusoid is small. Since the fit has consistent posteriors, the additional parameters needed for the Gaussian process fit do not seem warranted and we present the Keplerian fit as our final result.

\begin{deluxetable*}{lrr}
\tablecaption{Radial Velocity Fit Parameters\label{tab:rvfit}}
\tablehead{\colhead{Parameter} & \colhead{Name (Units)} & \colhead{Value}}
\startdata
\sidehead{\bf{Planet Parameters}}      
$P_{b}$ & Period (days)   &  $9.49073\pm 0.00015$ \\
$T\rm{conj}_{b}$ & Time of conjunction    & $2456361.805\pm 0.0006$\\
& (BJD$_{\rm TDB}$) & \\
$e_{b}$ & Eccentricity & $\equiv$ 0.0 \\
$\omega_{b}$ & Argument of periapse    & $\equiv$ 0.0 \\
& (radians) & \\
$K_{b}$ & Semi-amplitude (m s$^{-1}$)   &  $2.81\pm 0.15$ \\
$M_{b}$ & Mass (M$_\oplus$)   &  $7.81^{+0.55}_{-0.44}$\\
$\rho_{b}$ & Density (g cm$^{-3}$)   & $3.78^{+0.61}_{-0.51}$ \\
$P_{\rm activity}$ & Period (days)   &  $3652^{+130}_{-120}$ \\
$T\rm{conj}_{\rm activity}$ & Reference Time    & $2457605^{+100}_{-89}$ \\
& (BJD$_{\rm TDB}$) & \\
$e_{\rm activity}$ & Eccentricity & $\equiv$ 0.0 \\
$\omega_{\rm activity}$ & Argument of periapse    & $\equiv$ 0.0 \\
& (radians) & \\
$K_{\rm activity}$ & Semi-amplitude (m s$^{-1}$)   &  $1.96\pm 0.21$ \\
\hline 
\sidehead{\bf{Other Parameters}}      
$\gamma_{\rm HIRES}$ & Mean center-of-mass   & $-0.85\pm 0.17$ \\
& velocity (m s$^{-1}$) & \\
$\gamma_{\rm APF}$ & Mean center-of-mass   & $-0.42^{+0.33}_{-0.34}$ \\
& velocity (m s$^{-1}$) & \\
$\dot{\gamma}$ & Linear acceleration    & $\equiv$ 0.0 \\
& (m s$^{-1}$ day$^{^{-1}}$) & \\
$\ddot{\gamma}$ & Quadratic acceleration  & $\equiv$ 0.0 \\
& (m s$^{-1}$ day$^{-2}$)  & \\
$\sigma_{\rm HIRES}$ & Jitter ($\rm m\ s^{-1}$)   & $2.93^{+0.11}_{-0.1}$ \\
$\sigma_{\rm APF}$ & Jitter ($\rm m\ s^{-1}$)   &  $1.3^{+0.31}_{-0.35}$ \\
\enddata
\end{deluxetable*}

\begin{deluxetable}{lrrcc}
\tablecaption{Radial Velocities \label{tab:rvs}}
\tablehead{
  \colhead{Time} &   \colhead{RV} &   \colhead{RV Unc.} &   \colhead{S$_{\rm HK}$} &   \colhead{Instrument} \\  \colhead{($BJD_{TDB}$)} &   \colhead{(m s$^{-1}$)} &   \colhead{(m s$^{-1}$)} &   \colhead{} &   \colhead{}
}
\startdata
2458559.90624	& 4.41	&	1.03	& 0.175	&	HIRES	\\
2458559.90814	& 3.03	&	1.16	& 0.174	&	HIRES	\\
2458487.99980	& -5.97	&	2.19	& 0.185	&	APF	\\
2458508.86464	& -1.51	&	2.15	& 0.194	&	APF	
\enddata
\tablecomments{The full table is available in machine-readable form.
} 
\end{deluxetable}


\subsection{Stellar activity and rotation in context}
\label{sec:activity}
As described above, our Keck-HIRES spectra allow us to identify a 9.6~yr stellar activity cycle for  HD~97658.  In addition, we also estimate the star's rotation period by calculating a Lomb-Scargle periodogram of the Keck/HIRES and APF activity indices after removing the long-term variations induced by the stellar activity cycle. No single period dominates, but we see an excess of power from 32--36~d in both data sets. We therefore report a stellar rotation period of $P_\mathrm{rot} =34 \pm 2$~d, slightly lower than previously reported \citep{henry:2011}.

HD~97658's rotation period is typical for stars with similar photometric colors and activity levels \citep[as measured by
 $R^\prime_\mathrm{HK}$;][]{mascareno:2016}. The activity cycle is also typical, with HD~97658 falling nicely on the empirical relation between $P_\mathrm{rot}$ and $P_\mathrm{activity}/P_\mathrm{rot}$ \citep{baliunas:1996,mascareno:2016}. Both HD~97658's overall activity level ($R^\prime_\mathrm{HK}$ and $S_\mathrm{HK}$) and the 2~m~s$^{-1}$ RV variations induced by its activity mark it as a quieter-than-average star for its spectral type \citep{Isaacson2010,lovis:2011}.

Low mass stars (M or K~dwarfs) often have excessive X-ray and UV radiation from their chromosphere and corona, and these energetic emissions can drive photochemistry and ionization processes in atmospheres of the planets orbiting around them. \citet{Loyd2016} put together a catalog of 7 M~dwarfs and 4 K~dwarfs, including HD~97658, in their MUSCLES Treasury Survey, and obtained \Chandra\ or \XMM\ observations of each of them. An interline continuum in the FUV bandpass is detected at 6.3$\sigma$ significance in HD~97658. No observation of the X-ray bandpass was made on this star, although integrated X-ray flux was detected higher than $\rm 10^{-14} erg s^{-1} cm^{-2}$ with all other three K~dwarfs in the survey. Since UV and X-ray radiation is strongly related to the dissociation of atmosphere molecules including $\rm H_2O,~CH_4,~CO,~O_3$ and etc \citep{Rugheimer2013} and production of hazes \citep{Horst2018}, atmospheric models ought to take the effect of high-energy stellar radiation into account. And to understand the atmospheric compositions and evolutionary history of HD~97658b, more detailed stellar spectroscopy in full wavelength coverage needs to be conducted. 

\subsection{No Additional Planets Found}

We searched for additional planet candidates orbiting HD~97658 by applying an iterative periodogram algorithm to our radial velocity data. First, we define an orbital frequency/period grid over which to search, with sampling such that the difference in frequency between adjacent grid points is $({2\pi \tau})^{-1}$, where $\tau$ is the observational baseline. Using this grid, we compute a goodness-of-fit periodogram, by fitting a sinusoid with a fixed period to the data, for each period in the grid. We choose to measure goodness-of-fit as the change in the Bayesian Information Criterion (BIC) at each grid point between the best-fit 1-planet model with the given fixed period, and the BIC value of the 0-planet fit to the data. We then fit a power law to the noise histogram (50-95 percent) of the data, and accordingly extrapolate a BIC detection threshold corresponding to an empirical false-alarm probability of our choice (we choose 0.003). 

If one planet is detected, we perform a final fit to the one-planet model with all parameters free, including eccentricity, and record the BIC of that best-fit model. We then add a second planet to our RV model and conduct another grid search, leaving the parameters of the first planet free to converge to a more optimal solution. In this case, we compute the goodness-of-fit as the difference between the BIC of the best-fit one-planet model, and the BIC of the two-planet model at each fixed period in the grid. We set a detection threshold in the manner described above and continue this iterative search until the (n+1)th search rules out additional signals. This search method is known as a recursive periodogram, also described in \cite{Anglada2012}. Similar to our RV analysis described in the previous section, we use RadVel \citep{Fulton2017} to fit Keplerian orbits, and an implementation of this search algorithm known as RVSearch, to be released at a later date (Rosenthal et al. in prep). For HD 97658, we search from 1.5 days to five times the observational baseline, and detect no new planet candidates with significance higher than FAP=0.01.

\section{Atmospheric Properties}\label{sec:atm_retrieve}

We use two independent approaches, PLATON \citep{PLATON_Zhang2019} and ATMO \citep{Goyal2019, Tremblin2015}, to extract atmospheric parameter information from our transmission spectrum. In section \ref{sec:forward_model}, we present acceptable model ranges by comparing our data with forward models generated with PLATON.

We first test retrieving and forward modeling with only the \HST/\WFC3\ data, and then also test adding the \most\ ($\rm 0.5~\mu m$) and \spitzer\ observation results ($\rm 3.6~\mu m$ and $\rm 4.5~\mu m$) into our atmospheric property analysis process with PLATON and ATMO. Since  the mean transit depth of the \HST/\WFC3\ spectrum is uncertain due to its degeneracy with the \HST/\WFC3\ visit-long trend shape, we let the mean transit depth on the \HST/\WFC3\ bandpass float as a free parameter. The posteriors of atmospheric parameters are similar to what we see with only the \HST/\WFC3\ data in both cases, and there is almost no extra constraint from the  extra \most\ and \spitzer\ data points. In light of this, we only present the best-fit transmission spectrum model obtained with the combined \HST/\WFC3, \spitzer\ and \most\ data in the following section (Figure \ref{fig:PLATON_noSTIS_best}). 

\subsection{Retrieval with PLATON}\label{sec:platon}

PLATON (PLanetary Atmospheric Transmission for Observer Noobs) is a forward modeling and atmosphere retrieval tool for exoplanet transmission spectral analysis developed by \citet{PLATON_Zhang2019}. PLATON uses the same opacity files and part of the same algorithms as the widely used atmosphere forward modeling package Exo-Transmit \citep{Exo_transmit_Kempton2017}, but is 100-1000 times faster than Exo-Transmit, so that a Markov Chain Monte Carlo (MCMC hereafter) or a nested sampling retrieval method can be used to extract the posteriors of atmospheric parameters. 

We first retrieve the HD~97658b atmospheric parameters with the observed transmission spectrum on the \HST/\WFC3\ bandpass only. The input stellar radius and effective temperature are set to the latest published value in \Gaia\ DR2, with $R_{*}=0.746~R_{\odot}$ and $T_{\rm eff} = 5192$~K, and the input planet mass is set according to our radial velocity results (section \ref{sec:RV_ephemeris}) with $M_{\rm p} = 7.81~M_{\oplus}$. We let five atmospheric parameters -- planet radius $R_{\rm p}$, planet surface temperature $T_{\rm p}$ assuming an isothermal atmosphere, logarithmic metallicity $\log(Z)$, C/O ratio, and logarithimc cloud top pressure $\log({P)}$ -- and one hyper-parameter ``err\_multiple,'' which is applied to the spectroscopic error bars as a scaling factor, float as free parameters during the retrieval process. We apply a uniform prior on $T_{\rm p}$ and constrain it to between 550~K and 950~K, which are estimated assuming a 0 to 0.67 albedo and zero to full global heat redistribution. Rain-out condensation process is taken into account, and since our wavelength coverage from 1.1~$\mu$m to 1.7~$\mu$m is not adequate to identify the Mie scattering shape at the short wavelength limit, we set the base-10 logarithm of the scattering factor to its default value of 0, assuming a uniform opaque cloud deck where the cloud-top pressure is a free parameter in the atmosphere. We test using both MCMC and nested sampling methods to explore the parameter space and find that the nested sampling method is better suited for this task to prevent samples from being trapped in local minima and for faster convergence. 

\begin{figure*}
\begin{center}
\hspace*{-0.5cm}
	{%
	\includegraphics[width=0.98\textwidth]{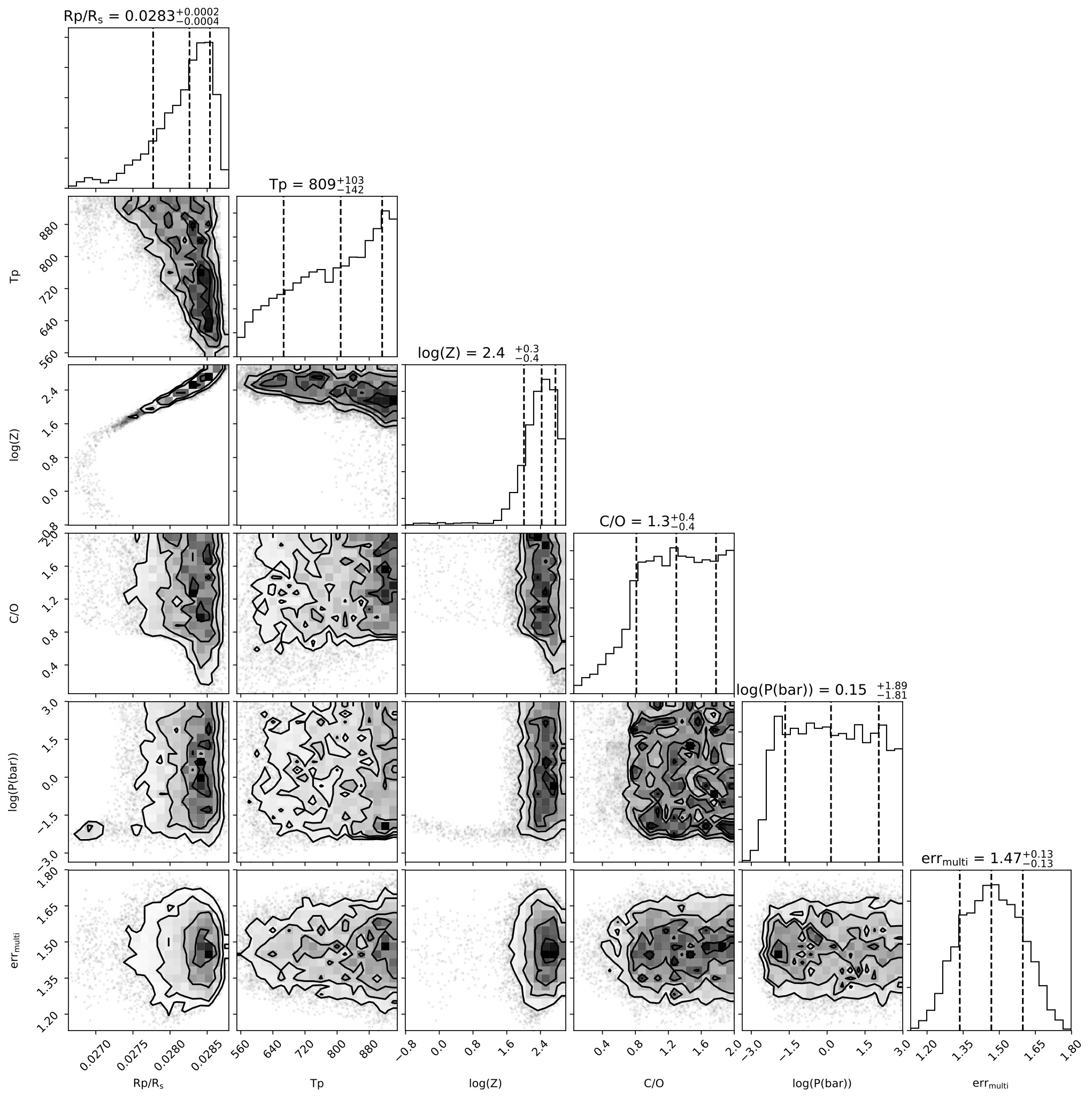}
	}%
\end{center}
    \caption{Posterior distributions of atmospheric parameters from PLATON retrieval on \HST/\WFC3\ transmission spectrum. $P$ represents cloud-top pressure, and $\rm err_{multi}$ is the uniform scaling factor applied to spectral error bars so that the best-fit model achieves $\chi_{\nu}^2$ of around 1. The planet radius posterior is consistent with previous studies of this planet, and the atmospheric metallicity posterior peaks around $\log(Z) = 2.4$. The result also shows that our transmission spectrum favors planet equilibrium temperature as high as 900~K, the C/O ratio is constrained to 0.8 or higher, and the cloud-top pressure is constrained to $0.01$~bar or higher.} 
\label{fig:PLATON_WFC3}
\bigskip
\end{figure*}

Figure \ref{fig:PLATON_WFC3} shows the posterior distributions of free parameters in the retrieval, and we present the input fixed parameter, priors, and the output posterior distribution center and 1$\sigma$ uncertainties in Table \ref{tab:retrieval_post}. 
Our fits constrain the cloud-top pressure to values greater than 0.01 bars and C/O ratios to above 0.8 (i.e., super-solar), but the posteriors for both parameters are effectively uniform within this preferred range. The planet radius posterior is consistent with our results from the \WFC3, \most, and  \spitzer\ data, the log($Z$) posterior peaks around $\log(Z) = 2.4$, and the equilibrium temperature posterior favors temperature as high as 900~K. Aside from these, the posterior of err\_multiple is constrained at $1.47_{-0.13}^{+0.13}$, indicating that the uncertainties in our \HST/\WFC3\ transmission spectrum may  be slightly larger than estimated, despite the conservative approach we have adopted when analyzing our uncertainty budget. 

\begin{table*}
\renewcommand*{\arraystretch}{1.2}
    \caption{Parameters for the PLATON atmospheric retrieval using our \WFC3\ spectrum}
    \centering
    \begin{tabularx}{0.65\textwidth}{ccccc}
    \hline\hline
    Parameter Names & median & lower error ($1\sigma$) & upper error ($1\sigma$) \\
    \hline\hline
     & & & Fixed Parameters \\
    \hline
    $R_{\rm s}$($R_{\odot}$) & 0.746 & -- & -- \\
    $M_{\rm p}$($M_{\oplus}$) & 7.81 & -- & -- \\
    $T_{\rm eff}$(K) & 5192 & -- & -- \\
    $\log(\rm scattering~factor)$ & 0.0 & -- & -- \\
    \hline
     & & & Fit Prioir \\
    \hline
    $R_{\rm p}$($R_{\rm Jup}$) & $\mathcal{U}(0.19,~0.22)$ & -- & -- \\
    $T_{\rm eq}$(K) & $\mathcal{U}(550,~950)$ & -- & -- \\
    $\log{Z}$ & $\mathcal{U}(-1.0,~3.0)$ & -- & -- \\
    C/O & $\mathcal{U}(0.05,~2.0)$ & -- & -- \\
    $\log{P(\rm bar)}$ & $\mathcal{U}(-8.99,~2.99)$ & -- & -- \\
    \hline
     & & & Fit Posteriors \\
    \hline
    $R_{\rm p}/R_{\rm s}$) & 0.0283 & 0.0004 & 0.0002 \\
    $T_{\rm eq}$(K) & 809 & 142 & 103 \\
    $\log{Z}$ & 2.4 & 0.4 & 0.3 \\
    C/O & 1.3 & 0.4 & 0.4 \\
    $\log{P(\rm bar)}$ & 0.15 & 1.81 & 1.89 \\
    error multiple & 1.47 & 0.13 & 0.13 \\
    \hline
    \end{tabularx}
    \label{tab:retrieval_post}
\end{table*}

\begin{figure*}
\begin{center}
\hspace*{-0.5cm}
	{%
	\includegraphics[width=0.98\textwidth]{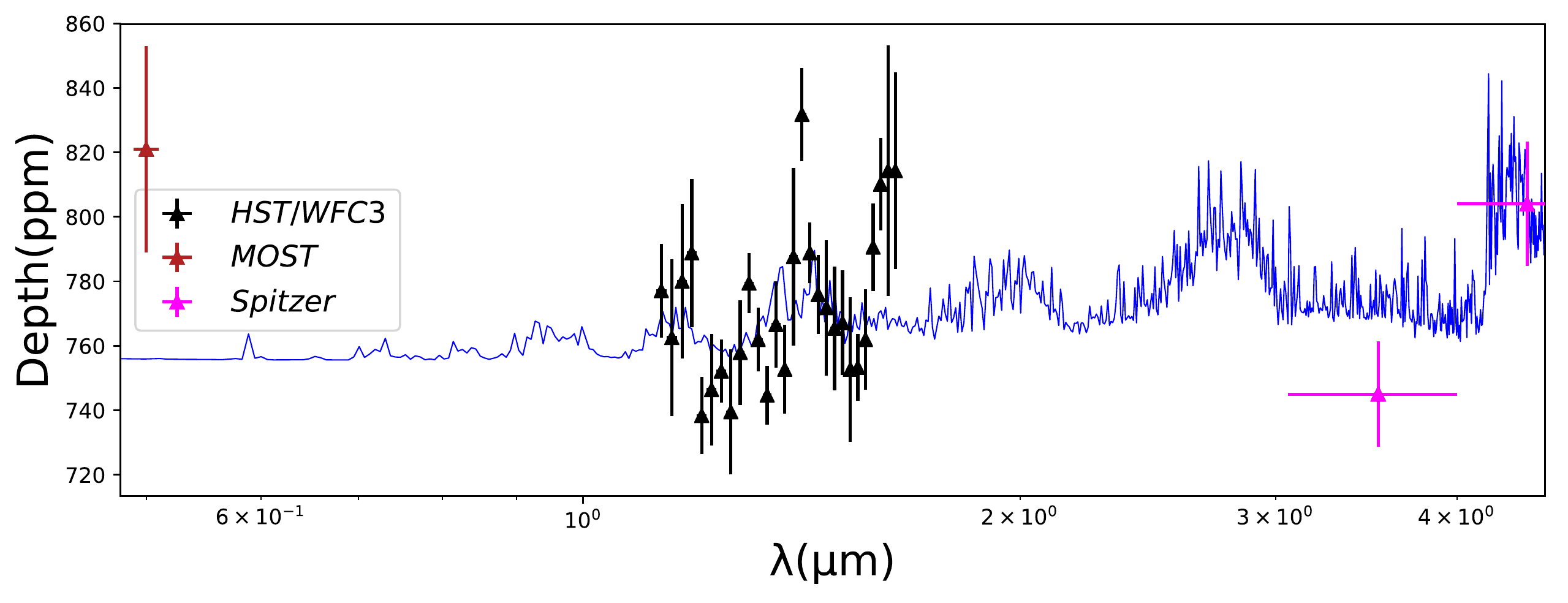}
	}%
\end{center}
\caption{Best-fit full range retrieval model ($T_{\rm eq}=770~$K, $\log(Z)=2.7$, C/O=1.0, and $\log{P(\rm Pa)}=3.6$, with a \WFC3\ spectrum shift of -78~ppm) against all transit depth results from \WFC3, \most, and \spitzer. The full range retrieval parameter posteriors are consistent with retrieval results using only the \WFC3\ dataset within 1$\sigma$ uncertainty. Horizontal error bars indicate the widths of the \most\ and \spitzer\ photometric bandpasses.} 
\label{fig:PLATON_noSTIS_best}
\bigskip
\end{figure*}

Additionally, we also combine all transit depth results from \WFC3, \most, and \spitzer\ to  perform a full  (0.5$\mu$m -- 4.5$\mu$m) transmission spectrum analysis using PLATON. During the fitting, the observed mean transit depth of the \WFC3\ spectrum is adjusted while the relative spectral shape is fixed to account for the uncertainty in visit-long trend model selection when analyzing the \WFC3\ data (section \ref{sec:spectral_lc}). The retrieved parameter posteriors are consistent with retrieval results using only the \WFC3\ dataset within 1$\sigma$ uncertainty. We plot the best-fit full range retrieval model ($T_{\rm eq}=770~$K, $\log(Z)=2.7$, C/O=1.0, and $\log{P(\rm Pa)}=3.6$, with a \WFC3\ spectrum shift of -78~ppm) against our transmission spectrum datasets in Figure \ref{fig:PLATON_noSTIS_best}. And since adding three data points (\most\ and \spitzer) in the spectrum does not change our atmospheric analysis result, we leave out the repeated process in the following sections and only present results using the \WFC3\ transmission spectrum. 

\subsection{Posterior Marginalization with ATMO}\label{sec:post_marginal}

\begin{figure*}
\begin{center}
\hspace*{-0.5cm}
    \subfigure
	{%
	\includegraphics[width=0.32\textwidth]{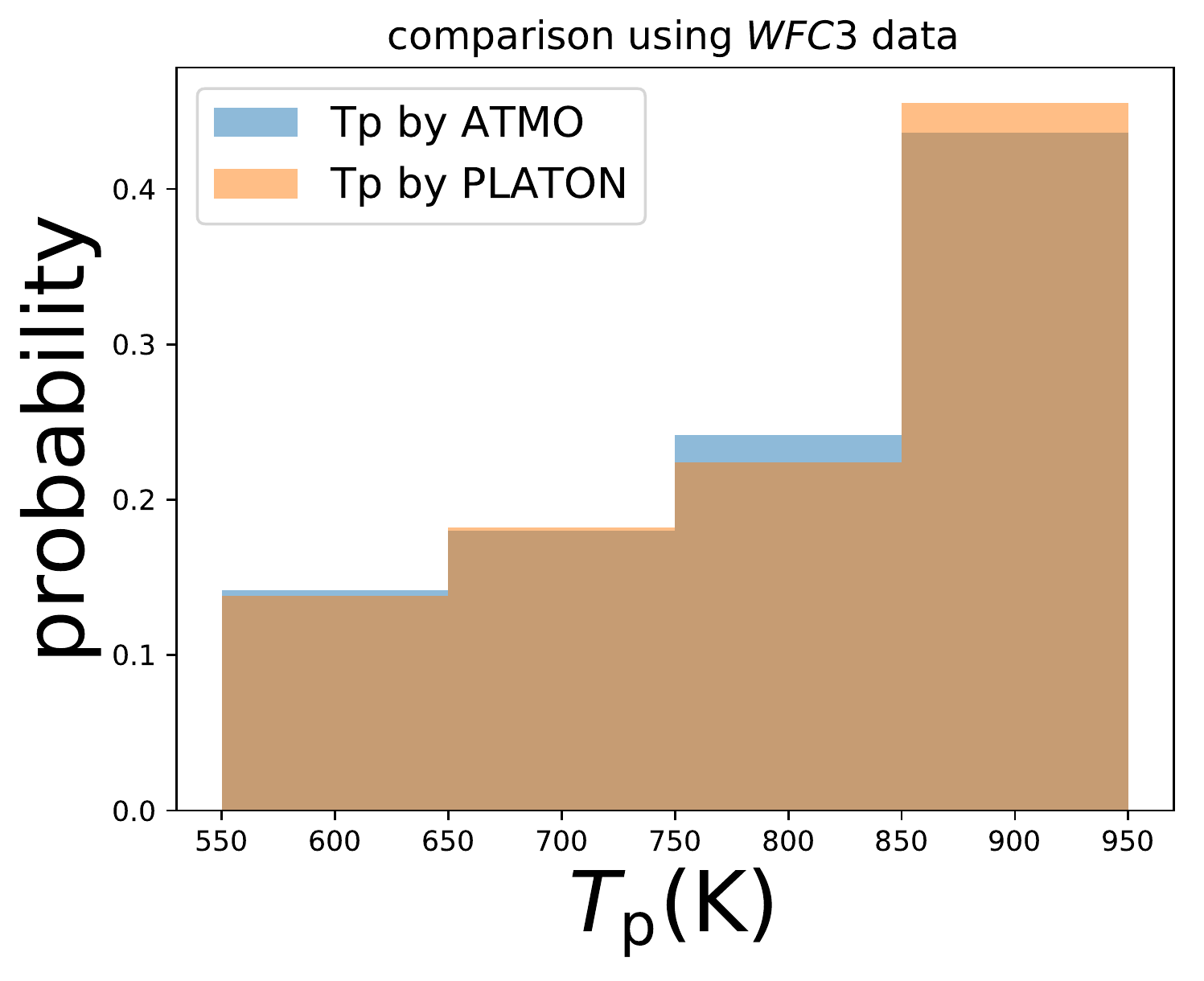}
	}
    \subfigure
	{%
	\includegraphics[width=0.32\textwidth]{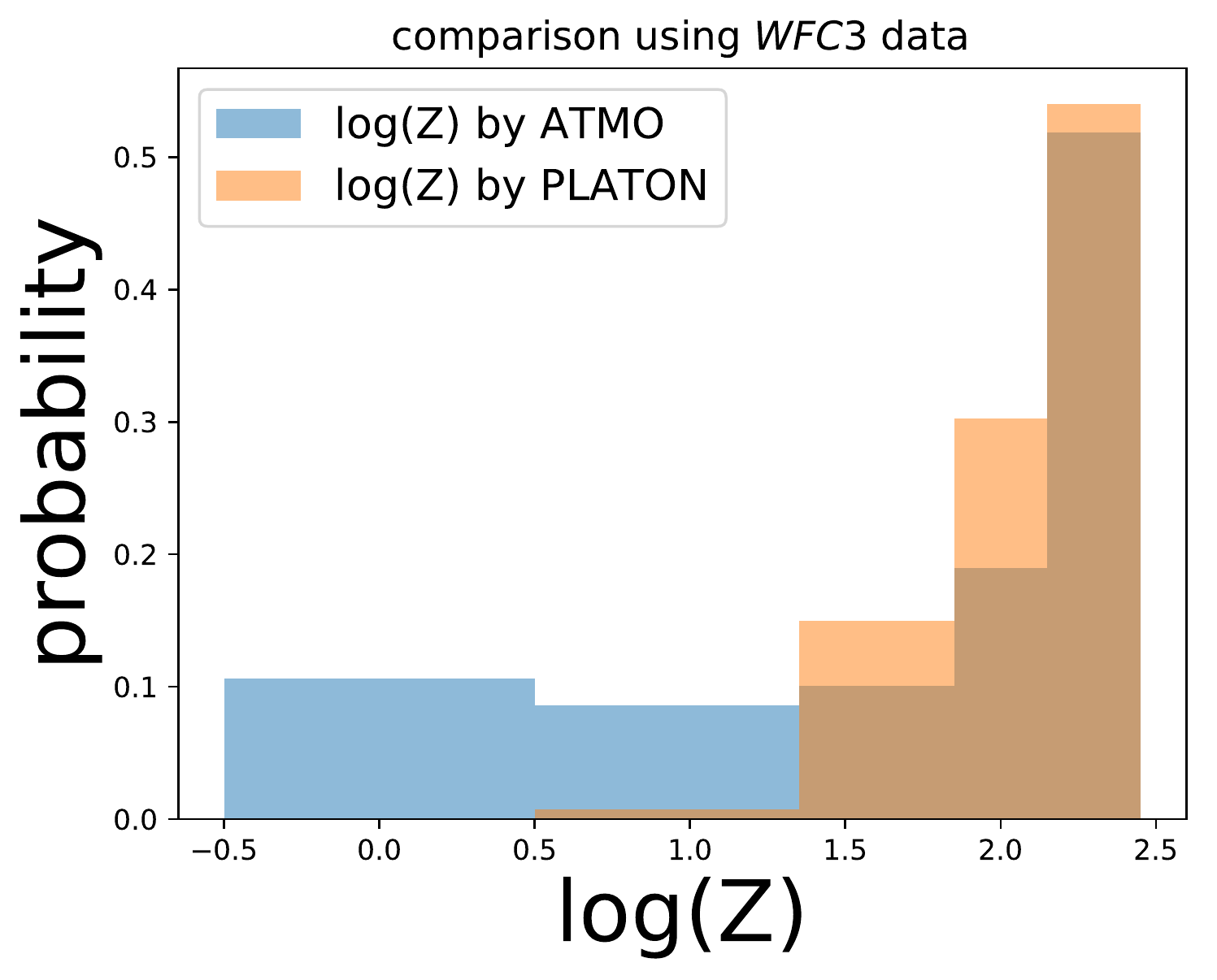}
	}
    \subfigure
	{%
	\includegraphics[width=0.32\textwidth]{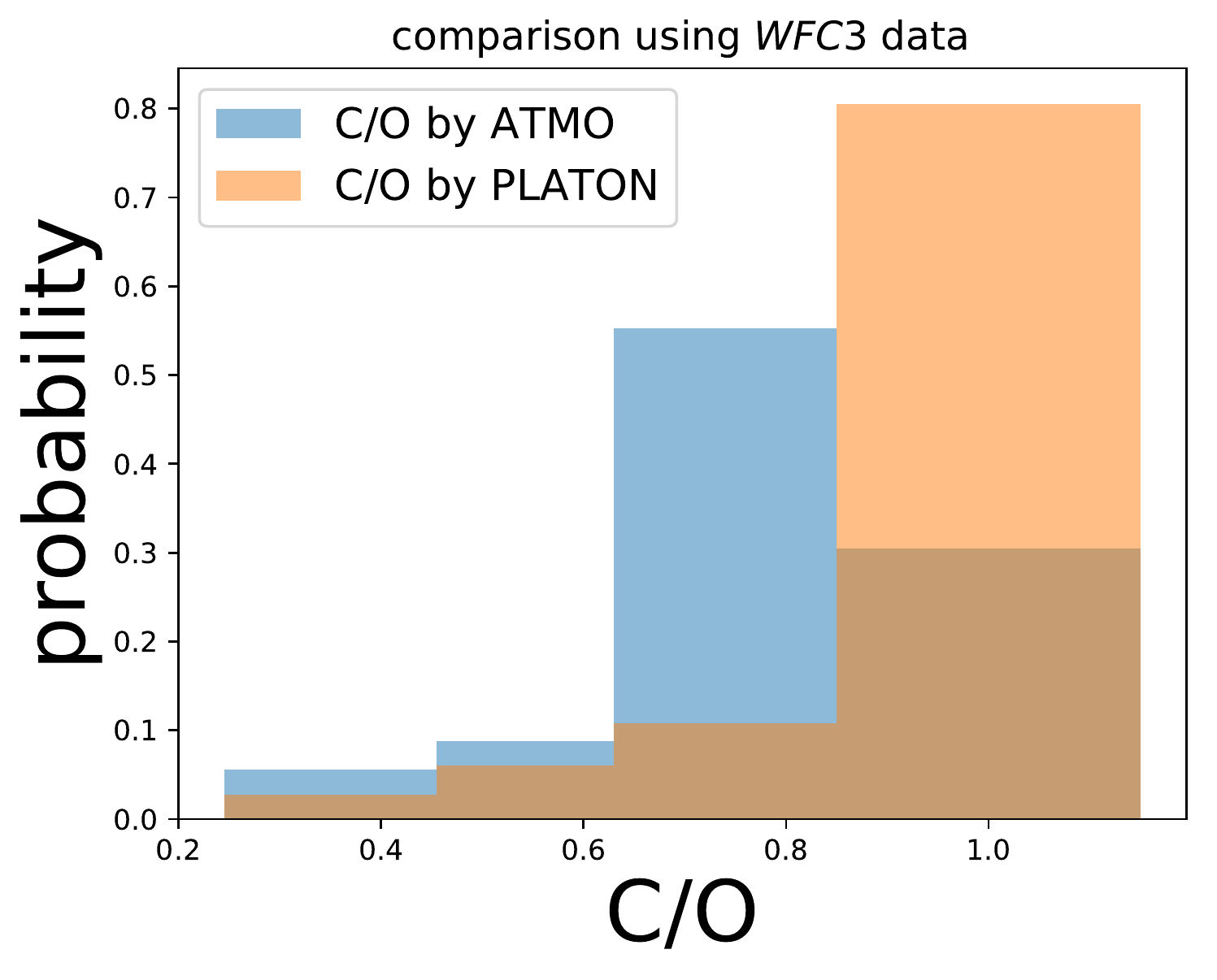}
	} 
\end{center}
\caption{Comparison of posteriors calculated using PLATON and ATMO models. We evaluate the likelihood of PLATON models on the same parameter grid as the ATMO models. The results show great consistency in the posterior of $T_{\rm p}$. For the posterior distribution of $\log(Z)$, ATMO produces a bigger tail in the low metallicity end, while they both peak at the high metallicity (200$\times$solar) end; and for the posterior distribution of C/O ratio, ATMO models favor $\rm C/O=0.7$ while the PLATON models favor values as high as 1.0.}
\label{fig:ATMO_compare}
\bigskip
\end{figure*}

In addition to PLATON, we use an existing generic forward model generated with ATMO \citep{Goyal2019, Tremblin2015} to further explore HD~97658b's atmospheric properties. We adopt this method based on PLATON's assumption that planetary atmospheres can be parameterized using only bulk metallicity enhancement, C/O ratio and equilibrium chemistry. We first take a subset of the entire grid, which spans across four planet equilibrium temperatures from 600~K to 900~K, two surface gravities 10 and 20~$\rm m/s^2$, five atmospheric metallicities (1$\times$ -- 200$\times$ solar), four C/O ratios (0.35--1.0), and four uniform cloud parameters. The equilibrium temperature grid span is constrained by our equilibrium temperature prior, while the other three dimension span are constrained by the original ATMO grid upper and lower limits. At each parameter setting, we also adjust the planet's photospheric radius, and find its best-fit value by maximizing the log-likelihood. Then we calculate the posterior distribution of each parameter by marginalizing maximum likelihoods over all the rest dimensions. In order for the comparison to be effective, we repeat the same evaluation process with PLATON's transmission spectra generation module on the same parameters grid, and the resulting posterior distribution comparisons of three critical atmospheric parameters are shown in Figure \ref{fig:ATMO_compare}. 

High consistency is achieved for $T_{\rm p}$ posterior distributions. For the distributions of $\log(Z)$ and C/O, ATMO and PLATON models give similar shapes. ATMO produces a bigger tail in the low metallicity end, but ATMO and PLATON both indicate high metallicity (200$\times$solar) peaks on the $\log(Z)$ plot; on the C/O ratio plot, ATMO models favor $\rm C/O=0.7$ while  PLATON models favor $\rm C/O=1.0$ or higher.

\subsection{Forward Modeling with PLATON}\label{sec:forward_model}

In the previous retrieval sections we showed that no clear peak value of the parameters are found even when  a large volume of atmospheric parameter space is explored. We now also pick out a small subset of forward models to compare with our data. 

\begin{figure*}
\includegraphics[width=\textwidth]{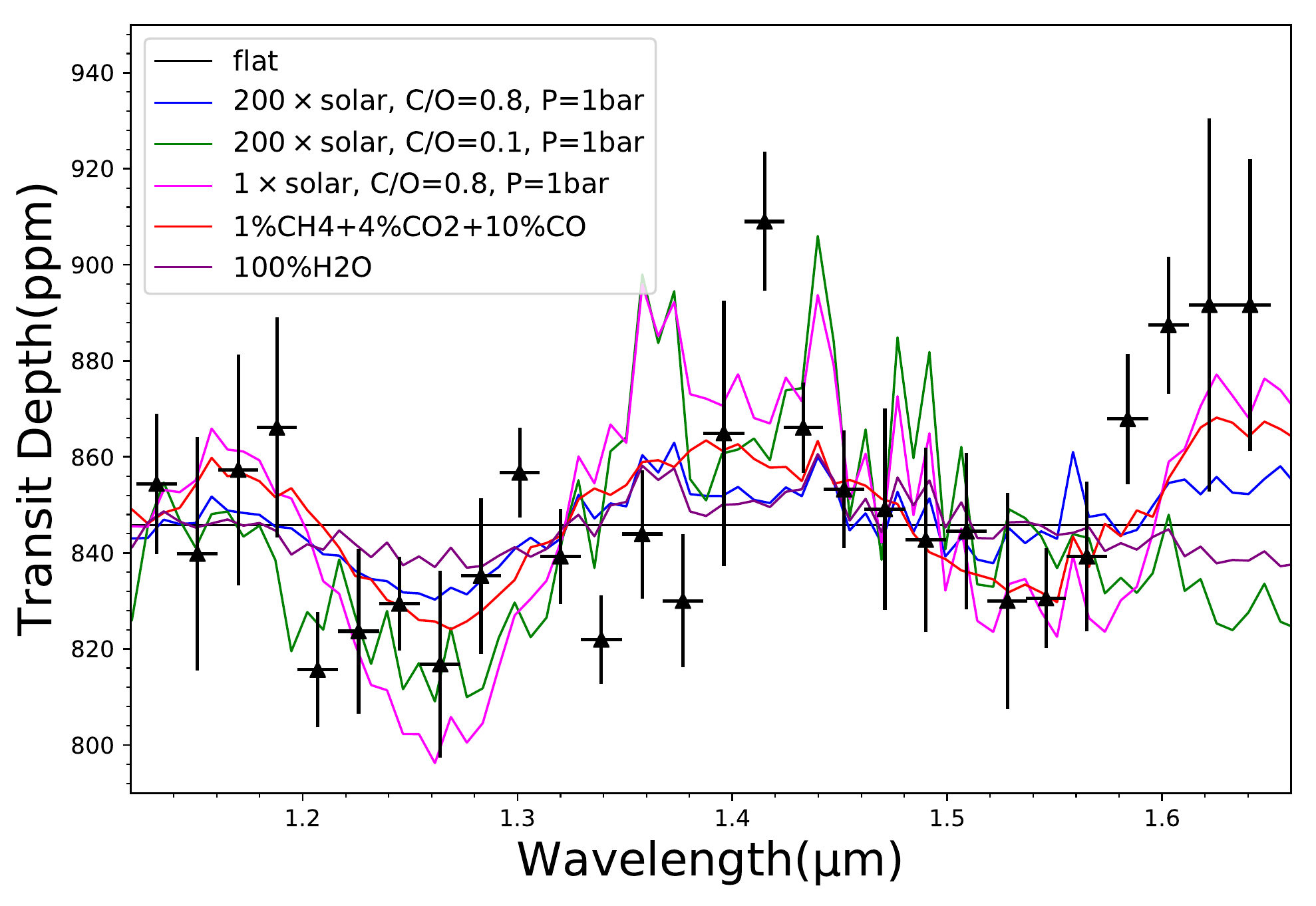}
\caption{Our extracted transmission spectrum as black data points, and five typical atmospheric models that we are fitting our transmission data with, along with a flat model that assumes no atmosphere. The transmission spectrum here is the same as the magenta spectrum reported in Figure \ref{fig:compareTrends_spectraShape}, except that in Figure \ref{fig:compareTrends_spectraShape}, the spectrum is shifted to have zero mean. For all atmospheric models plotted here, the equilibrium temperature $T_{\rm p}$ is 900~K and the cloud-top pressure is 1~bar. The model that describe the data the best is a $\rm H_2$ dominated atmosphere with 1\%CH4+4\%CO2+10\%CO in number fraction, although the model with $200\times$solar or $1000\times$solar metallicity and C/O=0.8, and the atmosphere composed of 100\% $\rm H_2O$ are not exluded.}
\label{fig:forwardModel}
\bigskip
\end{figure*}

As a baseline, we test similar scenarios that were considered in \cite{Knutson2014}, which includes a range of models with varying metallicity, cloud top pressure, and C/O ratio, and a range of models composed of only $\rm H_2$ and $\rm H_2O$ with varying $\rm H_2O$ number fractions. On top of previously proposed models, we also investigate the effect of adjusting the abundance of other molecular species in the atmosphere. Several typical atmospheric models that we investigated are presented in Figure \ref{fig:forwardModel}, along with a perfectly flat model (i.e., a wholly featureless transmission spectrum). The equilibrium temperature is set to $900$~K in all atmospheric models, and we adjust the mean transit depth of the model spectrum until finding a location with the maximum likelihood. Table \ref{tab:forwardModel} summarize all models that have been compared with our transmission spectrum data, as well as the best model retrieved with PLATON ($T_{\rm eq}=950~$K, $\log(Z)=2.2$, C/O=0.8, and $\log{P(\rm Pa)}=7.8$). Their reduced $\chi^2$ values and the confidence levels with which they can be ruled out are presented in the same table. 

The result shows that with this data set, even the best-fit atmospheric model is excluded with a 4$\sigma$ significance level, indicating that the uncertainties associated with the transmission spectrum are underestimated, as pointed out in section \ref{sec:platon}, despite the careful error budget treatments we implemented. 

As mentioned in section \ref{sec:platon}, the retrieval with PLATON shows that $\chi_{\nu}^2\approx 1.0$ can be achieved for best-fit models when the spectrum data uncertainties are scaled up by a factor of around 1.5. We apply this scaling factor to our data and recalculate the forward modeling $\chi_{\nu}^2$ and number of sigmas, and present the results in the last two columns of Table \ref{tab:forwardModel}. After applying the scale factor, we find that only models with low metallicity (1$\times$solar) or low C/O ratio (C/O=0.1) can be ruled out with around $3\sigma$ significance.

In light of the above analyses, we conclude that the existing atmospheric models can barely be distinguished with the current \HST/\WFC3\ data set, although the Z=200$\times$solar, C/O=0.8 and P=1~bar or 10~mbar models, which fall into high posterior regions presented in section \ref{sec:platon} and \ref{sec:post_marginal}, and a H-dominated atmosphere consisting of around 1\% of $\rm CH_4$, a few percents of $\rm CO_2$ and around 10\% of CO are favored over other metallcity, C/O ratio and cloud top pressure settings or other atmosphere molecular combinations. 

We propose that more observations of HD~97658b are needed in order to constrain its atmospheric features tighter, and more discussions are provided in section \ref{sec:discussion}.


\begin{table*}
\renewcommand*{\arraystretch}{1.2}
    \caption{Forward modeling of HD~97658b transmission spectrum in the \HST/\WFC3\ G141 bandpass}
    \centering
    \begin{tabularx}{1.0\textwidth}{cnnnn}
    \hline\hline
    Model & $\chi_{\nu}^2$ & rule-out confidence & $\chi_{\nu}^2$(error scaled up) & rule-out confidence(error scaled up) \\
    \hline
    Flat & 2.5 & 5.4$\sigma$ & 1.2 & 0.6$\sigma$\\
    Best model retrieved with PLATON & 2.5 & 4.9$\sigma$ & 1.2 & 0.5$\sigma$\\
    1$\times$solar, C/O=0.8, P=1~bar & 4.1 & 10.2$\sigma$ & 1.9 & 3.0$\sigma$\\
    200$\times$solar, C/O=0.8, P=1~bar & 2.5 & 5.0$\sigma$ & 1.2 & 0.5$\sigma$\\
    200$\times$solar, C/O=0.8, P=10~mbar & 2.6 & 5.3$\sigma$ & 1.2 & 0.7$\sigma$\\
    200$\times$solar, C/O=0.8, P=1~mbar & 3.3 & 7.7$\sigma$ & 1.5 & 1.7$\sigma$\\
    200$\times$solar, C/O=0.1, P=1~bar & 4.0 & 9.9$\sigma$ & 1.9 & 2.8$\sigma$\\
    1000$\times$solar, C/O=0.8, P=1~bar & 2.7 & 5.8$\sigma$ & 1.2 & 0.8$\sigma$\\
    1\%CH4+4\%CO2+10\%CO & 2.2 & 4.0$\sigma$ & 1.0 & 0.1$\sigma$\\
    4\%CO2+10\%CO & 3.3 & 8.1$\sigma$ & 1.5 & 1.8$\sigma$\\
    100\%H2O & 2.4 & 5.2$\sigma$ & 1.1 & 0.4$\sigma$\\
    50\%H2O & 2.6 & 5.6$\sigma$ & 1.2 & 0.7$\sigma$\\
    \hline
    \end{tabularx}
    \label{tab:forwardModel}
\bigskip
\end{table*}

\bigskip

\section{Summary and Discussion}\label{sec:discussion}

We analyzed four \HST/\WFC3\ observations on a bright ($V=7.7$) K1 dwarf HD~97658 using the RECTE ramp modeling method, measured their combined transmission spectrum consisting of 28 channels (from 1.1~$\mu$m to 1.7~$\mu$m) with the divide-by-white method, and achieved an uncertainty of around 20~ppm in each spectral channel. Of the four \HST/\WFC3\ observations, two have previously been analyzed and published in \citet{Knutson2014}, and our reanalysis in combination with the two new observations suggests that the slight upward slope reported in the original paper is likely systematic.

An atmospheric retrieval from the obtained transmission spectrum is attempted with the PLATON package. We found that the atmospheric metallicity posterior peaks around $\log(Z) = 2.4$, high C/O ratio ($\rm C/O\gtrsim 0.8$) is favored, and the cloud could be covering up to as high as a few millibar pressure, while a factor of 1.5 is suggested to be applied to the spectral data uncertainties so that the best-fit model has a $\chi_{\nu}^2$ of around 1. In a second experiment, marginalized likelihood distributions of atmospheric parameters calculated by fitting our transmission data to forward models generated with PLATON on a parameter grid is compared with those calculated with the generic forward models generated with ATMO in Figure \ref{fig:ATMO_compare}, and consistency is found between these two modeling tools. Subsequently, we generate a range of transmission models with various $\log{Z}$, C/O ratio, and cloud-top pressure $P$ values, or different molecular fraction combinations, and calculate their reduced-$\chi^2$ and significance with which they can be ruled out by our \HST/\WFC3\ data. When applying the 1.5 error bar scale factor revealed by the PLATON retrieval, we find that only models with low metallicity (1$\times$solar) or low C/O ratio (C/O=0.1) can be ruled out with around $3\sigma$ significance. 

The C/O ratio of an exoplanet depends on the environment from which it accretes its gaseous envelope. 
Previous work has reported the C/O ratio of the host star HD~97658 to be 0.45 with around 10\% uncertainty \citep{Hinkel2017}, slightly lower than the Solar C/O ratio of 0.54. \HL{If future observations like \JWST\ confirms HD~97658b's high C/O ratio hinted by the analysis from this work,} this planet may have a C/O ratio significantly discrepant from that of its host star.  
\citet{Oberg2011} predicted that in a core accretion model, planets formed between the $\rm H_2O$ and CO snowlines have a large fraction of oxygen preserved in icy form, leading to elevated C/O ratio in the atmosphere. However, a planet as small as HD~97658b is unlikely to have formed beyond the $\rm H_2O$ snowline, and more recent models suggest that planetesimal accretion by sub-Neptunes results in sub-stellar C/O ratios \citep{Cridland2019}. To further investigate the origin of a possible high C/O ratio in HD~97658b atmosphere, studies of this planet's formation history are needed. 

In addition to \HST/\WFC3\ transit analyses, we also analyzed eleven new transit observations of HD97658~b with \spitzer\ and eight transits (three old and five new) with \most. We updated their ephemeris, applied a linear fit on all mid-transit times, and updated the orbital period of HD97658~b to be $P=9.489295 \pm 0.000005$, with a best-fit initial mid-transit time at $T_0 = 2456361.8069 \pm 0.0004$~(BJD). The best-fit reduced-$\chi^2$ is 1.7, indicating that no TTV is detected. 

\HL{As a reference to future works, we summarize the stellar and planet parameters of the HD~97658 system in Table \ref{tab:summary} using the analysis results from this work and preferred values from previous works. }

\subsection{\HL{Transit Light Source Effect on HD~97658b}}

\HL{\citet{Rackham2018} proposed the transit light source effect, which describes the problem that spots and faculae of M~dwarf stars can produce contamination in the transmission spectra of nearby planets more than 10 times larger than the transit depth changes expected from planet atmospheric features. In \citet{Rackham2019}, the transit light source effect analysis was extended to F/G/K host stars. They found that while the stellar contamination can bias the transit depth measurement by as much as ~1\% for late G and K type dwarfs from UV to NIR, the offsets in transmission spectral features, including $\rm H_2O$, $\rm CH_4$, $\rm O_3$ and $\rm CO$ amplitudes, induced by F/G/K stellar contamination assuming both spots only and spots+faculae models on stellar surface are far smaller than atmospheric features expected in transmission spectra of planets around these stars. Therefore, the stellar contamination in the \HST/\WFC3\ wavelength range and the \spitzer\ IRAC bandpass is not problematic for transmission spectroscopy analysis for typically active F/G/K dwarfs. As is described in section \ref{sec:activity}, HD~97658b's activity cycle and rotation period are typical of early K type stars, and its overall activity level ($R^{'}_{\rm HK}$ and $S_{\rm HK}$) and $\rm 2~ms^{-1}$ RV variations indicate that HD~97658b is slightly quieter than average K type stars. These facts ensure the reliability of the transmission spectroscopy results presented in this work. Nonetheless, \citet{Rackham2019} shows that F/G/K stellar contamination may have larger impacts on transmission spectra at UV and visual wavelengths, in which TiO and VO display significant opacity. Therefore we note that stellar contamination should be taken care of when analyzing the transmission spectrum of HD~97658b at UV and visual wavelengths in future works.}

\subsection{Transmission Spectroscopy Metric of All Small Planets Cooler Than 1000~K}\label{sec:tsm}

We assess the transmission spectrum detectability of all currently confirmed small planets with future missions by calculating the Transmission Spectroscopy Metric (TSM hereafter) of each planet, as defined in \cite{Kempton2018}. The TSM is defined to approximate the signal-to-noise ratio ($SNR$) of one scale height in transmission spectra when observed with the \NIRISS\ instrument on \JWST\ for 10 hours; therefore, it is proportional to $R_{\rm p}H/R_{*}^2\times F$, where $H=kT_{\rm eq}/\mu g$ is the atmospheric scale height, $R_{\rm p}$ is the planet radius, $R_{*}$ is the stellar radius, and $F$ is the stellar flux received on detectors. For planets with $R_{\rm p}<2~R_{\oplus}$, a mean molecular weights of $\mu=18$ is chosen, representing pure water atmospheres, and for planets with $R_{\rm p}\geq2~R_{\oplus}$, a mean molecular weights of $\mu=2.3$ is chosen, representing $\rm H/He$-dominated atmospheres. The final formula of TSM is shown as equation~(1) of \cite{Kempton2018}
Applying the scale factors specified for different $R_{\rm p}$ bins in \cite{Kempton2018}, the TSM value represents near-realistic $SNR$ of 10-hour \JWST\ observations.

\begin{figure*}
\includegraphics[width=1.02\textwidth]{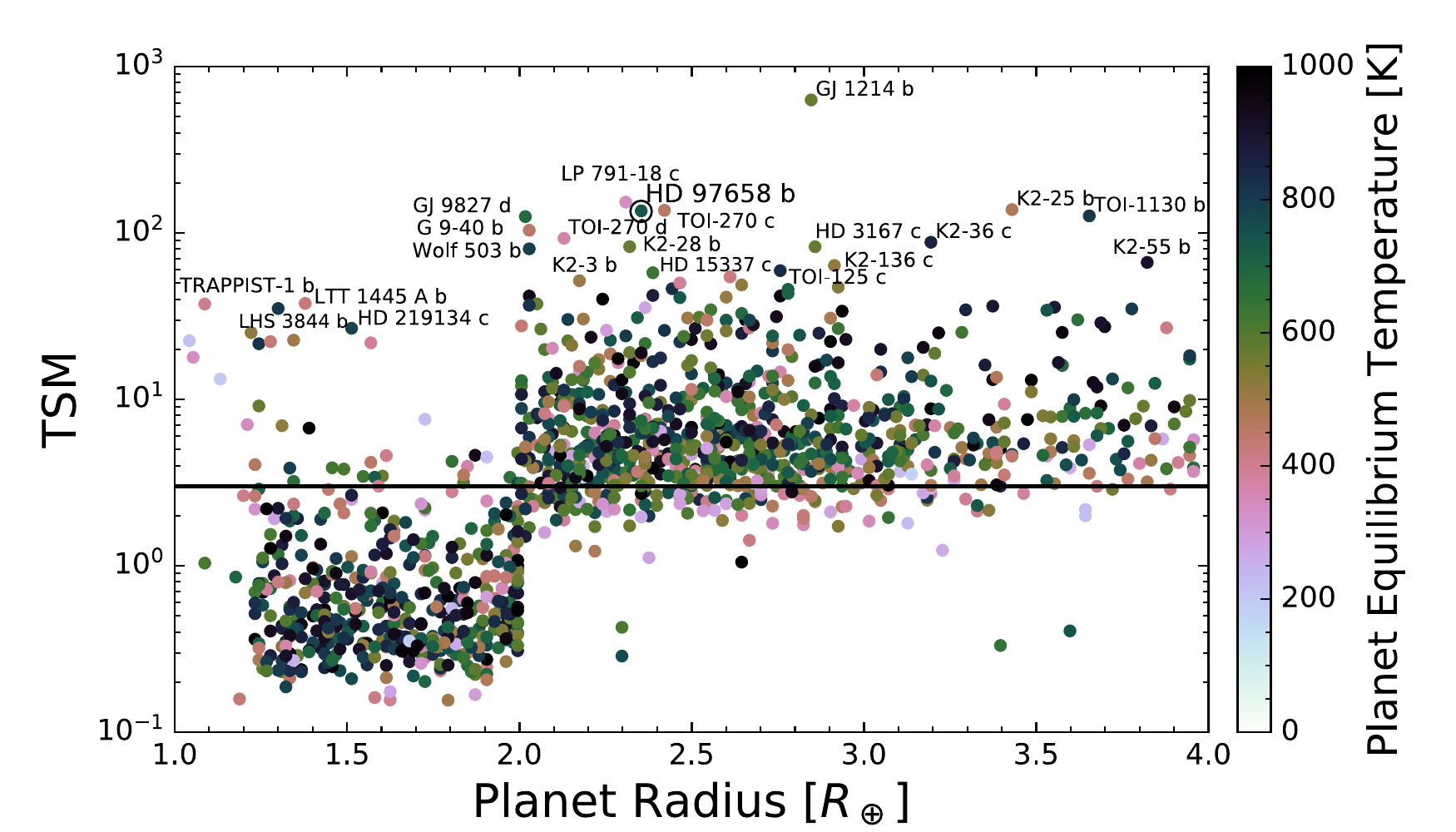}
\caption{The TSM distribution of all planets cooler than 1000~K and with sizes between 1~$\rm R_{\oplus}$ and 4~$\rm R_{\oplus}$. The color of data points represent the equilibrium temperature of each planet, and we mark the $\rm TSM=3.0$ limit with a horizontal line. And the abrupt transition at $2~R_{\oplus}$ is due to assumptions made for the atmospheric mean molecular weight when calculating the TSM of each planet.}
\label{fig:TSM}
\bigskip
\end{figure*}

Following the above procedure, we calculate TSM of all confirmed planets with $1~R_{\oplus}<R_{\rm p}<4~R_{\oplus}$ and equilibrium temperature $T_{\rm eq}<1000~$K, and rank them according to their TSM values. The twenty highest rank planets are listed in table \ref{tab:TSM}, and the full table will be available online. In Figure \ref{fig:TSM}, we show the TSM value of each planet versus their radius, and each data point is color coded with the planet equilibrium temperature. The names of the top 20 planets with the highest TSM values are labeled. The shift in TSM values that we observe between planets smaller than $2~R_{\oplus}$ and larger than $2~R_{\oplus}$ is caused by an artificial sudden jump of mean molecular weight values of these two groups of planets as described above, but it represents the actual TSM trend when we move from small planets to larger planets. 

\citet{Louie2018} simulates the transmission spectroscopy $SNR$ of 18 known planets with sizes between $\rm 0.5~R_{\oplus}$ and $\rm 4.0~R_{\oplus}$ assuming they are observed with \JWST/\NIRISS. 12 of the 18 planets are also in our sample. Their treatment of atmospheric mean molecular weight is similar to our method, except that they divide the exoplanets into 2 categories at $1.5~R_{\oplus}$ instead of $2.0~R_{\oplus}$, and as a result, there is a similar $SNR$ shift at $1.5~R_{\oplus}$ in Figure 5 of \citet{Louie2018}. Although we are measuring the detectability with different \JWST\ instruments, and the definition of TSM in this work is slightly different from the simulated $SNR$ in \citet{Louie2018}, the resulting TSM and the $SNR$ from \citet{Louie2018} of the overlapping planets are highly consistent. 10 of the 12 overlapping planets are ordered the same in this work as in \citet{Louie2018}, and the TSM values of all 12 planets are between 0.5 to 2 times of their $SNR$ values in \citet{Louie2018}, showing that our analysis is reliable. 

Among all planets in our sample, GJ~1214b has the highest TSM value, but the transmission spectrum of GJ~1214b has been revealed to be featureless from 1.1~$\mu$m to 1.6~$\mu$m \citep{Kreidberg2014a}, indicating a cloudy atmosphere or no atmosphere. Nevertheless, opacity of clouds may vary across a broader wavelength range, and emission and reflection spectra could also contain additional features. Therefore, GJ~1214b could still be a valuable target for \JWST\ observation. Ranked second is K2-25b, a Neptune-sized planet orbiting a M4.5 dwarf in the 650-Myr-old Hyades cluster \citep{Mann2016, ChiaThao2019}. The unusually large size ($3.43~R_{\oplus}$) of K2-25b in comparison to other planets with similar orbital periods (3.485 days) and the fact that it is orbiting a young star suggest that K2-25b could represent an early or intermittent phase of planetary evolution, where young Neptunes are in the process of losing their atmospheres \citep{Mann2016}. These features make K2-25b a target of high scientific value, although we have to be careful that a young M~dwarf like K2-25 could have large spot and faculae covering fractions, which may generate stellar contamination as large as several times the transit depth changes expected in the transmission spectrum of K2-25b due to its atmospheric features \citep{Rackham2018}. Ranked fifth is HD~97658b, the planet analyzed in this work. Since we are not able to precisely determine the atmospheric composition of HD~97658b with the current \HST/\WFC3\ data as described in previous sections, we simulate \JWST\ observations of HD~97658b using its Near InfraRed Spectrograph (\NIRSpec) instrument in the following subsection, and analyze quantitatively how well we can characterize the atmosphere of HD~97658b with \JWST\ observations.

Out of all 1404 planets in Figure \ref{fig:TSM}, 515 have $\rm TSM>5.0$ and 820 have $\rm TSM>3.0$. At least a third of small planets cooler than 1000~K can be well characterized using \JWST, and more valuable targets will be added to the pool with the ongoing \TESS\ mission.

\subsection{\JWST\ Simulation of HD~97658b}\label{sec:JWSTsimu}

\begin{figure*}
\includegraphics[width=\textwidth]{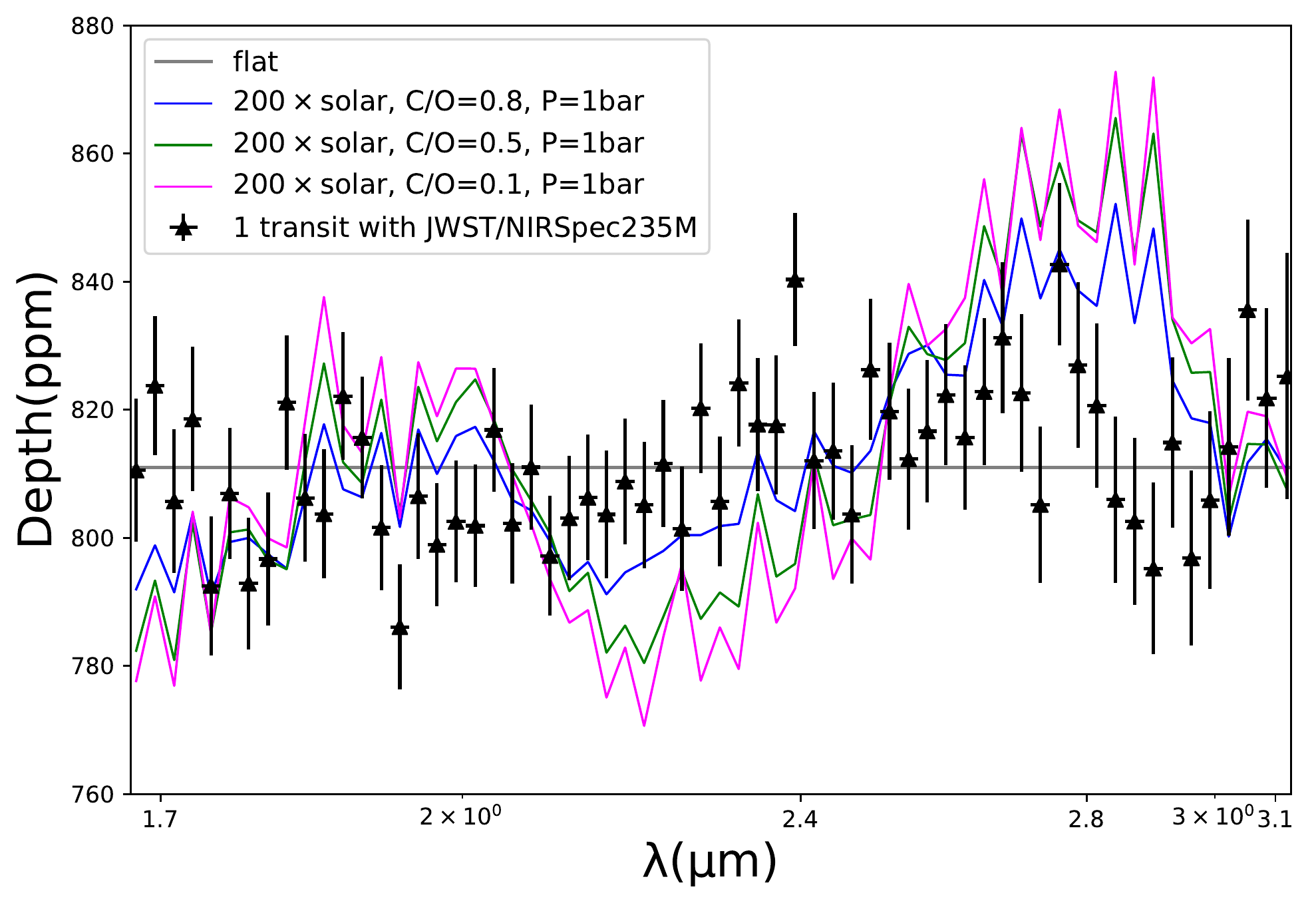}
\caption{HD~97658b transmission spectrum simulated with \JWST/\NIRSpec\ G235M filter assuming one transit observation and a 200$\times$solar metallicity atmosphere with 1~bar cloud-top pressure and $\rm C/O=0.8$. In comparison, two other atmospheric models with different $\rm C/O$ values and a flat spectrum are also shown.}
\label{fig:JWSTsimu}
\bigskip
\end{figure*}

The upcoming James Webb Space Telescope (\JWST) mission, with a 6.5 meter diameter primary mirror and four near/mid-infrared instruments covering wavelength range from 0.6~$\mu$m to 28.5~$\mu$m will provide unprecedented opportunities to characterize exoplanets of all sizes and environments. Here we simulate a transmission spectrum of HD~97658b as observed by \JWST's Near InfraRed Spectrograph (\NIRSpec) with it's G235M filter (1.6~$\mu$m to 3.2~$\mu$m) using the PANDEXO package \citep{PandExo}. We assume one transit observation with 45\% in-transit time and 90\% saturation level. A reasonably optimistic noise floor of 30~ppm, 
is adopted according to \citet{Greene2016}. Figure \ref{fig:JWSTsimu} shows the simulated transmission spectrum observed by \JWST\ assuming a 200$\times$solar metallicity atmosphere with 1~bar cloud-top pressure and $\rm C/O=0.8$, and compares it with a flat spectrum and two other atmospheric models with $\rm C/O=0.5$ and 0.1 respectively. Our calculation shows that with only one transit observed by \JWST, we will be able to distinguish a $\rm C/O=0.8$ atmosphere from a $\rm C/O=0.7$ atmosphere, a $\rm C/O=0.5$ atmosphere and a $\rm C/O=0.1$ atmosphere with 5$\sigma$, 12$\sigma$ and 17$\sigma$ significance respectively, which will enable further study of the formation environment and formation process of HD~97658b. The same data will also be able to exclude the flat spectrum with 4$\sigma$ significance.

With this result, and considering the fact that HD~97658b is ranked the fifth best target for future transmission spectra observations from our previous calculations, we propose HD~97658b to be assigned top priority in \JWST\ exoplanet proposals.\\

\bigskip

\noindent
\textbf{Software Usage:} \\
BAsic Transit Model cAlculatioN (BATMAN) \citep{Kreidberg2015}; \\
Celerite \citep{celerite};\\
EXOFASTv2 \citep{Eas17};\\
RadVel \citep{Fulton2017};\\
PLATON \citep{PLATON_Zhang2019}; \\
PANDEXO \citep{PandExo}.

\bigskip

\section{Acknowledgements}

We acknowledge support for the analysis by NASA through grants under the HST-GO-13665 program.

I.J.M.C. acknowledges support from the NSF through grant AST-1824644 and through NASA and STScI through funding for program GO-13665.

D.D. acknowledges support provided by NASA through Hubble Fellowship grant HSTHF2-51372.001-A awarded by the Space Telescope Science Institute, which is operated by the Association of Universities for Research in Astronomy, Inc., for NASA, under contract NAS5-26555.

M.R.K acknowledges support from the NSF Graduate Research Fellowship, grant No. DGE 1339067.

G.W.H. acknowledges long-term support from NASA, NSF, Tennessee State University, and the State of Tennessee through its Centers of Excellence program.

A.W.H acknowledges NSF grant AST-1517655.

\bibliographystyle{aasjournal}

\begin{thebibliography}{}
\expandafter\ifx\csname natexlab\endcsname\relax\def\natexlab#1{#1}\fi
\providecommand{\url}[1]{\href{#1}{#1}}
\providecommand{\dodoi}[1]{doi:~\href{http://doi.org/#1}{\nolinkurl{#1}}}
\providecommand{\doeprint}[1]{\href{http://ascl.net/#1}{\nolinkurl{http://ascl.net/#1}}}
\providecommand{\doarXiv}[1]{\href{https://arxiv.org/abs/#1}{\nolinkurl{https://arxiv.org/abs/#1}}}

\bibitem[{{Agol} {et~al.}(2010){Agol}, {Cowan}, {Knutson}, {Deming}, {Steffen},
  {Henry}, \& {Charbonneau}}]{Agol2010}
{Agol}, E., {Cowan}, N.~B., {Knutson}, H.~A., {et~al.} 2010, \apj, 721, 1861,
  \dodoi{10.1088/0004-637X/721/2/1861}

\bibitem[{{Anglada-Escud{\'e}} \& {Tuomi}(2012)}]{Anglada2012}
{Anglada-Escud{\'e}}, G., \& {Tuomi}, M. 2012, Astronomy and Astrophysics, 548,
  A58, \dodoi{10.1051/0004-6361/201219910}

\bibitem[{{Baliunas} {et~al.}(1996){Baliunas}, {Nesme-Ribes}, {Sokoloff}, \&
  {Soon}}]{baliunas:1996}
{Baliunas}, S.~L., {Nesme-Ribes}, E., {Sokoloff}, D., \& {Soon}, W.~H. 1996,
  \apj, 460, 848, \dodoi{10.1086/177014}

\bibitem[{Batalha {et~al.}(2017)Batalha, Mandell, Pontoppidan, Stevenson,
  Lewis, Kalirai, Earl, Greene, Albert, \& Nielsen}]{PandExo}
Batalha, N.~E., Mandell, A., Pontoppidan, K., {et~al.} 2017, Publications of
  the Astronomical Society of the Pacific, 129, 064501

\bibitem[{{Bell} {et~al.}(2017){Bell}, {Nikolov}, {Cowan}, {Barstow}, {Barman},
  {Crossfield}, {Gibson}, {Evans}, {Sing}, {Knutson}, {Kataria}, {Lothringer},
  {Benneke}, \& {Schwartz}}]{Bell2017}
{Bell}, T.~J., {Nikolov}, N., {Cowan}, N.~B., {et~al.} 2017, \apjl, 847, L2,
  \dodoi{10.3847/2041-8213/aa876c}

\bibitem[{{Benneke} {et~al.}(2019){Benneke}, {Knutson}, {Lothringer},
  {Crossfield}, {Moses}, {Morley}, {Kreidberg}, {Fulton}, {Dragomir}, {Howard},
  {Wong}, {D{\'e}sert}, {McCullough}, {Kempton}, {Fortney}, {Gilliland },
  {Deming}, \& {Kammer}}]{Benneke2019}
{Benneke}, B., {Knutson}, H.~A., {Lothringer}, J., {et~al.} 2019, Nature
  Astronomy, 377, \dodoi{10.1038/s41550-019-0800-5}

\bibitem[{{Berta} {et~al.}(2012){Berta}, {Charbonneau}, {D{\'e}sert},
  {Miller-Ricci Kempton}, {McCullough}, {Burke}, {Fortney}, {Irwin}, {Nutzman},
  \& {Homeier}}]{Berta2012}
{Berta}, Z.~K., {Charbonneau}, D., {D{\'e}sert}, J.-M., {et~al.} 2012, \apj,
  747, 35, \dodoi{10.1088/0004-637X/747/1/35}

\bibitem[{{Butler} {et~al.}(1996){Butler}, {Marcy}, {Williams}, {McCarthy},
  {Dosanjh}, \& {Vogt}}]{Butler1996}
{Butler}, R.~P., {Marcy}, G.~W., {Williams}, E., {et~al.} 1996, \pasp, 108,
  500, \dodoi{10.1086/133755}

\bibitem[{{Chia Thao} {et~al.}(2020){Chia Thao}, {Mann}, {Johnson}, {Newton},
  {Guo}, {Kain}, {Rizzuto}, {Charbonneau}, {Dalba}, {Gaidos}, {Irwin}, \&
  {Kraus}}]{ChiaThao2019}
{Chia Thao}, P., {Mann}, A.~W., {Johnson}, M.~C., {et~al.} 2020, \aj, 159, 32,
  \dodoi{10.3847/1538-3881/ab579b}

\bibitem[{{Claret} \& {Bloemen}(2011)}]{Claret2011}
{Claret}, A., \& {Bloemen}, S. 2011, \aap, 529, A75,
  \dodoi{10.1051/0004-6361/201116451}

\bibitem[{{Cridland} {et~al.}(2019){Cridland}, {van Dishoeck}, {Alessi}, \&
  {Pudritz}}]{Cridland2019}
{Cridland}, A.~J., {van Dishoeck}, E.~F., {Alessi}, M., \& {Pudritz}, R.~E.
  2019, \aap, 632, A63, \dodoi{10.1051/0004-6361/201936105}

\bibitem[{{Crossfield} \& {Kreidberg}(2017)}]{Crossfield2017}
{Crossfield}, I.~J.~M., \& {Kreidberg}, L. 2017, \aj, 154, 261,
  \dodoi{10.3847/1538-3881/aa9279}

\bibitem[{{Damiano} {et~al.}(2017){Damiano}, {Morello}, {Tsiaras}, {Zingales},
  \& {Tinetti}}]{Damiano2017}
{Damiano}, M., {Morello}, G., {Tsiaras}, A., {Zingales}, T., \& {Tinetti}, G.
  2017, \aj, 154, 39, \dodoi{10.3847/1538-3881/aa738b}

\bibitem[{{Deming} {et~al.}(2015){Deming}, {Knutson}, {Kammer}, {Fulton},
  {Ingalls}, {Carey}, {Burrows}, {Fortney}, {Todorov}, {Agol}, {Cowan},
  {Desert}, {Fraine}, {Langton}, {Morley}, \& {Showman}}]{Deming2015}
{Deming}, D., {Knutson}, H., {Kammer}, J., {et~al.} 2015, \apj, 805, 132,
  \dodoi{10.1088/0004-637X/805/2/132}

\bibitem[{{Demory} {et~al.}(2015){Demory}, {Ehrenreich}, {Queloz}, {Seager},
  {Gilliland}, {Chaplin}, {Proffitt}, {Gillon}, {G{\"u}nther}, {Benneke},
  {Dumusque}, {Lovis}, {Pepe}, {S{\'e}gransan}, {Triaud}, \&
  {Udry}}]{Demory2015}
{Demory}, B.-O., {Ehrenreich}, D., {Queloz}, D., {et~al.} 2015, \mnras, 450,
  2043, \dodoi{10.1093/mnras/stv673}

\bibitem[{{Dragomir} {et~al.}(2014){Dragomir}, {Benneke}, {Crossfield},
  {Howard}, \& {Knutson}}]{Dragomir2013SpitzerProp}
{Dragomir}, D., {Benneke}, B., {Crossfield}, I., {Howard}, A., \& {Knutson}, H.
  2014, {A Comparative Study of Super-Earth Atmospheres}, Spitzer Proposal

\bibitem[{{Dragomir} {et~al.}(2013){Dragomir}, {Matthews}, {Eastman},
  {Cameron}, {Howard}, {Guenther}, {Kuschnig}, {Moffat}, {Rowe}, {Rucinski},
  {Sasselov}, \& {Weiss}}]{Dragomir2013}
{Dragomir}, D., {Matthews}, J.~M., {Eastman}, J.~D., {et~al.} 2013, \apjl, 772,
  L2, \dodoi{10.1088/2041-8205/772/1/L2}

\bibitem[{{Eastman}(2017)}]{Eas17}
{Eastman}, J. 2017, {EXOFASTv2: Generalized publication-quality exoplanet
  modeling code}, Astrophysics Source Code Library.
\newblock \doeprint{1710.003}

\bibitem[{{Evans} {et~al.}(2013){Evans}, {Pont}, {Sing}, {Aigrain}, {Barstow},
  {D{\'e}sert}, {Gibson}, {Heng}, {Knutson}, \& {Lecavelier des
  Etangs}}]{Evans2013}
{Evans}, T.~M., {Pont}, F., {Sing}, D.~K., {et~al.} 2013, \apjl, 772, L16,
  \dodoi{10.1088/2041-8205/772/2/L16}

\bibitem[{{Evans} {et~al.}(2016){Evans}, {Sing}, {Wakeford}, {Nikolov},
  {Ballester}, {Drummond}, {Kataria}, {Gibson}, {Amundsen}, \&
  {Spake}}]{Evans2016}
{Evans}, T.~M., {Sing}, D.~K., {Wakeford}, H.~R., {et~al.} 2016, \apjl, 822,
  L4, \dodoi{10.3847/2041-8205/822/1/L4}

\bibitem[{{Evans} {et~al.}(2017){Evans}, {Sing}, {Kataria}, {Goyal}, {Nikolov},
  {Wakeford}, {Deming}, {Marley}, {Amundsen}, {Ballester}, {Barstow},
  {Ben-Jaffel}, {Bourrier}, {Buchhave}, {Cohen}, {Ehrenreich}, {Garc{\'\i}a
  Mu{\~n}oz}, {Henry}, {Knutson}, {Lavvas}, {Lecavelier Des Etangs}, {Lewis},
  {L{\'o}pez-Morales}, {Mandell}, {Sanz-Forcada}, {Tremblin}, \&
  {Lupu}}]{Evans2017}
{Evans}, T.~M., {Sing}, D.~K., {Kataria}, T., {et~al.} 2017, \nat, 548, 58,
  \dodoi{10.1038/nature23266}

\bibitem[{{Evans} {et~al.}(2018){Evans}, {Sing}, {Goyal}, {Nikolov}, {Marley},
  {Zahnle}, {Henry}, {Barstow}, {Alam}, {Sanz-Forcada}, {Kataria}, {Lewis},
  {Lavvas}, {Ballester}, {Ben-Jaffel}, {Blumenthal}, {Bourrier}, {Drummond},
  {Garc{\'\i}a Mu{\~n}oz}, {L{\'o}pez-Morales}, {Tremblin}, {Ehrenreich},
  {Wakeford}, {Buchhave}, {Lecavelier des Etangs}, {H{\'e}brard}, \&
  {Williamson}}]{Evans2018}
{Evans}, T.~M., {Sing}, D.~K., {Goyal}, J.~M., {et~al.} 2018, \aj, 156, 283,
  \dodoi{10.3847/1538-3881/aaebff}

\bibitem[{{Foreman-Mackey} {et~al.}(2017){Foreman-Mackey}, {Agol}, {Angus}, \&
  {Ambikasaran}}]{celerite}
{Foreman-Mackey}, D., {Agol}, E., {Angus}, R., \& {Ambikasaran}, S. 2017,
  ArXiv.
\newblock \url{https://arxiv.org/abs/1703.09710}

\bibitem[{{Fressin} {et~al.}(2013){Fressin}, {Torres}, {Charbonneau}, {Bryson},
  {Christiansen}, {Dressing}, {Jenkins}, {Walkowicz}, \&
  {Batalha}}]{Fressin2013}
{Fressin}, F., {Torres}, G., {Charbonneau}, D., {et~al.} 2013, \apj, 766, 81,
  \dodoi{10.1088/0004-637X/766/2/81}

\bibitem[{{Fu} {et~al.}(2017){Fu}, {Deming}, {Knutson}, {Madhusudhan},
  {Mandell}, \& {Fraine}}]{fu2017}
{Fu}, G., {Deming}, D., {Knutson}, H., {et~al.} 2017, \apjl, 847, L22,
  \dodoi{10.3847/2041-8213/aa8e40}

\bibitem[{{Fulton} \& {Petigura}(2017)}]{Fulton2017}
{Fulton}, B., \& {Petigura}, E. 2017, {Radvel: Radial Velocity Fitting
  Toolkit}, v0.9.1,  Zenodo, \dodoi{10.5281/zenodo.580821}

\bibitem[{{Gaia Collaboration} {et~al.}(2018){Gaia Collaboration}, {Brown},
  {Vallenari}, {Prusti}, {de Bruijne}, {Babusiaux}, {Bailer-Jones}, {Biermann},
  {Evans}, {Eyer}, {Jansen}, {Jordi}, {Klioner}, {Lammers}, {Lindegren},
  {Luri}, {Mignard}, {Panem}, {Pourbaix}, {Randich}, {Sartoretti}, {Siddiqui},
  {Soubiran}, {van Leeuwen}, {Walton}, {Arenou}, {Bastian}, {Cropper},
  {Drimmel}, {Katz}, {Lattanzi}, {Bakker}, {Cacciari}, {Casta{\~n}eda},
  {Chaoul}, {Cheek}, {De Angeli}, {Fabricius}, {Guerra}, {Holl}, {Masana},
  {Messineo}, {Mowlavi}, {Nienartowicz}, {Panuzzo}, {Portell}, {Riello},
  {Seabroke}, {Tanga}, {Th{\'e}venin}, {Gracia-Abril}, {Comoretto},
  {Garcia-Reinaldos}, {Teyssier}, {Altmann}, {Andrae}, {Audard},
  {Bellas-Velidis}, {Benson}, {Berthier}, {Blomme}, {Burgess}, {Busso},
  {Carry}, {Cellino}, {Clementini}, {Clotet}, {Creevey}, {Davidson}, {De
  Ridder}, {Delchambre}, {Dell'Oro}, {Ducourant},
  {Fern{\'a}ndez-Hern{\'a}ndez}, {Fouesneau}, {Fr{\'e}mat}, {Galluccio},
  {Garc{\'\i}a-Torres}, {Gonz{\'a}lez-N{\'u}{\~n}ez}, {Gonz{\'a}lez-Vidal},
  {Gosset}, {Guy}, {Halbwachs}, {Hambly}, {Harrison}, {Hern{\'a}ndez},
  {Hestroffer}, {Hodgkin}, {Hutton}, {Jasniewicz}, {Jean-Antoine-Piccolo},
  {Jordan}, {Korn}, {Krone-Martins}, {Lanzafame}, {Lebzelter}, {L{\"o}ffler},
  {Manteiga}, {Marrese}, {Mart{\'\i}n-Fleitas}, {Moitinho}, {Mora}, {Muinonen},
  {Osinde}, {Pancino}, {Pauwels}, {Petit}, {Recio-Blanco}, {Richards},
  {Rimoldini}, {Robin}, {Sarro}, {Siopis}, {Smith}, {Sozzetti}, {S{\"u}veges},
  {Torra}, {van Reeven}, {Abbas}, {Abreu Aramburu}, {Accart}, {Aerts},
  {Altavilla}, {{\'A}lvarez}, {Alvarez}, {Alves}, {Anderson}, {Andrei},
  {Anglada Varela}, {Antiche}, {Antoja}, {Arcay}, {Astraatmadja}, {Bach},
  {Baker}, {Balaguer-N{\'u}{\~n}ez}, {Balm}, {Barache}, {Barata}, {Barbato},
  {Barblan}, {Barklem}, {Barrado}, {Barros}, {Barstow}, {Bartholom{\'e}
  Mu{\~n}oz}, {Bassilana}, {Becciani}, {Bellazzini}, {Berihuete}, {Bertone},
  {Bianchi}, {Bienaym{\'e}}, {Blanco-Cuaresma}, {Boch}, {Boeche}, {Bombrun},
  {Borrachero}, {Bossini}, {Bouquillon}, {Bourda}, {Bragaglia}, {Bramante},
  {Breddels}, {Bressan}, {Brouillet}, {Br{\"u}semeister}, {Brugaletta},
  {Bucciarelli}, {Burlacu}, {Busonero}, {Butkevich}, {Buzzi}, {Caffau},
  {Cancelliere}, {Cannizzaro}, {Cantat-Gaudin}, {Carballo}, {Carlucci},
  {Carrasco}, {Casamiquela}, {Castellani}, {Castro-Ginard}, {Charlot},
  {Chemin}, {Chiavassa}, {Cocozza}, {Costigan}, {Cowell}, {Crifo}, {Crosta},
  {Crowley}, {Cuypers}, {Dafonte}, {Damerdji}, {Dapergolas}, {David}, {David},
  {de Laverny}, {De Luise}, {De March}, {de Martino}, {de Souza}, {de Torres},
  {Debosscher}, {del Pozo}, {Delbo}, {Delgado}, {Delgado}, {Di Matteo},
  {Diakite}, {Diener}, {Distefano}, {Dolding}, {Drazinos}, {Dur{\'a}n},
  {Edvardsson}, {Enke}, {Eriksson}, {Esquej}, {Eynard Bontemps}, {Fabre},
  {Fabrizio}, {Faigler}, {Falc{\~a}o}, {Farr{\`a}s Casas}, {Federici},
  {Fedorets}, {Fernique}, {Figueras}, {Filippi}, {Findeisen}, {Fonti},
  {Fraile}, {Fraser}, {Fr{\'e}zouls}, {Gai}, {Galleti}, {Garabato},
  {Garc{\'\i}a-Sedano}, {Garofalo}, {Garralda}, {Gavel}, {Gavras}, {Gerssen},
  {Geyer}, {Giacobbe}, {Gilmore}, {Girona}, {Giuffrida}, {Glass}, {Gomes},
  {Granvik}, {Gueguen}, {Guerrier}, {Guiraud}, {Guti{\'e}rrez-S{\'a}nchez},
  {Haigron}, {Hatzidimitriou}, {Hauser}, {Haywood}, {Heiter}, {Helmi}, {Heu},
  {Hilger}, {Hobbs}, {Hofmann}, {Holland}, {Huckle}, {Hypki}, {Icardi},
  {Jan{\ss}en}, {Jevardat de Fombelle}, {Jonker}, {Juh{\'a}sz}, {Julbe},
  {Karampelas}, {Kewley}, {Klar}, {Kochoska}, {Kohley}, {Kolenberg},
  {Kontizas}, {Kontizas}, {Koposov}, {Kordopatis}, {Kostrzewa-Rutkowska},
  {Koubsky}, {Lambert}, {Lanza}, {Lasne}, {Lavigne}, {Le Fustec}, {Le
  Poncin-Lafitte}, {Lebreton}, {Leccia}, {Leclerc}, {Lecoeur-Taibi},
  {Lenhardt}, {Leroux}, {Liao}, {Licata}, {Lindstr{\o}m}, {Lister}, {Livanou},
  {Lobel}, {L{\'o}pez}, {Managau}, {Mann}, {Mantelet}, {Marchal}, {Marchant},
  {Marconi}, {Marinoni}, {Marschalk{\'o}}, {Marshall}, {Martino}, {Marton},
  {Mary}, {Massari}, {Matijevi{\v{c}}}, {Mazeh}, {McMillan}, {Messina},
  {Michalik}, {Millar}, {Molina}, {Molinaro}, {Moln{\'a}r}, {Montegriffo},
  {Mor}, {Morbidelli}, {Morel}, {Morris}, {Mulone}, {Muraveva}, {Musella},
  {Nelemans}, {Nicastro}, {Noval}, {O'Mullane}, {Ord{\'e}novic},
  {Ord{\'o}{\~n}ez-Blanco}, {Osborne}, {Pagani}, {Pagano}, {Pailler},
  {Palacin}, {Palaversa}, {Panahi}, {Pawlak}, {Piersimoni}, {Pineau}, {Plachy},
  {Plum}, {Poggio}, {Poujoulet}, {Pr{\v{s}}a}, {Pulone}, {Racero}, {Ragaini},
  {Rambaux}, {Ramos-Lerate}, {Regibo}, {Reyl{\'e}}, {Riclet}, {Ripepi}, {Riva},
  {Rivard}, {Rixon}, {Roegiers}, {Roelens}, {Romero-G{\'o}mez}, {Rowell},
  {Royer}, {Ruiz-Dern}, {Sadowski}, {Sagrist{\`a} Sell{\'e}s}, {Sahlmann},
  {Salgado}, {Salguero}, {Sanna}, {Santana-Ros}, {Sarasso}, {Savietto},
  {Schultheis}, {Sciacca}, {Segol}, {Segovia}, {S{\'e}gransan}, {Shih},
  {Siltala}, {Silva}, {Smart}, {Smith}, {Solano}, {Solitro}, {Sordo}, {Soria
  Nieto}, {Souchay}, {Spagna}, {Spoto}, {Stampa}, {Steele},
  {Steidelm{\"u}ller}, {Stephenson}, {Stoev}, {Suess}, {Surdej}, {Szabados},
  {Szegedi-Elek}, {Tapiador}, {Taris}, {Tauran}, {Taylor}, {Teixeira},
  {Terrett}, {Teyssand ier}, {Thuillot}, {Titarenko}, {Torra Clotet}, {Turon},
  {Ulla}, {Utrilla}, {Uzzi}, {Vaillant}, {Valentini}, {Valette}, {van Elteren},
  {Van Hemelryck}, {van Leeuwen}, {Vaschetto}, {Vecchiato}, {Veljanoski},
  {Viala}, {Vicente}, {Vogt}, {von Essen}, {Voss}, {Votruba}, {Voutsinas},
  {Walmsley}, {Weiler}, {Wertz}, {Wevers}, {Wyrzykowski}, {Yoldas},
  {{\v{Z}}erjal}, {Ziaeepour}, {Zorec}, {Zschocke}, {Zucker}, {Zurbach}, \&
  {Zwitter}}]{GaiaDR2}
{Gaia Collaboration}, {Brown}, A.~G.~A., {Vallenari}, A., {et~al.} 2018, \aap,
  616, A1, \dodoi{10.1051/0004-6361/201833051}

\bibitem[{{Gibson}(2014)}]{Gibson2014}
{Gibson}, N.~P. 2014, \mnras, 445, 3401, \dodoi{10.1093/mnras/stu1975}

\bibitem[{{Gibson} {et~al.}(2012){Gibson}, {Aigrain}, {Pont}, {Sing},
  {D{\'e}sert}, {Evans}, {Henry}, {Husnoo}, \& {Knutson}}]{Gibson2012HD189733b}
{Gibson}, N.~P., {Aigrain}, S., {Pont}, F., {et~al.} 2012, \mnras, 422, 753,
  \dodoi{10.1111/j.1365-2966.2012.20655.x}

\bibitem[{{Gilliland} {et~al.}(1999){Gilliland}, {Goudfrooij}, \&
  {Kimble}}]{Gilliland1999}
{Gilliland}, R.~L., {Goudfrooij}, P., \& {Kimble}, R.~A. 1999, \pasp, 111,
  1009, \dodoi{10.1086/316407}

\bibitem[{{Goyal} {et~al.}(2019){Goyal}, {Wakeford}, {Mayne}, {Lewis},
  {Drummond}, \& {Sing}}]{Goyal2019}
{Goyal}, J.~M., {Wakeford}, H.~R., {Mayne}, N.~J., {et~al.} 2019, \mnras, 482,
  4503, \dodoi{10.1093/mnras/sty3001}

\bibitem[{Greene {et~al.}(2016)Greene, Line, Montero, Fortney, Lustig-Yaeger,
  \& Luther}]{Greene2016}
Greene, T.~P., Line, M.~R., Montero, C., {et~al.} 2016, The Astrophysical
  Journal, 817, 17

\bibitem[{{Guo} {et~al.}(2019){Guo}, {Ballard}, {Dragomir}, {Werner},
  {Livingston}, \& {Gorjian}}]{Guo2018}
{Guo}, X., {Ballard}, S., {Dragomir}, D., {et~al.} 2019, \apj, 880, 64,
  \dodoi{10.3847/1538-4357/ab24be}

\bibitem[{{Henry}(1999)}]{Henry1999}
{Henry}, G.~W. 1999, \pasp, 111, 845, \dodoi{10.1086/316388}

\bibitem[{{Henry} {et~al.}(2011){Henry}, {Howard}, {Marcy}, {Fischer}, \&
  {Johnson}}]{henry:2011}
{Henry}, G.~W., {Howard}, A.~W., {Marcy}, G.~W., {Fischer}, D.~A., \&
  {Johnson}, J.~A. 2011, ArXiv e-prints.
\newblock \doarXiv{1109.2549}

\bibitem[{Hilbert(2012)}]{WFC3_Badpixel_table}
Hilbert, B. 2012, Instrument Science Report WFC3, 10

\bibitem[{Hinkel {et~al.}(2017)Hinkel, Mamajek, Turnbull, Osby, Shkolnik,
  Smith, Klimasewski, Somers, \& Desch}]{Hinkel2017}
Hinkel, N.~R., Mamajek, E.~E., Turnbull, M.~C., {et~al.} 2017, The
  Astrophysical Journal, 848, 34

\bibitem[{{H{\"o}rst} {et~al.}(2018){H{\"o}rst}, {He}, {Lewis}, {Kempton},
  {Marley}, {Morley}, {Moses}, {Valenti}, \& {Vuitton}}]{Horst2018}
{H{\"o}rst}, S.~M., {He}, C., {Lewis}, N.~K., {et~al.} 2018, Nature Astronomy,
  2, 303, \dodoi{10.1038/s41550-018-0397-0}

\bibitem[{{Howard} {et~al.}(2010){Howard}, {Johnson}, {Marcy}, {Fischer},
  {Wright}, {Bernat}, {Henry}, {Peek}, {Isaacson}, {Apps}, {Endl}, {Cochran},
  {Valenti}, {Anderson}, \& {Piskunov}}]{Howard2010}
{Howard}, A.~W., {Johnson}, J.~A., {Marcy}, G.~W., {et~al.} 2010, \apj, 721,
  1467, \dodoi{10.1088/0004-637X/721/2/1467}

\bibitem[{{Howard} {et~al.}(2011){Howard}, {Johnson}, {Marcy}, {Fischer},
  {Wright}, {Henry}, {Isaacson}, {Valenti}, {Anderson}, \&
  {Piskunov}}]{Howard2011}
---. 2011, \apj, 730, 10, \dodoi{10.1088/0004-637X/730/1/10}

\bibitem[{{Howard} {et~al.}(2012){Howard}, {Marcy}, {Bryson}, {Jenkins},
  {Rowe}, {Batalha}, {Borucki}, {Koch}, {Dunham}, {Gautier}, {Van Cleve},
  {Cochran}, {Latham}, {Lissauer}, {Torres}, {Brown}, {Gilliland}, {Buchhave},
  {Caldwell}, {Christensen-Dalsgaard}, {Ciardi}, {Fressin}, {Haas}, {Howell},
  {Kjeldsen}, {Seager}, {Rogers}, {Sasselov}, {Steffen}, {Basri},
  {Charbonneau}, {Christiansen}, {Clarke}, {Dupree}, {Fabrycky}, {Fischer},
  {Ford}, {Fortney}, {Tarter}, {Girouard}, {Holman}, {Johnson}, {Klaus},
  {Machalek}, {Moorhead}, {Morehead}, {Ragozzine}, {Tenenbaum}, {Twicken},
  {Quinn}, {Isaacson}, {Shporer}, {Lucas}, {Walkowicz}, {Welsh}, {Boss},
  {Devore}, {Gould}, {Smith}, {Morris}, {Prsa}, {Morton}, {Still}, {Thompson},
  {Mullally}, {Endl}, \& {MacQueen}}]{Howard2012}
{Howard}, A.~W., {Marcy}, G.~W., {Bryson}, S.~T., {et~al.} 2012, \apjs, 201,
  15, \dodoi{10.1088/0067-0049/201/2/15}

\bibitem[{{Isaacson} \& {Fischer}(2010)}]{Isaacson2010}
{Isaacson}, H., \& {Fischer}, D. 2010, \apj, 725, 875,
  \dodoi{10.1088/0004-637X/725/1/875}

\bibitem[{{Kempton} {et~al.}(2017){Kempton}, {Lupu}, {Owusu-Asare}, {Slough},
  \& {Cale}}]{Exo_transmit_Kempton2017}
{Kempton}, E.~M.-R., {Lupu}, R., {Owusu-Asare}, A., {Slough}, P., \& {Cale}, B.
  2017, \pasp, 129, 044402, \dodoi{10.1088/1538-3873/aa61ef}

\bibitem[{Kempton {et~al.}(2018)Kempton, Bean, Louie, Deming, Koll, Mansfield,
  Christiansen, L{\'o}pez-Morales, Swain, Zellem, {et~al.}}]{Kempton2018}
Kempton, E. M.-R., Bean, J.~L., Louie, D.~R., {et~al.} 2018, Publications of
  the Astronomical Society of the Pacific, 130, 114401

\bibitem[{{Knutson} {et~al.}(2014){Knutson}, {Dragomir}, {Kreidberg},
  {Kempton}, {McCullough}, {Fortney}, {Bean}, {Gillon}, {Homeier}, \&
  {Howard}}]{Knutson2014}
{Knutson}, H.~A., {Dragomir}, D., {Kreidberg}, L., {et~al.} 2014, \apj, 794,
  155, \dodoi{10.1088/0004-637X/794/2/155}

\bibitem[{{Kreidberg}(2015)}]{Kreidberg2015}
{Kreidberg}, L. 2015, \pasp, 127, 1161, \dodoi{10.1086/683602}

\bibitem[{{Kreidberg} {et~al.}(2014{\natexlab{a}}){Kreidberg}, {Bean},
  {D{\'e}sert}, {Benneke}, {Deming}, {Stevenson}, {Seager}, {Berta-Thompson},
  {Seifahrt}, \& {Homeier}}]{Kreidberg2014a}
{Kreidberg}, L., {Bean}, J.~L., {D{\'e}sert}, J.-M., {et~al.}
  2014{\natexlab{a}}, \nat, 505, 69, \dodoi{10.1038/nature12888}

\bibitem[{{Kreidberg} {et~al.}(2014{\natexlab{b}}){Kreidberg}, {Bean},
  {D{\'e}sert}, {Line}, {Fortney}, {Madhusudhan}, {Stevenson}, {Showman},
  {Charbonneau}, {McCullough}, {Seager}, {Burrows}, {Henry}, {Williamson},
  {Kataria}, \& {Homeier}}]{Kreidberg2014b}
---. 2014{\natexlab{b}}, \apjl, 793, L27, \dodoi{10.1088/2041-8205/793/2/L27}

\bibitem[{{Line} {et~al.}(2013){Line}, {Knutson}, {Deming}, {Wilkins}, \&
  {Desert}}]{Line2013}
{Line}, M.~R., {Knutson}, H., {Deming}, D., {Wilkins}, A., \& {Desert}, J.-M.
  2013, \apj, 778, 183, \dodoi{10.1088/0004-637X/778/2/183}

\bibitem[{{Lothringer} {et~al.}(2018){Lothringer}, {Benneke}, {Crossfield},
  {Henry}, {Morley}, {Dragomir}, {Barman}, {Knutson}, {Kempton}, {Fortney},
  {McCullough}, \& {Howard}}]{Lothringer2018}
{Lothringer}, J.~D., {Benneke}, B., {Crossfield}, I. J.~M., {et~al.} 2018, \aj,
  155, 66, \dodoi{10.3847/1538-3881/aaa008}

\bibitem[{{Louie} {et~al.}(2018){Louie}, {Deming}, {Albert}, {Bouma}, {Bean},
  \& {Lopez-Morales}}]{Louie2018}
{Louie}, D.~R., {Deming}, D., {Albert}, L., {et~al.} 2018, \pasp, 130, 044401,
  \dodoi{10.1088/1538-3873/aaa87b}

\bibitem[{{Lovis} {et~al.}(2011){Lovis}, {Dumusque}, {Santos}, {Bouchy},
  {Mayor}, {Pepe}, {Queloz}, {S{\'e}gransan}, \& {Udry}}]{lovis:2011}
{Lovis}, C., {Dumusque}, X., {Santos}, N.~C., {et~al.} 2011, arXiv e-prints,
  arXiv:1107.5325.
\newblock \doarXiv{1107.5325}

\bibitem[{Loyd {et~al.}(2016)Loyd, France, Youngblood, Schneider, Brown, Hu,
  Linsky, Froning, Redfield, Rugheimer, {et~al.}}]{Loyd2016}
Loyd, R.~P., France, K., Youngblood, A., {et~al.} 2016, The Astrophysical
  Journal, 824, 102

\bibitem[{{Mann} {et~al.}(2016){Mann}, {Gaidos}, {Mace}, {Johnson}, {Bowler},
  {LaCourse}, {Jacobs}, {Vanderburg}, {Kraus}, {Kaplan}, \& {Jaffe}}]{Mann2016}
{Mann}, A.~W., {Gaidos}, E., {Mace}, G.~N., {et~al.} 2016, \apj, 818, 46,
  \dodoi{10.3847/0004-637X/818/1/46}

\bibitem[{{Matthews} {et~al.}(2004){Matthews}, {Kuschnig}, {Guenther},
  {Walker}, {Moffat}, {Rucinski}, {Sasselov}, \& {Weiss}}]{Mat04}
{Matthews}, J.~M., {Kuschnig}, R., {Guenther}, D.~B., {et~al.} 2004, Nature,
  430, 51, \dodoi{10.1038/nature02671}

\bibitem[{{McCullough} \& {MacKenty}(2012)}]{McCullough2012}
{McCullough}, P., \& {MacKenty}, J. 2012, {Considerations for using Spatial
  Scans with WFC3}, Tech. rep.

\bibitem[{{\"O}berg {et~al.}(2011){\"O}berg, Murray-Clay, \&
  Bergin}]{Oberg2011}
{\"O}berg, K.~I., Murray-Clay, R., \& Bergin, E.~A. 2011, The Astrophysical
  Journal Letters, 743, L16

\bibitem[{{Rackham} {et~al.}(2018){Rackham}, {Apai}, \&
  {Giampapa}}]{Rackham2018}
{Rackham}, B.~V., {Apai}, D., \& {Giampapa}, M.~S. 2018, \apj, 853, 122,
  \dodoi{10.3847/1538-4357/aaa08c}

\bibitem[{{Rackham} {et~al.}(2019){Rackham}, {Apai}, \&
  {Giampapa}}]{Rackham2019}
---. 2019, \aj, 157, 96, \dodoi{10.3847/1538-3881/aaf892}

\bibitem[{{Radovan} {et~al.}(2014){Radovan}, {Lanclos}, {Holden}, {Kibrick},
  {Allen}, {Deich}, {Rivera}, {Burt}, {Fulton}, {Butler}, \&
  {Vogt}}]{Radovan2014}
{Radovan}, M.~V., {Lanclos}, K., {Holden}, B.~P., {et~al.} 2014, in \procspie,
  Vol. 9145, Ground-based and Airborne Telescopes V, 91452B,
  \dodoi{10.1117/12.2057310}

\bibitem[{{Rodriguez} {et~al.}(2017){Rodriguez}, {Zhou}, {Vanderburg},
  {Eastman}, {Kreidberg}, {Cargile}, {Bieryla}, {Latham}, {Irwin}, {Mayo},
  {Calkins}, {Esquerdo}, \& {Mink}}]{Rodriguez2017}
{Rodriguez}, J.~E., {Zhou}, G., {Vanderburg}, A., {et~al.} 2017, \aj, 153, 256,
  \dodoi{10.3847/1538-3881/aa6dfb}

\bibitem[{{Rowe} {et~al.}(2006){Rowe}, {Matthews}, {Seager}, {Kuschnig},
  {Guenther}, {Moffat}, {Rucinski}, {Sasselov}, {Walker}, \& {Weiss}}]{Row06}
{Rowe}, J.~F., {Matthews}, J.~M., {Seager}, S., {et~al.} 2006, \apj, 646, 1241,
  \dodoi{10.1086/504252}

\bibitem[{{Rowe} {et~al.}(2008){Rowe}, {Matthews}, {Seager}, {Miller-Ricci},
  {Sasselov}, {Kuschnig}, {Guenther}, {Moffat}, {Rucinski}, {Walker}, \&
  {Weiss}}]{Row08}
---. 2008, \apj, 689, 1345, \dodoi{10.1086/591835}

\bibitem[{Rugheimer {et~al.}(2013)Rugheimer, Kaltenegger, Zsom, Segura, \&
  Sasselov}]{Rugheimer2013}
Rugheimer, S., Kaltenegger, L., Zsom, A., Segura, A., \& Sasselov, D. 2013,
  Astrobiology, 13, 251

\bibitem[{{Southworth} {et~al.}(2017){Southworth}, {Mancini}, {Madhusudhan},
  {Molli{\`e}re}, {Ciceri}, \& {Henning}}]{Southworth2017}
{Southworth}, J., {Mancini}, L., {Madhusudhan}, N., {et~al.} 2017, \aj, 153,
  191, \dodoi{10.3847/1538-3881/aa6477}

\bibitem[{{Su{\'a}rez Mascare{\~n}o} {et~al.}(2016){Su{\'a}rez Mascare{\~n}o},
  {Rebolo}, \& {Gonz{\'a}lez Hern{\'a}ndez}}]{mascareno:2016}
{Su{\'a}rez Mascare{\~n}o}, A., {Rebolo}, R., \& {Gonz{\'a}lez Hern{\'a}ndez},
  J.~I. 2016, \aap, 595, A12, \dodoi{10.1051/0004-6361/201628586}

\bibitem[{{Tremblin} {et~al.}(2015){Tremblin}, {Amundsen}, {Mourier},
  {Baraffe}, {Chabrier}, {Drummond}, {Homeier}, \& {Venot}}]{Tremblin2015}
{Tremblin}, P., {Amundsen}, D.~S., {Mourier}, P., {et~al.} 2015, \apjl, 804,
  L17, \dodoi{10.1088/2041-8205/804/1/L17}

\bibitem[{{Tsiaras} {et~al.}(2016){Tsiaras}, {Rocchetto}, {Waldmann}, {Venot},
  {Varley}, {Morello}, {Damiano}, {Tinetti}, {Barton}, {Yurchenko}, \&
  {Tennyson}}]{Tsiaras2016}
{Tsiaras}, A., {Rocchetto}, M., {Waldmann}, I.~P., {et~al.} 2016, \apj, 820,
  99, \dodoi{10.3847/0004-637X/820/2/99}

\bibitem[{{Tsiaras} {et~al.}(2018){Tsiaras}, {Waldmann}, {Zingales},
  {Rocchetto}, {Morello}, {Damiano}, {Karpouzas}, {Tinetti}, {McKemmish},
  {Tennyson}, \& {Yurchenko}}]{Tsiaras2018}
{Tsiaras}, A., {Waldmann}, I.~P., {Zingales}, T., {et~al.} 2018, \aj, 155, 156,
  \dodoi{10.3847/1538-3881/aaaf75}

\bibitem[{{Van Grootel} {et~al.}(2014){Van Grootel}, {Gillon}, {Valencia},
  {Madhusudhan}, {Dragomir}, {Howe}, {Burrows}, {Demory}, {Deming},
  {Ehrenreich}, {Lovis}, {Mayor}, {Pepe}, {Queloz}, {Scuflaire}, {Seager},
  {Segransan}, \& {Udry}}]{VanGrootel2014}
{Van Grootel}, V., {Gillon}, M., {Valencia}, D., {et~al.} 2014, \apj, 786, 2,
  \dodoi{10.1088/0004-637X/786/1/2}

\bibitem[{{Vogt} {et~al.}(1994){Vogt}, {Allen}, {Bigelow}, {Bresee}, {Brown},
  {Cantrall}, {Conrad}, {Couture}, {Delaney}, {Epps}, {Hilyard}, {Hilyard},
  {Horn}, {Jern}, {Kanto}, {Keane}, {Kibrick}, {Lewis}, {Osborne},
  {Pardeilhan}, {Pfister}, {Ricketts}, {Robinson}, {Stover}, {Tucker}, {Ward},
  \& {Wei}}]{Vogt1994}
{Vogt}, S.~S., {Allen}, S.~L., {Bigelow}, B.~C., {et~al.} 1994, in \procspie,
  Vol. 2198, Instrumentation in Astronomy VIII, ed. D.~L. {Crawford} \& E.~R.
  {Craine}, 362, \dodoi{10.1117/12.176725}

\bibitem[{{Vogt} {et~al.}(2014){Vogt}, {Radovan}, {Kibrick}, {Butler},
  {Alcott}, {Allen}, {Arriagada}, {Bolte}, {Burt}, {Cabak}, {Chloros},
  {Cowley}, {Deich}, {Dupraw}, {Earthman}, {Epps}, {Faber}, {Fischer}, {Gates},
  {Hilyard}, {Holden}, {Johnston}, {Keiser}, {Kanto}, {Katsuki}, {Laiterman},
  {Lanclos}, {Laughlin}, {Lewis}, {Lockwood}, {Lynam}, {Marcy}, {McLean},
  {Miller}, {Misch}, {Peck}, {Pfister}, {Phillips}, {Rivera}, {Sandford},
  {Saylor}, {Stover}, {Thompson}, {Walp}, {Ward}, {Wareham}, {Wei}, \&
  {Wright}}]{Vogt2014}
{Vogt}, S.~S., {Radovan}, M., {Kibrick}, R., {et~al.} 2014, \pasp, 126, 359,
  \dodoi{10.1086/676120}

\bibitem[{{Wakeford} {et~al.}(2016){Wakeford}, {Sing}, {Evans}, {Deming}, \&
  {Mandell}}]{Wakeford2016}
{Wakeford}, H.~R., {Sing}, D.~K., {Evans}, T., {Deming}, D., \& {Mandell}, A.
  2016, \apj, 819, 10, \dodoi{10.3847/0004-637X/819/1/10}

\bibitem[{{Wakeford} {et~al.}(2013){Wakeford}, {Sing}, {Deming}, {Gibson},
  {Fortney}, {Burrows}, {Ballester}, {Nikolov}, {Aigrain}, {Henry}, {Knutson},
  {Lecavelier des Etangs}, {Pont}, {Showman}, {Vidal-Madjar}, \&
  {Zahnle}}]{Wakeford2013}
{Wakeford}, H.~R., {Sing}, D.~K., {Deming}, D., {et~al.} 2013, \mnras, 435,
  3481, \dodoi{10.1093/mnras/stt1536}

\bibitem[{{Walker} {et~al.}(2003){Walker}, {Matthews}, {Kuschnig}, {Johnson},
  {Rucinski}, {Pazder}, {Burley}, {Walker}, {Skaret}, {Zee}, {Grocott},
  {Carroll}, {Sinclair}, {Sturgeon}, \& {Harron}}]{Wal03}
{Walker}, G., {Matthews}, J., {Kuschnig}, R., {et~al.} 2003, \pasp, 115, 1023,
  \dodoi{10.1086/377358}

\bibitem[{{Zhang} {et~al.}(2019){Zhang}, {Chachan}, {Kempton}, \&
  {Knutson}}]{PLATON_Zhang2019}
{Zhang}, M., {Chachan}, Y., {Kempton}, E. M.~R., \& {Knutson}, H.~A. 2019,
  \pasp, 131, 034501, \dodoi{10.1088/1538-3873/aaf5ad}

\bibitem[{{Zhou} {et~al.}(2017){Zhou}, {Apai}, {Lew}, \&
  {Schneider}}]{Zhou2017}
{Zhou}, Y., {Apai}, D., {Lew}, B.~W.~P., \& {Schneider}, G. 2017, \aj, 153,
  243, \dodoi{10.3847/1538-3881/aa6481}

\end{thebibliography}

\begin{deluxetable*}{ccccccc}
\renewcommand*{\arraystretch}{1.2}
\tablecaption{\HL{Summary of the Stellar and Planet Parameter Values of the HD~97658 System}\label{tab:summary}}
\tablehead{
  \colhead{Parameter} & 
  \colhead{Symbol} & 
  \colhead{Value} & 
  \colhead{Unit} &
  \colhead{Source}
}
\startdata
\sidehead{\bf{Stellar Parameters}}
  Stellar Mass & $M_*$ & 0.77$\pm$ 0.05 & $M_{\odot}$ & \cite{VanGrootel2014} \\
  Stellar Radius & $R_*$ & $0.746_{-0.034}^{+0.016}$ & $R_{\odot}$ & \citet{GaiaDR2} \\
  Effective Temperature & $T_{\rm eff}$ & $5192_{-55}^{+122}$ & K & \citet{GaiaDR2} \\
  Luminosity & $L_*$ & 0.4384$\pm$ 0.0007 & $L_{\odot}$ & \citet{GaiaDR2} \\
  Activity Cycle & $P_{\rm activity}$ & $3652_{-120}^{+130}$ & days & this work \\
  Rotation Period & $P_{\rm rot}$ & 34$\pm$ 2 & days & this work \\
\hline
\sidehead{\bf{Planet Parameters}}
 Ratio of Planet to Stellar Radius & $R_{\rm p}/R_*$ & $0.0283_{-0.0004}^{+0.0002}$ &  & this work \\
 Planet Radius & $R_{\rm p}$ & $2.303^{+0.052}_{-0.110}$ & $R_\oplus$ & this work \\
 Semi-major Axis Ratio & $a/R_{\rm *}$ & 26.7$\pm$ 0.4 &  & this work \\
 Orbital Period & $P$ & 9.489295$\pm$ 0.000005 & days & this work \\
 Mid-transit Time & $T_0$ & 2456361.80690$\pm$ 0.00038 & BJD & this work \\
 Eccentricity & $e$ & $0.030^{+0.034}_{-0.021}$ &  & this work \\
 Inclination & $i$ & 89.6$\pm$ 0.1 &  & this work \\ 
 Planet Mass & $M_{\rm p}\sin i$ & $7.81_{-0.44}^{+0.55}$ & $M_{\oplus}$ & this work \\
 Planet Density & $\rho_{\rm p}$ & $3.78_{-0.51}^{+0.61}$ & g cm$^{-3}$ & this work \\
 Equilibrium Temperature & $T_{\rm eq}$ & $809_{-142}^{+103}$ & K & this work \\
 Atmospheric Metallicity & $\log{Z}$ & $2.4_{-0.4}^{+0.3}$ &  & this work \\
 Carbon to Oxygen Ratio & C/O & 1.3$\pm$ 0.4 &  & this work \\
 Cloud-top Pressure & $\log{P(\rm bar)}$ & $0.15_{-1.81}^{+1.89}$ &  & this work \\
\enddata
\end{deluxetable*}

\begin{table*}
\renewcommand*{\arraystretch}{1.2}
    \caption{TSM of confirmed planets with $1~R_{\oplus}<R_{\rm p}<4~R{\oplus}$, and cooler than 1000~K\footnote{(The first twenty rows of this table is shown here, and the full table will be available online.)}}
    \begin{tabularx}{\textwidth}{cnnnnnnnn}
    \hline\hline
    Planet Name & $R_{\rm s}$ & $T_{\rm eff}$ & $J$ mag & $R_{\rm p}(R_{\oplus})$ & $M_{\rm p}(M_{\oplus})$ & $T_{\rm eq}$ & $P$ & TSM \\
    \hline
   GJ 1214 b &   0.22 &  3026 &  9.750 & 2.85 &  6.26 & 576 &  1.580405 & 630.0 \\
 LP 791-18 c &   0.17 &  2960 & 11.559 & 2.31 &  5.96 & 343 &  4.989963 & 153.2 \\
     K2-25 b &   0.29 &  3180 & 11.303 & 3.42 & 11.67 & 478 &  3.484552 & 138.1 \\
   TOI-270 c &   0.38 &  3386 &  9.099 & 2.42 &  6.46 & 463 &  5.660172 & 136.7 \\
  HD 97658 b &   0.74 &  5175 &  6.203 & 2.35 &  9.54 & 738 &  9.490900 & 135.7 \\
  TOI-1130 b &   0.69 &  4250 &  9.055 & 3.65 & 13.00 & 812 &  4.066499 & 126.6 \\
   GJ 9827 d &   0.60 &  4340 &  7.984 & 2.02 &  4.04 & 685 &  6.201470 & 125.4 \\
    G 9-40 b &   0.31 &  3348 & 10.058 & 2.03 &  4.78 & 458 &  5.746007 & 103.8 \\
   TOI-270 d &   0.38 &  3386 &  9.099 & 2.13 &  5.19 & 371 & 11.380140 &  92.8 \\
     K2-36 c &   0.72 &  4916 & 10.034 & 3.20 &  7.80 & 865 &  5.340888 &  87.9 \\
     K2-28 b &   0.29 &  3214 & 11.695 & 2.32 &  6.01 & 570 &  2.260455 &  82.7 \\
   HD 3167 c &   0.87 &  5528 &  7.548 & 2.86 &  8.56 & 579 & 29.838320 &  82.7 \\
  Wolf 503 b &   0.69 &  4716 &  8.324 & 2.03 &  4.78 & 790 &  6.001180 &  80.3 \\
     K2-55 b &   0.63 &  4456 & 11.230 & 3.82 & 14.03 & 913 &  2.849258 &  66.5 \\
    K2-136 c &   0.66 &  4499 &  9.096 & 2.91 &  8.85 & 511 & 17.307137 &  63.8 \\
   TOI-125 c &   0.85 &  5320 &  9.466 & 2.76 &  6.63 & 828 &  9.150590 &  59.4 \\
  HD 15337 c &   0.86 &  5125 &  7.553 & 2.39 &  8.11 & 643 & 17.178400 &  57.7 \\
    GJ 143 b &   0.69 &  4640 &  6.081 & 2.62 & 22.70 & 424 & 35.612530 &  54.5 \\
      K2-3 b &   0.56 &  3896 &  9.421 & 2.17 &  5.38 & 506 & 10.054490 &  51.6 \\
Kepler-445 c &   0.21 &  3157 & 13.542 & 2.47 &  6.66 & 391 &  4.871229 &  50.0 \\
    \hline
    \end{tabularx}
    
    \label{tab:TSM}
\bigskip
\end{table*}

\end{document}